\documentclass[traditabstract]{aa} 
\usepackage{graphicx}
\usepackage{subfig}
\usepackage{txfonts}
\usepackage{natbib}
\bibpunct{(}{)}{;}{a}{}{,} 
\usepackage{enumitem}
\usepackage{multirow}
\usepackage{hyperref}
\usepackage{color}
\usepackage{verbatim}

\begin{document}
\title{Controversial Age Spreads from the Main Sequence Turn-Off and Red Clump in Intermediate-Age Clusters in the LMC   
   \thanks{Based on observations made with the NASA/ESA Hubble Space Telescope, and obtained from the Hubble Legacy Archive, which is a collaboration between the Space Telescope Science Institute (STScI/NASA), the Space Telescope European Coordinating Facility (ST-ECF/ESA) and the Canadian Astronomy Data Centre (CADC/NRC/CSA).}}

   \author{F. Niederhofer 
   			\inst{1,2}   			
   			\and
   			N. Bastian
   			\inst{3}
   			\and
   			V. Kozhurina-Platais
   			\inst{4}
   			\and
   			M. Hilker
   			\inst{1,5} 
   			\and
   			S. E. de Mink
   			\inst{6}
   			\and
   			I. Cabrera-Ziri
   			\inst{3,5}  
   			\and 
   			C. Li
   			\inst{7,8,9}	
   			\and 
   			B. Ercolano
   			\inst{1,2}		
            }

   \institute{Excellence Cluster Origin and Structure of the Universe, Boltzmannstr. 2, D-85748 Garching bei M\"unchen, Germany\\
   \email{niederhofer@usm.lmu.de}
          \and
  Universit\"ats-Sternwarte M\"unchen, Scheinerstra\ss e 1, D-81679 M\"unchen, Germany
          \and
   Astrophysics Research Institute, Liverpool John Moores University, 146 Brownlow Hill, Liverpool L3 5RF, UK
           \and
   Space Telescope Science Institute, 3700 San Martin Drive, Baltimore, MD 21218, USA
           \and
   European Southern Observatory, Karl-Schwarzschild-Stra\ss e 2, D-85748 Garching bei M\"unchen, Germany
           \and
   Astronomical Institute Anton Pannekoek, Amsterdam University,
Science Park 904, 1098 XH, Amsterdam, The Netherlands
         \and
Purple Mountain Observatory, Chinese Academy of Sciences, Nanjing, 210008, China
       \and
Kavli Institute for Astronomy and Astrophysics, Peking University, Beijing, 100871, China
       \and
Department of Astronomy, Peking University, Beijing, 100871, China          
          }


  \abstract
{
Most star clusters at an intermediate age (1-2 Gyr) in the Large and Small Magellanic Clouds show a puzzling feature in their color-magnitude diagrams (CMD) that is not in agreement with a simple stellar population. The main sequence turn-off of these clusters is much broader than would be expected from photometric uncertainties. One interpretation of this feature is that age spreads of the order 200-500 Myr exist within individual clusters, although this interpretation is highly debated. Such large age spreads should affect other parts of the CMD, which are sensitive to age, as well. In this study, we analyze the CMDs of a sample of 12 intermediate-age clusters in the Large Magellanic Cloud that all show an extended turn-off using archival optical data taken with the Hubble Space Telescope. 
We fit the star formation history of the turn-off region and the red clump region independently with two different theoretical isochrone models. 
We find that in most of the cases, the age spreads inferred from the red clumps are smaller than the ones resulting from the turn-off region. However, the age ranges resulting from the red clump region are broader than would be expected for a single age. Only two out of 12 clusters in our sample show a red clump which seems to be consistent with a single age. As our results are not unambiguous, we can not ultimately tell if the extended main sequence turn-off feature is due to an age spread, or not, by fitting the star formation histories to the red clump regions.  
However, we find that the width of the extended main sequence turn-off feature is correlated with the age of the clusters in a way
which would be unexplained in the "age spread" interpretation, but which may be expected if stellar rotation is the cause of the spread at the turn-off.
}

   \keywords{galaxies: star clusters: general - galaxies: individual: LMC - Hertzsprung-Russel and C-M diagrams - stars: evolution}
\titlerunning{MSTO and RC}
   \maketitle
%
\section{Introduction\label{sec:intro}}
The traditional view of stellar clusters is that all stars in a cluster are born at the same time and have the same metallicity. This makes them ideal for stellar evolutionary models as all stars lie on a single isochrone in the color-magnitude diagram (CMD). However, high precision photometric and spectroscopic data of many clusters reveal features that are not in agreement with a simple stellar population (SSP). In the CMDs of globular clusters (GCs), broadened or double main sequences (MS) have been found that are caused by chemical anomalies (e.g. \citealt{Gratton12, Piotto12, Milone12a, Milone13a}). Different scenarios have been put forward to explain these abundance spreads inside GCs. The models propose either self-enrichment of the cluster caused by multiple bursts of star formation (e.g. \citealt{D'Ercole08, Decressin09, deMink09, Conroy11}) or enrichment of low mass pre-MS stars by the ejecta of massive stars of the same generation \citep{Bastian13a}. However, all models proposed so far have severe shortcomings (e.g., \citealt{Larsen12, Cabrera-Ziri15, Bastian15}).

Additionally, a common feature in intermediate-age (1 - 2~Gyr) clusters in the Large and Small Magellanic Clouds (LMC and SMC) is a main sequence turn-off (MSTO) that is more extended than would be expected for an SSP (e.g. \citealt{MackeyNielsen07, Mackey08, Milone09, Goudfrooij09, Goudfrooij11a, Goudfrooij11b, Girardi13,Piatti13}). Many studies have interpreted this extended MSTO feature as an age spread of the order of 200 - 500 Myr inside the clusters (e.g. \citealt{Goudfrooij09, Goudfrooij11a, Goudfrooij11b, Rubele13, Correnti14}). However, the interpretation of star formation lasting for several hundred Myr is highly debated.
A natural consequence of a prolonged star formation history (SFH) would be a chemical enrichment of heavier elements in the younger generation of stars. To date, no significant abundance spread among stars in clusters that show an extended MSTO has been found (e.g. \citealt{Mucciarelli08, Mucciarelli11, Mucciarelli14}; Mackey et al., in prep.). 

To address the question of age spreads in intermediate-age massive LMC and SMC clusters, some studies have aimed to search for large age spreads or signs of ongoing star formation in young ($<$1~Gyr) massive clusters. \citet{Bastian13b} analyzed a sample of 130 galactic and extragalactic young (10~Myr - 1~Gyr) and massive ($10^4$ - $10^8$ $\mathrm{M_{\odot}}$) clusters and did not find evidence for active star formation (see also \citealt{Peacock13}). \citet{BastianSilva13} fitted the resolved stellar population of NGC 1856 and NGC 1866, two young (180~Myr and 280~Myr) clusters in the LMC and concluded that they are compatible with an SSP. \citet{Niederhofer15a} continued this study by analyzing the CMDs of eight more young massive LMC clusters with ages between 20~Myr and 1~Gyr. They found no evidence for a significant age spread in any cluster of their sample. All clusters studied by \citet{BastianSilva13} and \citet{Niederhofer15a} have properties similar to the extended MSTO intermediate-age clusters (cf. Figure 6 in \citealt{Niederhofer15a}). 
\citet{Cabrera-Ziri14} analyzed the integrated spectrum of NGC~34 cluster 1, a $\sim$100 Myr, 10$^7$ $\mathrm{M_{\odot}}$ cluster, and did not find evidence for multiple star forming bursts or an extended SFH.

\citet{Goudfrooij11b,Goudfrooij14} put forward a model to explain how a prolonged period of star formation could have happened in intermediate-age LMC clusters. The authors propose a formation of a cluster that takes several steps: In the first step a cluster that is much more massive than observed today forms within a single burst of star formation. This cluster is massive enough that it can retain the ejecta from evolved stars and is also able to accrete pristine gas from its surrounding that is mixed with the processed gas from the ejecta. In the next step, a second generation of stars is formed within the cluster out of the mixed gas. This second burst of star formation lasts for 300 to 600 Myr. In the meantime the cluster looses almost all of its first generation stars (which make up to $\sim$90 \% of the mass of the cluster). What is observed today is a cluster that is almost only composed of the second generation of stars that have age differences of the order 300 to 600 Myr and contains only $\sim$10 \% of its initial mass. 

In order to form subsequent generations of stars, clusters would need to re-accrete material from their surroundings.  However, it is currently unclear how a cluster would actually do this, as young massive clusters are observed to be gas free from an early ($<$1-3 Myr) age and remain so \citep{Hollyhead15,BastianStrader14}. Additionally, \citet{Cabrera-Ziri15} searched for gas in three massive ($>10^6 \mathrm{M_{\odot}}$) clusters in the Antennae merging galaxies with ages between 50-200 Myr, and did not detect any, calling into doubt whether young massive clusters can host the necessary gas/dust reservoirs to form further generations of stars.

To explain the extended MSTOs, alternative scenarios have been suggested. \citet{BastianDeMink09} proposed stellar rotation as being the cause for this phenomenon. This interpretation was later called into question by \citet{Girardi11} based on calculations of isochrones for moderately rotating stars. They found that the longer lifetimes of rotating stars may cancel the effect of stellar structure changes caused by rotation. However, \citet{Yang13} found that in fact rotation is able to explain the extended MSTO depending on the efficiency of rotational mixing. 
For their analysis they used the Yale Rotating Evolution Code (YREC,  \citealt{Pinsonneault89, YangBi07}).

If age spreads are the cause of the extended MSTO feature, it would be expected to affect other parts of the CMD that are sensitive to age, as well. 
\citet{Li14} analyzed the sub-giant branch (SGB) of NGC 1651, a $\sim$2 Gyr old cluster in the LMC that shows an extended MSTO. They showed that the SGB stars mostly follow the youngest isochrone that covers the spread in the MSTO (1.74 Gyr) with a spread in the SGB of at most 80 Myr. This spread is much smaller as would be inferred from the MSTO region ($\sim$450 Myr). \citet{Li14} therefore concluded that the extended MSTO is not due to a prolonged star formation. \citet{BastianNiederhofer15} extended the study of \citet{Li14} to NGC 1806 and NGC 1846, two $\sim$1.4~Gyr old LMC clusters, which also show an extended MSTO. They used similar techniques as \citet{Li14} but also took into account the morphology of the red clump. Their results are that the SGB and red clump morphology for both clusters is consistent with an SSP that follows the youngest isochrone through the MSTO region, in agreement with what was found by \citet{Li14} for NGC~1651. 

In this study we analyze the MSTO and red clump regions of a sample of 12 intermediate-age LMC clusters in a self-consistent way. We fit the SFH of the MSTO and the red clump region independently using the SFH code StarFISH \citep{Harris01}.

The paper is organized as follows: In Section \ref{sec:obs} we describe our data sample, the observations, reduction and further processing of the data. Section \ref{sec:models} we introduce our models and methods that we use for our analysis. In Section \ref{sec:res} we present the results. In Section \ref{sec:goudfrooij15} we analyze the extended MSTO widths as a function of the age of the cluster.
Final conclusions are drawn in Section~\ref{sec:disc}. 
\section{Observations and Data Processing\label{sec:obs}}

\subsection{The Data Sample\label{data}}

The 12 LMC clusters in our total sample were observed with the HST Wide Field Channel at the Advanced Camera for Surveys (ASC/WFC) or the Ultraviolet and VIsible (UVIS) channel of the Wide Field Camera 3 (WFC3) within the programs \mbox{GO-9891} (PI: G. Gilmore), \mbox{GO-10595} (PI : P.Goudfrooij) and \mbox{GO-12257} (PI: L. Girardi) respectively. Both, the ACS/WFC and WFC3/UVIS are composed of two CCD chips with 4096~$\times$~2048 pixels and 4096~$\times$~2051 pixels in size with a gap of about 50 pixels in between. The pixel-scale of ACS/WFC
is $\sim$0.05$\arcsec$ witch gives a field-of-view of 202$\arcsec$~$\times$~202$\arcsec$. The pixel-scale of WFC3/UVIS is $\sim$0.04$\arcsec$ resulting in a smaller total field-of-view of 162$\arcsec$~$\times$~162$\arcsec$. The observations
in the programs \mbox{GO-9891} and \mbox{GO-10595} were taken 
through the F435W, F555W and F814W ACS/WFC filters and the data in run GO-12257 was observed through the F475W and F814W UVIS filters.  

\paragraph{The First Data Set:}
The photometry of two clusters NGC~1846 and NGC~1806 are combined from GO-9891 (PI: G. Gilmore) and GO-10595
(PI: P. Goudfrooij) and have been observed with the ACS/WFC. They are provided by \citet{Milone09}. We made use of the already reduced and calibrated catalogs for our analysis. The full description of the observations and the photometric reduction is described in detail in \citet{Milone09}. The photometry  of these clusters has already been cleaned for contamination of field stars, using a statistical subtraction method (see Section \ref{sec:fieldstar} for details). 

\paragraph{The Second Data Set:}
Our second data set includes the clusters Hodge 2,  
NGC~1651, NGC~1718, NGC~2173, NGC~2203 and NGC~2213 which have been observed with the WFC3/UVIS in GO-12257 (PI: L. Girardi). The data reduction procedure is described in \citet{Goudfrooij14}. As with the first data set, we obtained the catalogs from the authors of a previous work, in this case \citet{Goudfrooij14}.

\paragraph{The Third Data Set:} 
The last data set consists of 4 LMC clusters, namely LW~431, NGC~1783, NGC~1987 and NGC~2108. They are from the HST AR-12246 proposal (PI: V. Kozhurina-Platais). The observations for these clusters are from GO-10595
and the photometry has been re-examined from ACS/WFC images corrected for Charge Transfer Inefficiency. In the next Section, we will outline the steps of the reductions.

\subsection{Observations and Data Reduction\label{sec:reduction}}

The photometry for 4 LMC clusters has been obtained from the
HST AR-12246 proposal (PI: V. Kozhurina-Platais, Co-PI: A. Dotter). These clusters 
were observed with the F435W, F555W and F814W HST ACS/WFC filters 
in GO-10551 (PI: P. Goudfrooij). The photometry of the selected clusters in AR-12642 have been derived from ACS/WFC images corrected for Charge Transfer Inefficiency with a high-precision effective point-spread function (ePSF) fitting technique.

{\bf CTE correction}. The instrumental degradation with time known as imperfect Charge Transfer Efficiency (CTE) of CCDs is the most influential effect on the photometric accuracy and precision.  Typical photometric losses for ACS/WFC are in the range from $\sim$1-5\% in 2003-2006 and grow with time. A recently developed CTE correction \citep{AndersonBedin10} is the pixel-based correction, an empirical algorithm of the ACS/WFC image-restoration process, recreating the observed pixel values into original pixel values. As result it renovates PSF flux, positions and shape. This empirical pixel-based correction is implemented into the HST ACS/WFC pipeline to correct ACS/WFC images for the imperfect CTE. Thus, all images for the selected 4 clusters were corrected for the imperfect CTE in the HST pipe-line and simultaneously calibrated for bias, dark, low-frequency flats. 

{\bf Photometry}. Stellar photometry for each cluster was derived with the {\it effective PSF} or {\it ePSF} fitting technique developed 
by \citet{AndersonKing06} which represents a spatial variation with an array of 9$\times$5 fiducial PSFs across each ACS/WFC CCD. The software img2xym\_WFC.09x10.F uses these ACS/WFC PSF to find and measure stars in the images. The central 5$\times$5 pixels were fit with the local PSF to determine positions and flux. The output from the routine is the list of high-precision X\&Y positions, flux and the parameter {\it q} as quality of the fit (the residuals to the PSF fit)  for each star. The X\&Y positions are corrected for ACS/WFC geometric distortion \citep{AndersonKing06} and are accurate to the level of 0.01 ACS/WFC pixels.  X\&Y positions from each exposure were matched to the long exposure in the F435W ACS/WFC filter with POS-TARG 0:0. A tolerance in matching of $<$ 0.2 ACS/WFC pixel in both X\&Y positions allowed to select a good measured star and eliminate any spurious/false, cosmic rays and/or hot pixels detected on the CTE corrected 
images.  In order to transform instrumental magnitudes from ePSF measurements
into the VEGAMAG system, three photometric corrections were applied:
\\ 
{\it (i)\/}
the aperture correction from 5 pixel to 10 pixels ( 0.5$\arcsec$, which is the aperture used for ACS flux calibration, \citealt{Sirianni05}) the difference between ePSF-fitting photometry and the aperture photometry with aperture radius of 10 pixels for bright isolated stars on drizzled images.
\\
{\it (ii)\/} the aperture correction from 10 pixels (or 0.5$\arcsec$) aperture to an infinite aperture (Table 5, \citealt{Sirianni05}); 
\\
{\it(iii)\/} the VEGAMAG zero point (Table 10, \citealt{Sirianni05}) updated for different observing dates \footnote {\url{http://www.stsci.edu/hst/acs/analysis/zeropoints}}. 
\\
The final photometry for each star in each cluster (GO-10595) was determined from a weighted average of all three measurements in each filter. The photometric errors were calculated as {\it RMS\/} deviation of the independent measurements in the different exposures. If the star had only one measurement (due to the {\it POS-TARG} offset or stars rejected as being saturated from long exposures), the magnitude was assigned as single measurement and the photometric error was assigned as 0.001.

\begin{figure}
   \resizebox{\hsize}{!}{\includegraphics{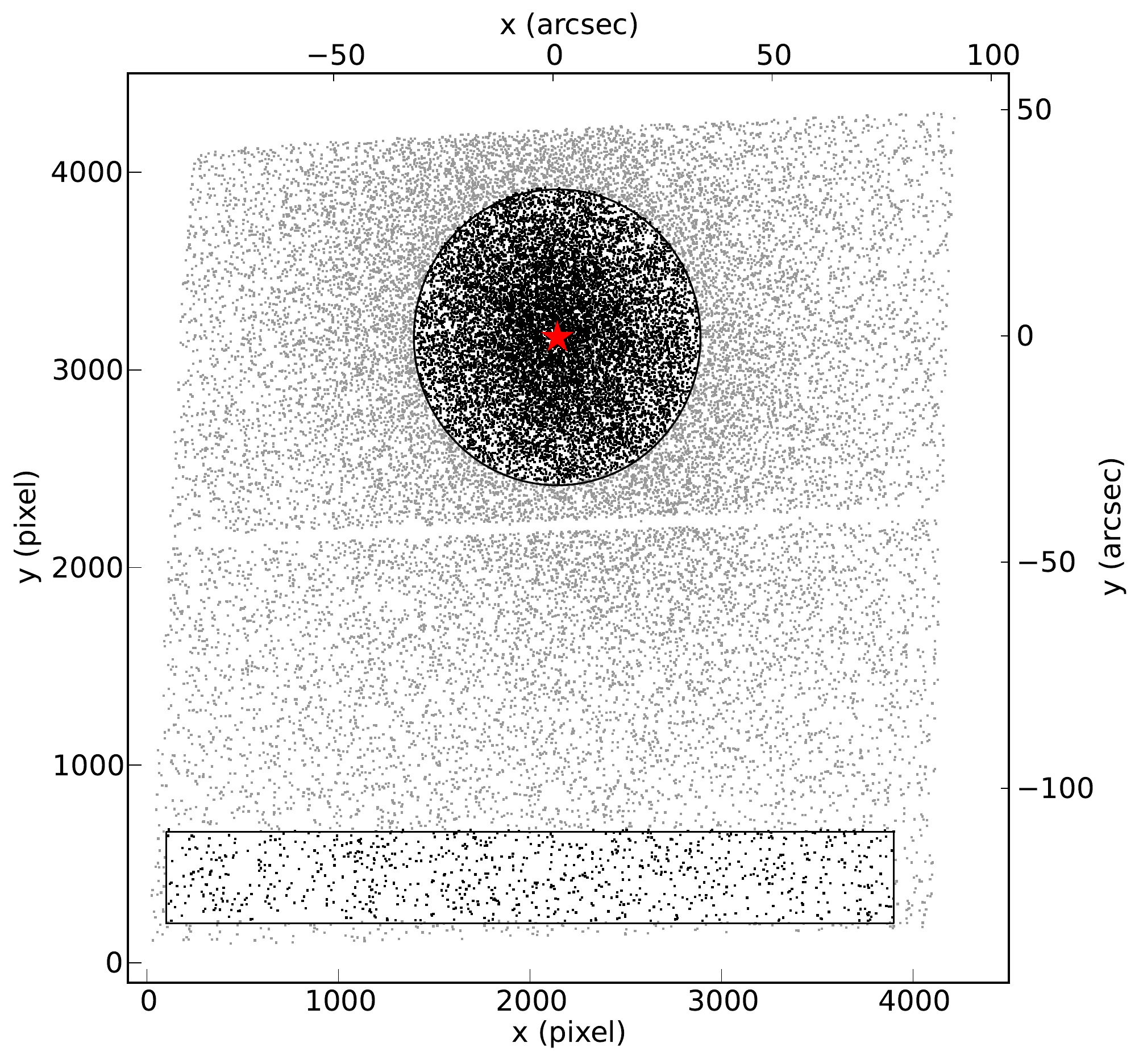}}
   \caption{Spatial positions of all detected stars  in NGC 1783 field on the ACS/WFC chip. The circle  indicates the core radius of the cluster and the center of the cluster is marked with a (red) asterisk symbol. We used the marked rectangle at the bottom of the chip as our reference field region for the statistical removal of field stars from the cluster region. The white line containing no stars is due to the gap between the two detector chips.}
              \label{fig:ngc1783_xy}
\end{figure}

\begin{figure*}
\centering
\begin{tabular}{ccc}
\includegraphics[width=6cm]{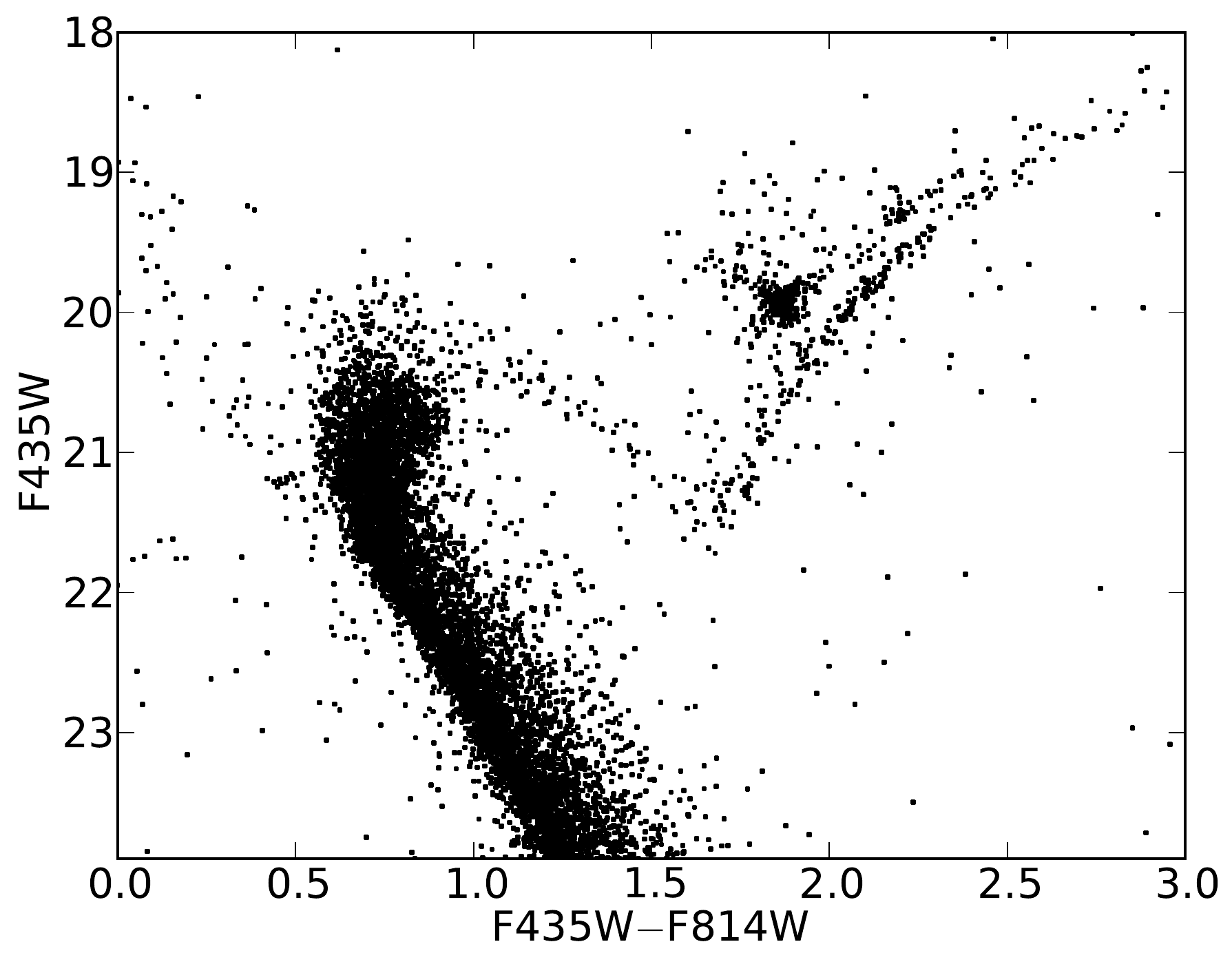} & \includegraphics[width=6cm]{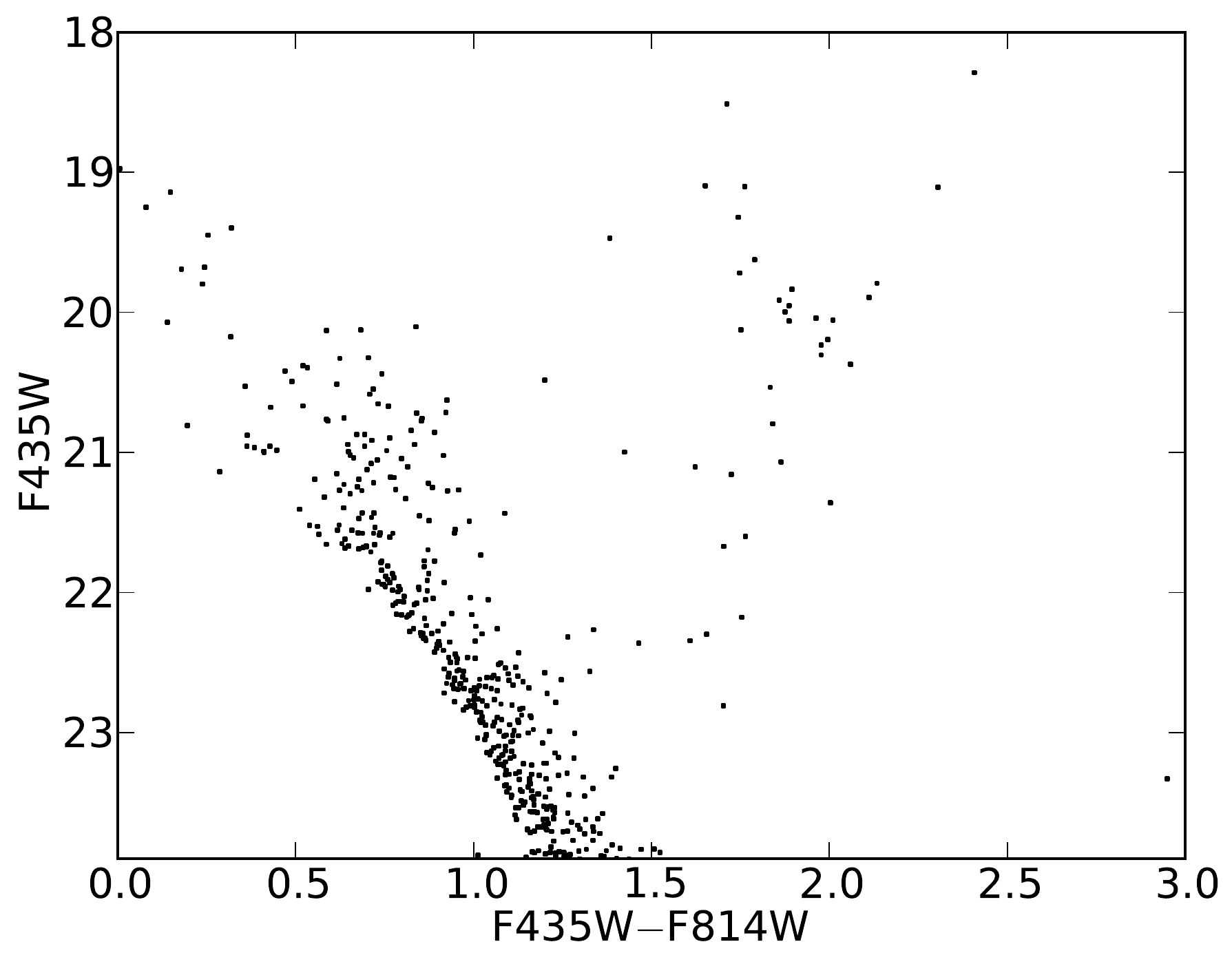} & 
\includegraphics[width=6cm]{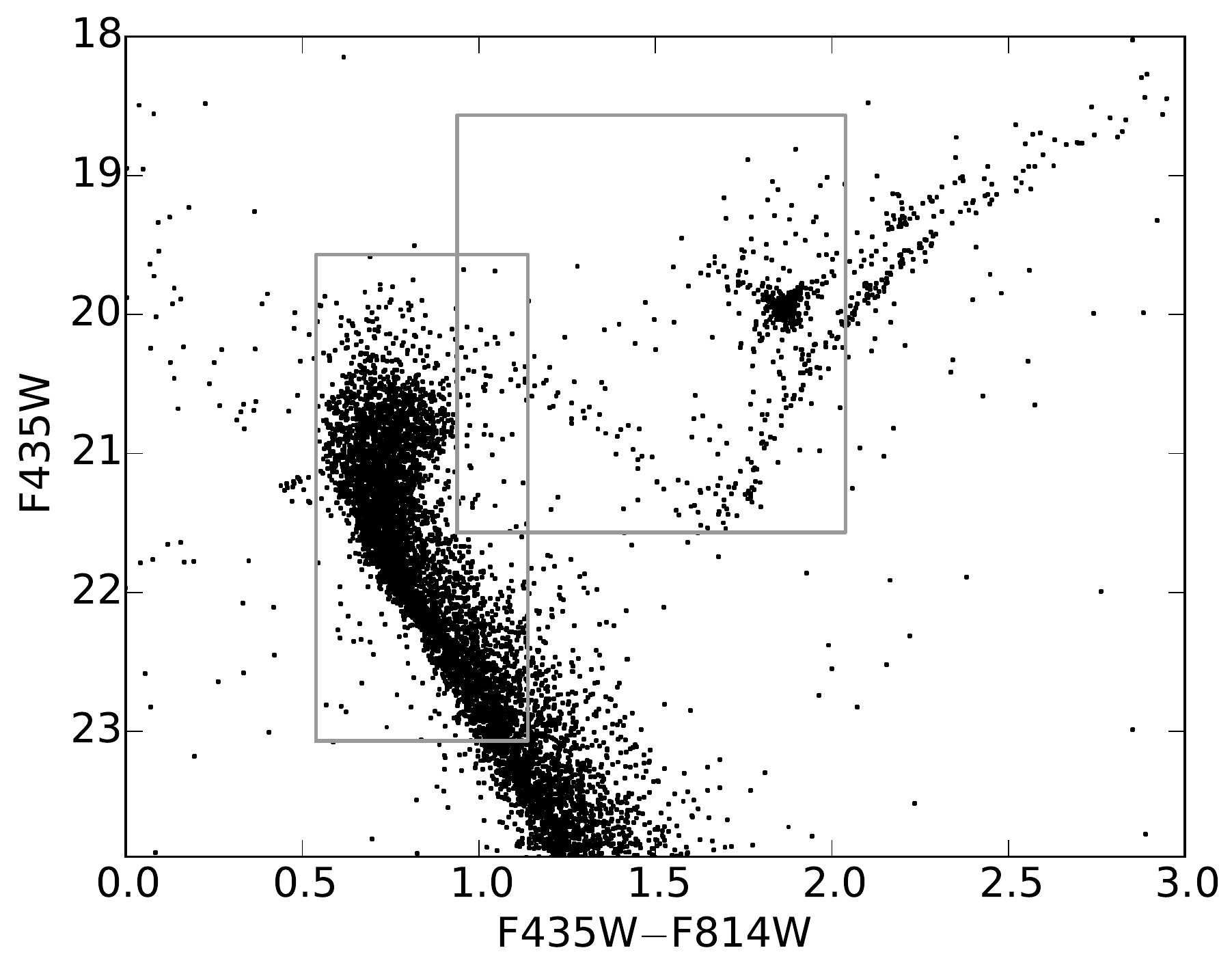}
\\
\end{tabular}
\caption{\textit{Left panel}: CMD of all stars within the core area of NGC 1783; \textit{Middle panel}: CMD of the selected field star region (rectangular area in Figure \ref{fig:ngc1783_xy}); \textit{Right panel}: CMD of the core area of NGC 1783 decontaminated from the field star population. However, it is not fully decontaminated as there still are some non-cluster stars left in the cleaned CMD. The two gray rectangles indicate the boxes used for fitting the SFH of the turn-off and red clump region. Note that the limits of the boxes are shifted to the not extinction corrected data.}
\label{fig:cmds_cluster_field_ngc1783}
\end{figure*}

\begin{figure*}
\centering
\begin{tabular}{ccc}
\includegraphics[width=6cm]{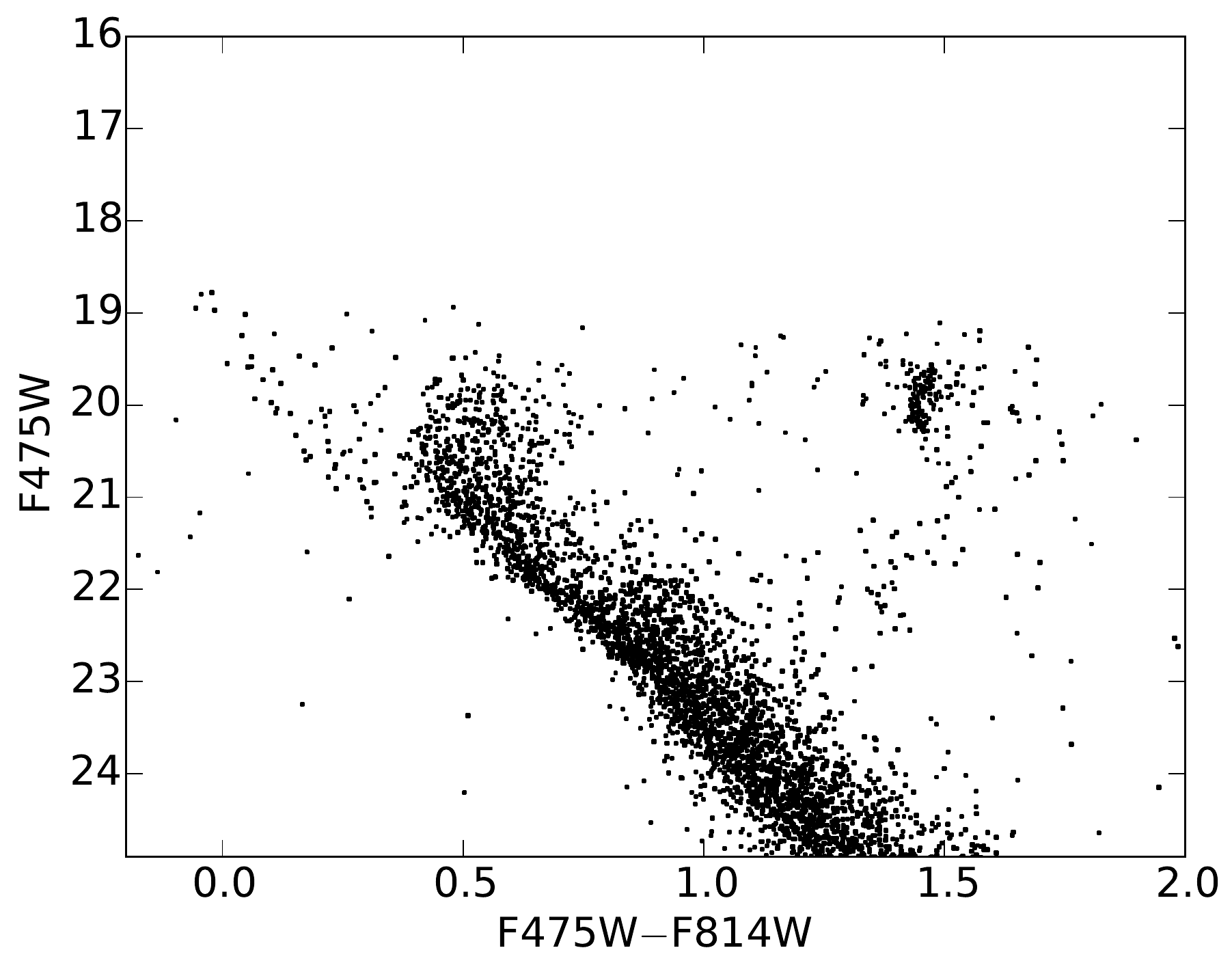} & 
\includegraphics[width=6cm]{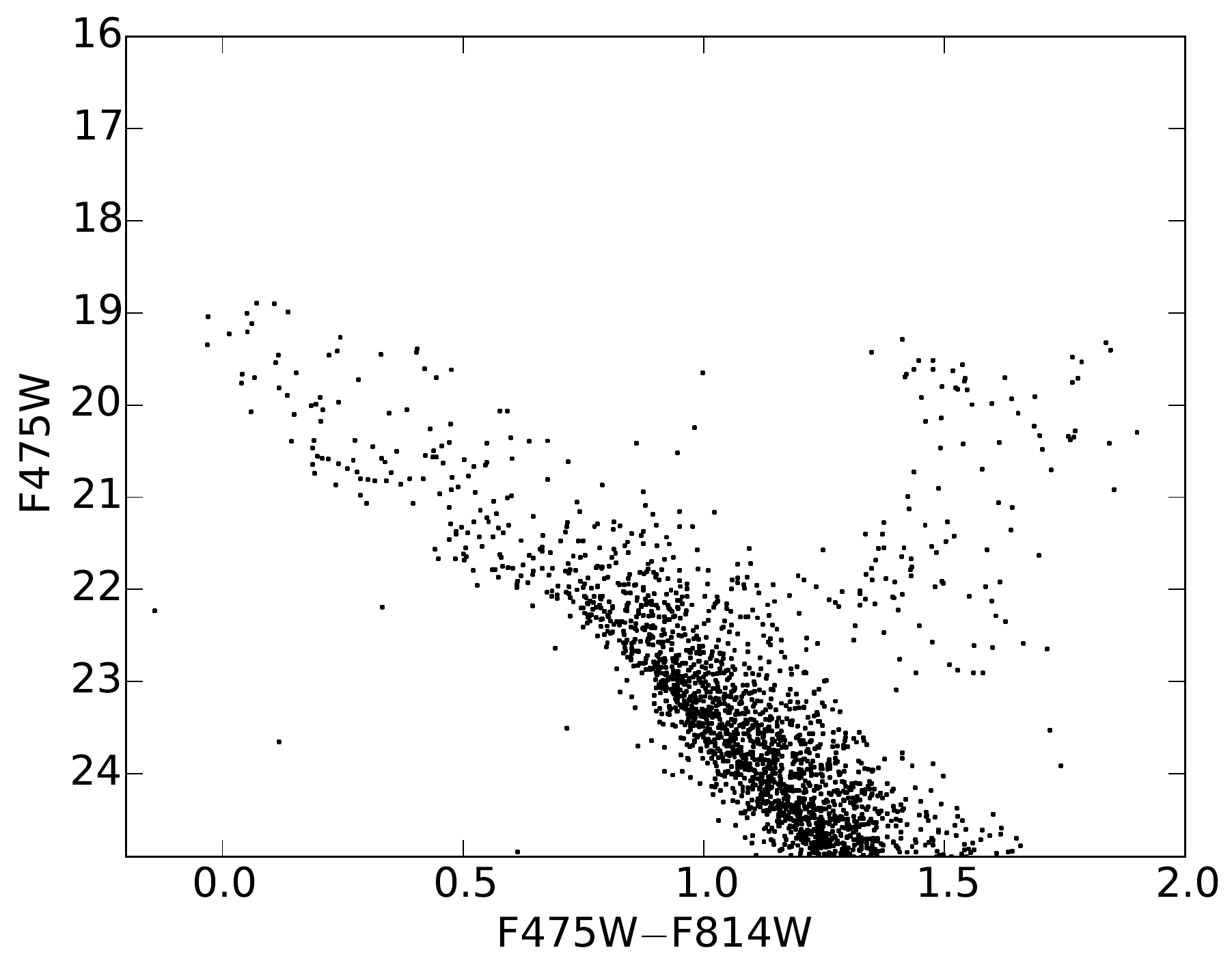} & 
\includegraphics[width=6cm]{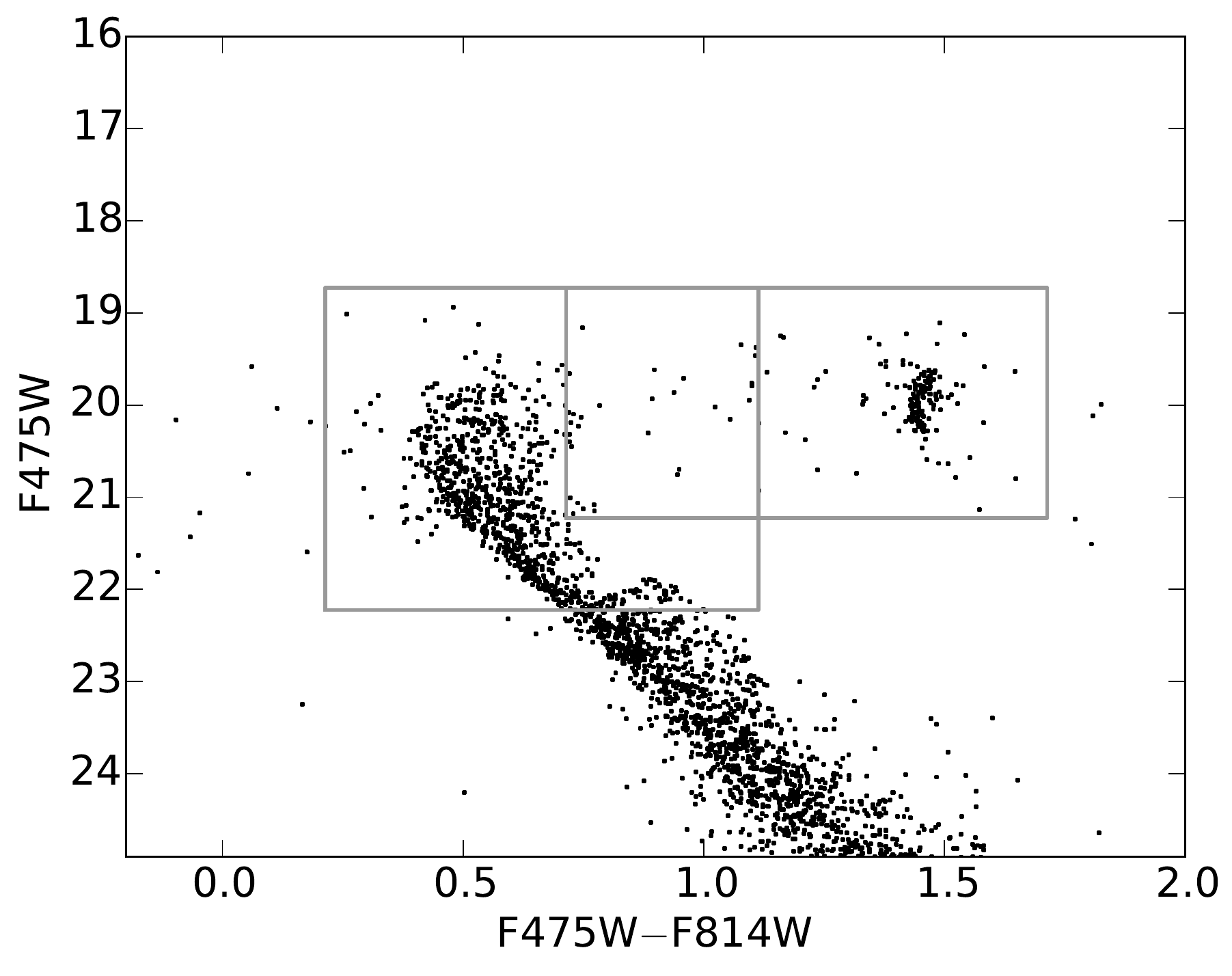}
\\
\end{tabular}
\caption{\textit{Left panel}: CMD of all stars within the core area of Hodge 2; \textit{Middle panel}: CMD of the selected field star region; \textit{Right panel}: CMD of the core area of Hodge 2 decontaminated from the field star population. As in Figure \ref{fig:cmds_cluster_field_ngc1783} the gray boxes indicate the areas in the CMD where we fit the SFH.}
\label{fig:cmds_cluster_field_hodge2}
\end{figure*}

\subsection{Field Star Subtraction\label{sec:fieldstar}}

Before we started to analyze the SFH of the clusters we cleaned their CMDs for field stars that do not belong to the clusters themselves. We started by determining the centers of the clusters on the detector chip. We first created an artificial blurred image of the cluster by assigning each detected star's position on the chip the measured flux of the star. Then we convolved the images with a Gaussian kernel such that we get a smooth brightness distribution across the cluster. We chose Gaussian width's ranging between 40 and 80 pixels, depending on the cluster. We then fitted elliptical isophotes to the image using the IRAF\footnote{IRAF is distributed by the National Optical Astronomy Observatories, which is operated by the Association of Universities for Research in Astronomy, Inc., under cooperative agreement with the national Science Foundation.} task \texttt{ellipse}. As the cluster's center we took the center of the innermost isophote. After we determined the center of the clusters we performed a statistical field star subtraction. For this we selected all stars in an area around the center of a cluster, the cluster field, and chose a rectangular area at the bottom of the detector chip, the reference field, that covers the same area as the cluster field (shown for NGC~1783 in Figure \ref{fig:ngc1783_xy}). 
As cluster fields, we selected the areas that are within 1-2 times the core radii of the clusters. For the size we had to find a compromise: On the one hand we want to chose the region large enough to have as many cluster stars as possible for our analysis, but on the other hand the region must be small enough to perform a reasonable field star subtraction as the distance between the reference and the cluster field is limited. For most of the clusters we selected the area within two times the core radius $\mathrm{R_{core}}$ (radius at which the density is half the central density) as our cluster field. Only for NGC~1783 and NGC~2203 which have large core radii (cf. Table \ref{tab:Cluster_Param}), and NGC~1987, which has a very numerous underlying field star population, we selected the area within one $\mathrm{R_{core}}$.

After defining the cluster and the reference field, we constructed CMDs for both fields. For every star in the reference CMD we removed the star in the cluster CMD that is closest to the reference field star in color-magnitude space (cf. \citealt{Niederhofer15a}). As two examples, Figures \ref{fig:cmds_cluster_field_ngc1783} and \ref{fig:cmds_cluster_field_hodge2} show the CMDs of all stars in the cluster field of NGC 1783 and Hodge 2 (left panels), the CMDs of the reference field (middle panels) and the field subtracted cluster CMDs (right panels). We note that a rich underlying old population which has a MSTO at $m_{F475W} \sim$22 is present in the CMD of Hodge 2 (Figure \ref{fig:cmds_cluster_field_hodge2} middle panel). This population is not completely subtracted and is still visible in the cleaned CMD (right panel) where it causes the 'bump' feature in the MS at $m_{F475W} \sim$22. The method of the statistical field star subtraction relies on two assumptions: First, that the field stars are distributed equally over the entire field-of-view; second, that there are no cluster stars left in the reference field. However, in our case, the second assumption is not entirely fulfilled, as we do not have an additional external field exposure for the clusters but rather performed the decontamination on the cluster exposure itself. The tidal radii $\mathrm{R_{tidal}}$ (radius at which the gravitational force of the host galaxy becomes larger than that of the cluster itself) of the clusters are at least 5 times larger as their core radii \citep{Goudfrooij11a,Goudfrooij14} and there are still some cluster stars left in our reference fields. We might therefore also subtract cluster stars.

\subsection{Differential Extinction Correction\label{sec:diffext}}

\begin{figure}
 \centering
 \begin{tabular}{c}
  \includegraphics[width=8cm, height=5cm]{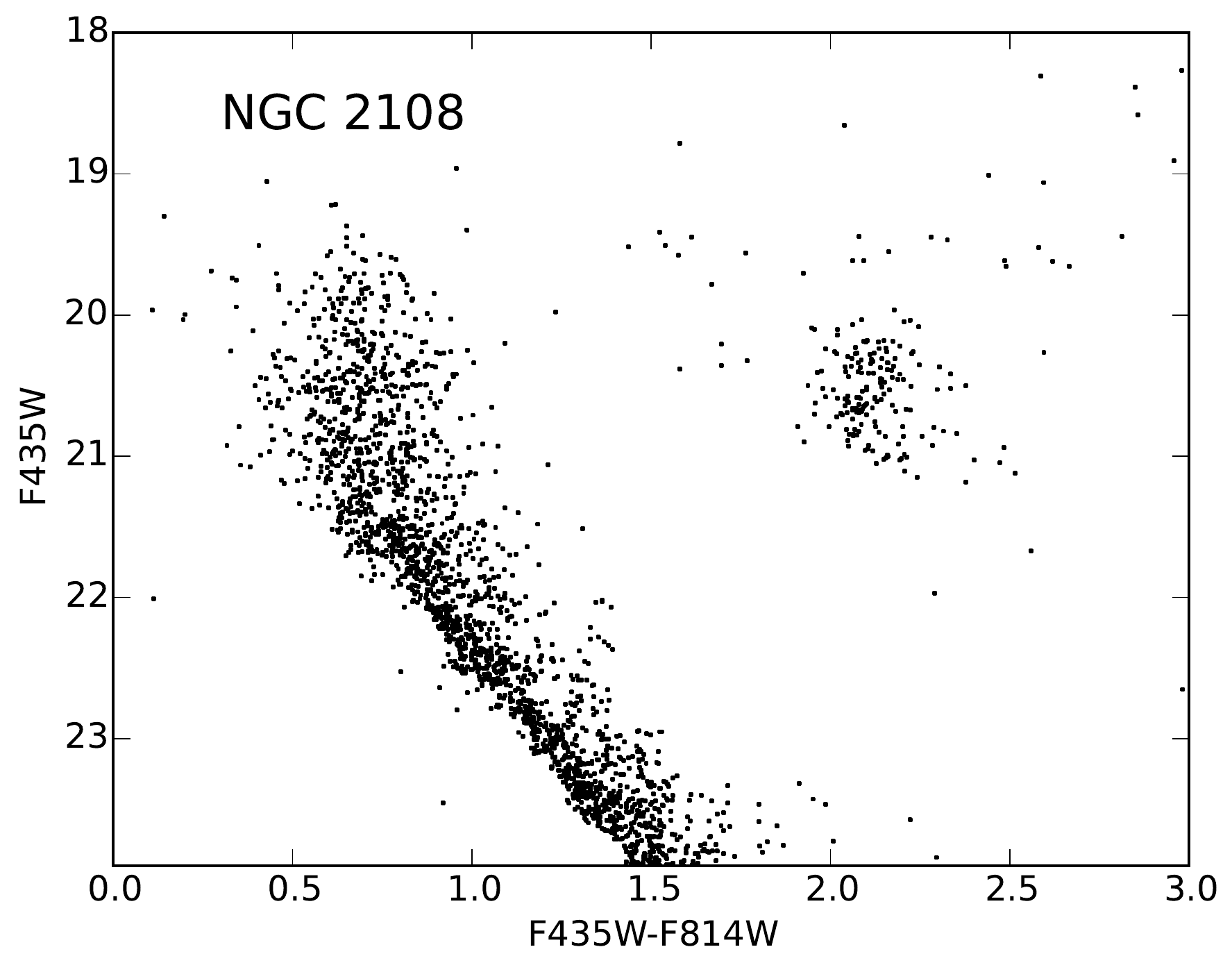}
  \\ 
  \includegraphics[width=8cm, height=5cm]{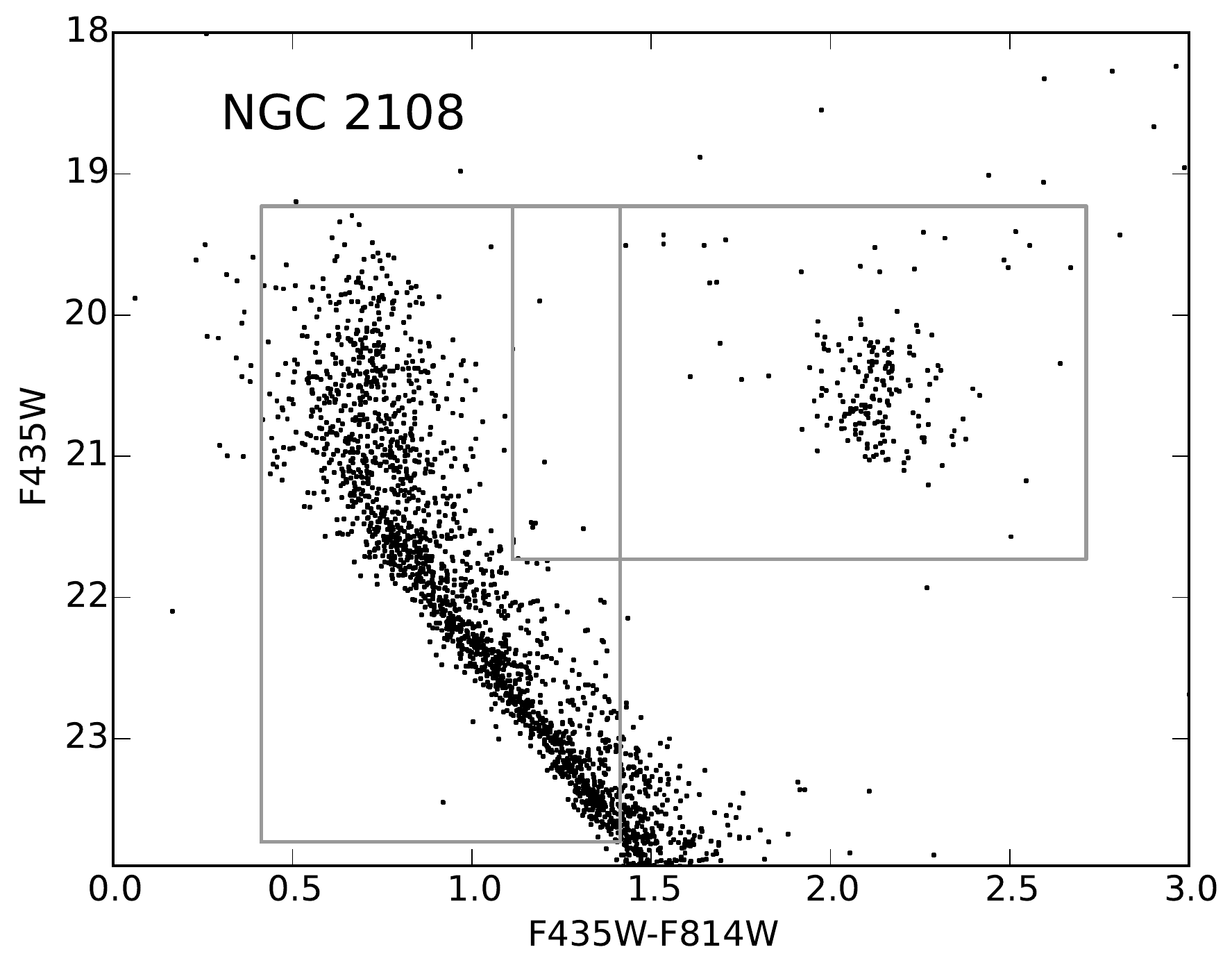}
   \\
  \end{tabular}            
  \caption{CMDs of NGC 2108 before and after the correction for differential extinction. \textit{Upper panel:} The original CMD; \textit{Lower panel:} The corrected CMD. The gray boxes show the areas in the CMD used to fit the SFH of the MSTO and the red clump. Also here, the limits have been shifted to match the data that has not been corrected for the mean extinction.
  }
   \label{fig:ngc2108_cmds_diff_ext_corr}
\end{figure}

\citet{Milone09} and \citet{Goudfrooij11a} reported in their works that NGC 2108 might be affected by differential extinction. We take into account this effect and correct NGC 2108 for differential reddening before carrying out the analysis. We followed the procedure that is described in \citet{Milone12b}. We recap here only the main steps of the correction: We rotated the CMD of NGC 2108 such that the reddening vector is horizontal. Then we defined a fiducial line in the rotated CMD along the MS well below the TO region by quadratically interpolating the median $x$-values of the MS stars in the rotated CMD in bins of 0.25 in $y$-units. For each star on the MS we calculated the median horizontal displacement from the fiducial line for the closest 10-20 (depending on the spatial position of the star in the cluster) neighbors (not including the star itself). This median displacement, as it is parallel to the extinction vector, is a measure for the local reddening. As a final step, we took the spatial median of the local extinction over a bin size of about 175$\times$175 pixel (as a trade-off between spatial resolution and numbers of stars per bin) and corrected each cluster star for its local reddening. We find a differential reddening across the cluster that has a standard deviation of about 0.07 mag in $A_V$ with a maximum value of 0.21 mag. Figure \ref{fig:ngc2108_cmds_diff_ext_corr} shows the CMD of NGC 2108 before (top panel) and after (lower panel) the correction for differential extinction. Although the differences are very moderate, the CMD of NGC 2108 has improved. The main sequence and the turn-off are smoother and better defined. Also the red clump has now a more compact shape. In the further analysis we will use the corrected photometry.

\section{Models and Methods\label{sec:models}}

\subsection{Models}
In this study we analyze the CMDs of a sample of 12 intermediate-age LMC clusters which all show the extended MSTO phenomenon. \citet{Li14} and \citet{BastianNiederhofer15} compared the distributions of stars that populate the SGB and the red clump with what would be expected for an SSP and an extended SFH. 
We continue on these studies by performing a quantitative study of the SFH of the clusters in our sample.  We perform the SFH fitting on two regions of the cluster CMDs, first on the extended main sequence turn-off and secondly on the post-main sequence region that includes the sub-giant branch, lower red giant branch and red clump.
For this, we fitted the SFH in the MSTO region and the post-MS region including the SGB and the red clump, independently. We used for our analysis the StarFISH package \citep{Harris01}. This code uses theoretical isochrones to create a library of synthetic single age CMDs. It takes into account the photometric errors of the observations and requires an a-priory knowledge (or assumption) of the metallicity Z, the distance modulus and extinction. Below the MSTO, StarFISH linearly interpolates along the isochrones due to the sparser sampling. However, the code does not interpolate between isochrones at various ages.
To reconstruct the SFH out of the observed CMD, the StarFISH code linearly combines the synthetic CMDs and searches for the best match between model and data by performing a $\chi^2$ minimization.

As theoretical models for the fitting, we used the newest PARSEC 1.2S isochrones \citep{Bressan12} from the Padova isochrone set\footnote{\url{http://stev.oapd.inaf.it/cgi-bin/cmd_2.7}} at a fixed metallicity Z of 0.008 (Z$_{\odot}$=0.0152 for this set of isochrones) and a He abundance Y of 0.25. These isochrones are computed from evolutionary tracks with a five times finer mass grid ($\Delta$M = 0.01 M$_{\odot}$) than the older set of \citet{Marigo08} from the same group. A metallicity of 0.008 is a typical value of LMC clusters in this age interval. For all the clusters, we fitted the SFH between log(age)=8.70 and log(age)=9.50 in steps of 0.03 (in log).

\subsection{Isochrone Fitting\label{sec:Isofit}}

\begin{table*} 
\caption{Parameters of the LMC clusters\label{tab:Cluster_Param}}
\centering
\begin{tabular}{c c c c c c c c} 
\hline\hline
\noalign{\smallskip}
Cluster & $\mathrm{R_{core}}$(arcsec) & $\mathrm{R_{tidal}}$(arcsec)  & Z & $\mathrm{A_V}$ & $(\mathrm{m-M})_0$ & Age (Gyr) & $\Delta \mathrm{Age}_\mathrm{RC}$ (Myr)
\\
(1) & (2) & (3) & (4) & (5) & (6) & (7) & (8)
\\
\noalign{\smallskip}
\hline
\noalign{\smallskip}
NGC 1783 & 36.7 & 346.8\tablefootmark{2}& 0.008 & 0.08 & 18.41 & 1.75 & $\sim$500
\\
LW 431 & 25.0 & 185.9\tablefootmark{2} & 0.008 & 0.20 & 18.40 & 1.75 & $\sim$300
\\
NGC 1987 & 23.2 & 270\tablefootmark{2} & 0.008 & 0.16 & 18.37 & 1.15 & $<$80
\\
NGC 2203 & 32.9\tablefootmark{1} & 161.2\tablefootmark{1}& 0.008 & 0.19 & 18.41 & 1.75 & $\sim$200 
\\
NGC 1718 & 15.4\tablefootmark{1} & 114.0\tablefootmark{1} & 0.008 & 0.53 & 18.54 & 1.75 & $<$100
\\
NGC 1651 & 18.8\tablefootmark{1} &  545.2\tablefootmark{1} & 0.008 & 0.23 & 18.48 & 2.00 & $\sim$200
\\
NGC 2213 & 11.1\tablefootmark{1} & 66.6\tablefootmark{1} & 0.008 & 0.16 & 18.40 & 1.75 & $\sim$200
\\
NGC 2173 & 14.5\tablefootmark{1} & 165.3\tablefootmark{1} & 0.008 & 0.23 & 18.37 & 1.70 & $\sim$400
\\
Hodge 2 & 11.0\tablefootmark{1} & 473\tablefootmark{1} & 0.008  & 0.19 & 18.45 & 1.45 & $\sim$400 
\\
NGC 2108 & 24.2 & 132.8\tablefootmark{2} & 0.008 & 0.54 & 18.40 & 1.00 & $\sim$300
\\
NGC 1806 & 29.3 & 196\tablefootmark{2}& 0.008 & 0.11 & 18.54 & 1.70 & $\sim$500
\\
NGC 1846 & 37.2 & 161.2\tablefootmark{3} & 0.008 & 0.11 & 18.44 & 1.75 & $\sim$200
\\
\noalign{\smallskip}
\hline
\end{tabular}\\

\tablefoot{Parameters of the clusters in our sample. Column (1): Name of the cluster, Column (2): Core radius $\mathrm{R_{core}}$ of the cluster, Column (3): Tidal radius $\mathrm{R_{tidal}}$ of the cluster, Column (4): Metallicity Z of the cluster, Column (5): Visual extinction $A_V$ to the cluster, Column (6): Distance modulus $(m-M)_0$ to the cluster, Column (7): Mean fitted age of the cluster in Gyr, Column (8): Maximum age spread in Myr as inferred from the post-MS phases.
\\
\tablefoottext{1}{\citet{Goudfrooij14}}; \tablefoottext{2}{\citet{Goudfrooij11a}}; \tablefoottext{3}{\citet{Goudfrooij09}}}

\end{table*}

In order to determine the basic parameters of the clusters in our sample that are needed to fit the SFH (distance modulus $(m-M)_0$, extinction $A_V$), we fitted theoretical models to the CMDs. 
To calculate from the visual extinction $A_V$ the reddening in each filter, we used the following conversion factors: For the ACS/WFC filter system: $A_{F435W}=1.351~A_V$, $A_{F555W}=1.026~A_V$ and $A_{F814W}=0.586~A_V$ \citep{Goudfrooij09}. And for the WFC3/UVIS filter set: $A_{F475W}=1.192~A_V$ and $A_{F814W}=0.593~A_V$ \citep{Goudfrooij14}. The filter dependent extinction conversions were derived using the ACS/WFC and WFC3/UVIS filter curves and a \citet{Cardelli89} extinction law with $R_V=3.1$. To determine the best fit parameters, we selected isochrones with ages between 500 Myr and 2 Gyr. They were then adjusted by eye in both distance modulus and $A_V$ in order to match the main sequence and the red clump. In most cases the red clump is well defined, which leads to small uncertainties in both parameters. For the younger clusters, however, which have more extended red clumps, there is more uncertainty in both parameters, extinction and distance modulus. However, our results are not dependent on the exact choice of the parameters as we are interested in the relative differences between the turn-off and the red clump of the same cluster. The values of the reddening $A_V$ and distance modulus that are found in this study and are used for the further analysis can be found in Table \ref{tab:Cluster_Param}. 
We note that we determined the distance moduli of all but two clusters to be smaller than the canonical distance modulus of the LMC center (18.50 mag $\hat{=}$50~kpc). The only exceptions are NGC~1718 and NGC ~1806 for which we found a distance modulus of 18.54 mag ($\hat{=}$51~kpc). 
Most of the clusters are located well in front of the LMC, up to 3 kpc if the nominal distance to the LMC is adopted. However, we note that the LMC distance was not derived consistently with the methods used here (i.e. there may be systematic uncertainties that account for the different distances).

\begin{table*} 
\caption{Limits of the Boxes where the SFH is fitted\label{tab:Fitboxes}}
\centering
\begin{tabular}{c c c c c} 
\hline\hline
\noalign{\smallskip}
Cluster & \multicolumn{2}{c}{MSTO Box} & \multicolumn{2}{c}{Post-MS Box}
\\
&Color Limits&Magnitude Limits&Color Limits&Magnitude Limits
\\
\noalign{\smallskip}
\hline
\noalign{\smallskip}
NGC 1783 &$0.5\leq(F435W-F814W)\leq1.1$ &$19.5\leq F435W\leq23.0$ & $0.9\leq(F435W-F814W)\leq2.0$&$18.5\leq F435W\leq21.5$
\\
LW 431 & $0.4\leq(F435W-F814W)\leq1.0$& $19.5\leq F435W\leq23.5$& $1.0\leq(F435W-F814W)\leq2.3$&$19.0\leq F435W\leq21.5$
\\
NGC 1987 & $0.1\leq(F435W-F814W)\leq1.0$& $18.5\leq F435W\leq23.0$& $0.7\leq(F435W-F814W)\leq2.0$&$18.5\leq F435W\leq21.0$
\\
NGC 2203 &$0.2\leq(F475W-F814W)\leq1.0$ & $19.0\leq F475W\leq23.0$& $0.8\leq(F475W-F814W)\leq1.5$&$19.0\leq F475W\leq21.5$
\\
NGC 1718 & $0.3\leq(F475W-F814W)\leq0.9$&$19.5\leq F475W\leq24.0$ &$0.8\leq(F475W-F814W)\leq1.6$ &$18.5\leq F475W\leq21.5$
\\
NGC 1651 & $0.4\leq(F475W-F814W)\leq1.0$& $20.0\leq F475W\leq24.0$& $0.8\leq(F475W-F814W)\leq1.6$&$19.0\leq F475W\leq21.5$
\\
NGC 2213 & $0.3\leq(F475W-F814W)\leq0.9$& $19.0\leq F475W\leq24.0$& $0.8\leq(F475W-F814W)\leq1.6$&$18.5\leq F475W\leq21.0$
\\
NGC 2173 &$0.3\leq(F475W-F814W)\leq0.9$ &$19.0\leq F475W\leq24.0$ & $0.7\leq(F475W-F814W)\leq1.7$&$18.0\leq F475W\leq21.5$
\\
Hodge 2 &$0.1\leq(F475W-F814W)\leq1.0$ & $18.5\leq F475W\leq22.0$& $0.6\leq(F475W-F814W)\leq1.6$& $18.5\leq F475W\leq21.0$
\\
NGC 2108 & $0.0\leq(F435W-F814W)\leq1.0$& $18.5\leq F435W\leq23.0$&$0.7\leq(F435W-F814W)\leq2.3$ &$18.5\leq F435W\leq21.0$
\\
NGC 1806 & $0.4\leq(F435W-F814W)\leq1.0$& $19.5\leq F435W\leq23.0$&$0.9\leq(F435W-F814W)\leq2.1$ &$19.0\leq F435W\leq21.5$
\\
NGC 1846 & $0.4\leq(F435W-F814W)\leq1.0$& $19.5\leq F435W\leq23.0$& $1.0\leq(F435W-F814W)\leq.21$&$19.0\leq F435W\leq21.5$
\\
\noalign{\smallskip}
\hline
\end{tabular}\\

\end{table*}

\subsection{Fitting of the Star Formation History\label{sec:fitting}}

For the fitting of the SFH we used the StarFISH software package, assuming: i) a \citet{Salpeter55} initial mass function; ii) values of $A_V$ and $(m-M)_0$ determined in the previous section; iii) a binary fraction of 0.3 for each cluster. \citet{Goudfrooij14} simulated the CMDs of a sample of 11 intermediate age LMC and SMC clusters leaving the binary fraction as a free parameter. For most of the clusters they found best fitting values between 0.22 and 0.33. In order to test how the assumed fraction of binaries affects the results for the SFH, we tried different values of the binary fraction between 0.1 and 0.5 and found that the overall shape of the SFH both of the MSTO and red clump region does not change significantly for values between 0.2 and 0.4.
 
Furthermore, the SFH code requires a model of the photometric errors and the completeness fraction.
We have photometric error estimates derived from a series of artificial star tests for the cluster sample from \citet{Goudfrooij14}. For the other clusters we created an analytical model for the photometric errors within StarFISH based on the following estimates. We assumed a photometric error of 0.01 in the F435W filter for the brightest stars in the clusters which is comparable to the errors inferred from the artificial star tests. The errors increase exponentially to about 0.03 to 0.045 at F435W magnitudes between 23.5 and 25, depending on the cluster. We estimated the errors from the width of the main sequence (well below the MSTO). 
From the artificial star tests we find a 90\% completeness limit at a magnitude between 25 and 23.5 in F475W, depending on the cluster. Only Hodge~2 has a lower 90\% completeness limit at about 21.2 mag in F475W due to the high crowding in its field. The limits should be comparable for the clusters from the ACS/WFC sample. However, incompleteness will not affect our analysis by much as we are interested in the morphology of the MSTO and the red clump which both are above the 90\% completeness limit in all cases.
We tested the reliability of our assumed error model by applying it to a cluster from the \citet{Goudfrooij14} data set and fitted the SFH of the cluster once again using the analytical error model. 
The similarity of the results suggests that our adopted error model is reasonable.

We fit the SFH in two separate parts of the CMD, one centered on the extended MSTO and one including the red clump, SGB and lower red giant branch. The limits of the boxes that contain the regions we used for the fitting are listed in Table~\ref{tab:Fitboxes}. We binned the MSTO region into grid boxes that are 0.05$\times$0.05 mag in size. For the red clump region we choose bins that are 0.025$\times$0.025 mag. Only for LW~431 which has a very sparsely populated red clump we use a bin size of 0.05$\times$0.05 mag for the red clump. 

\begin{figure*}
 \centering
 \begin{tabular}{cc}
\includegraphics[width=9cm]{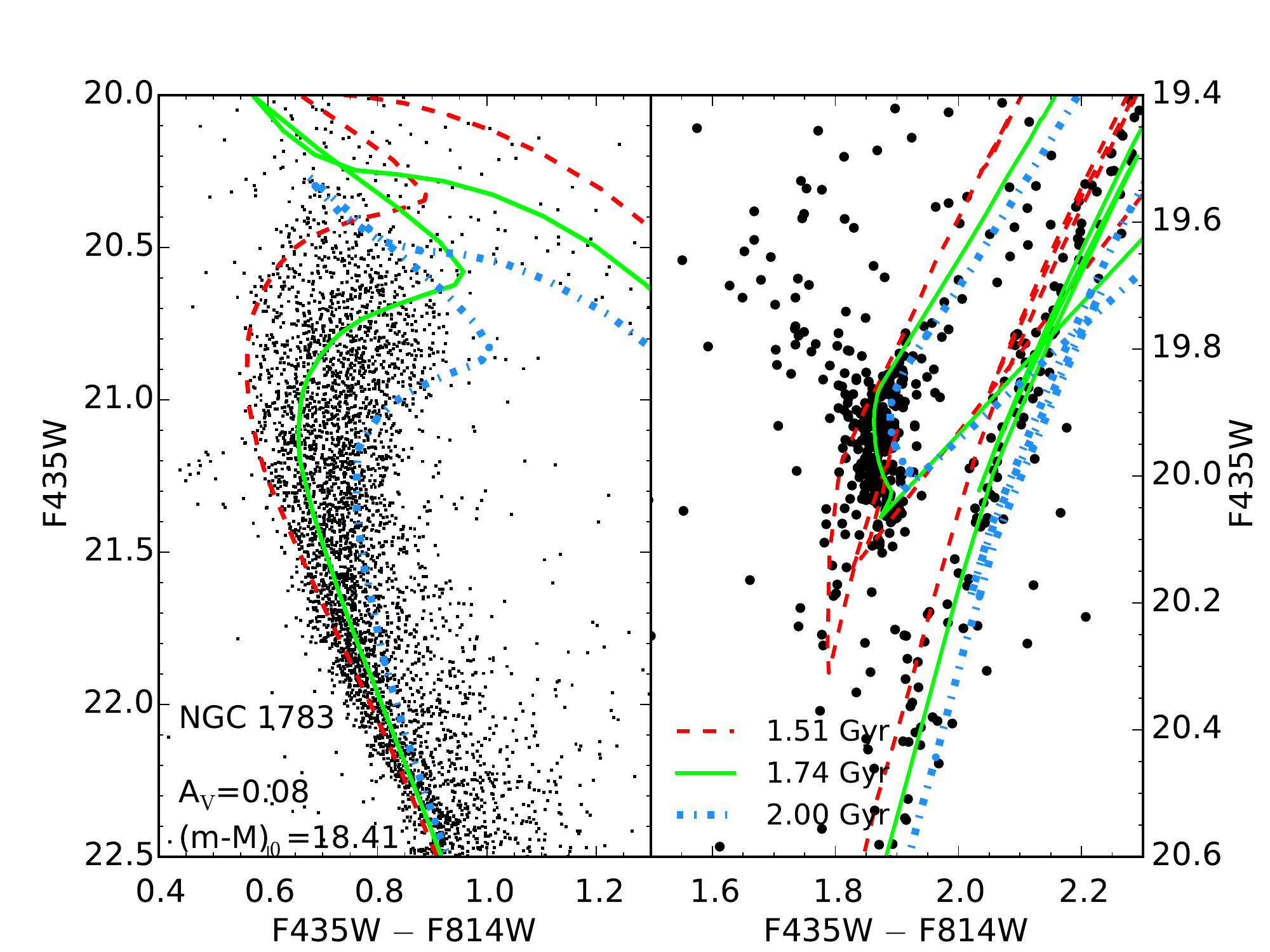} &
\includegraphics[width=9cm]{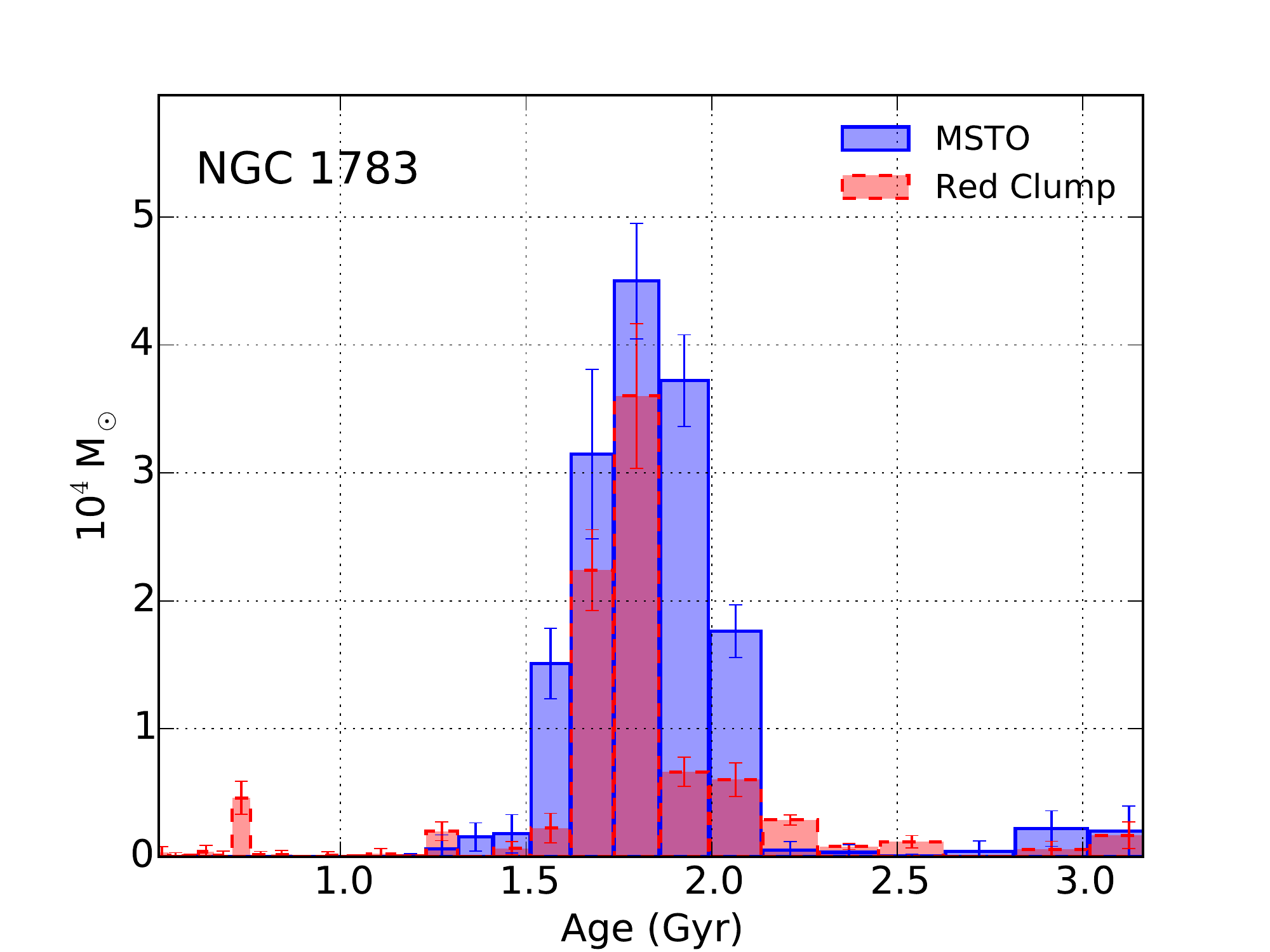}
   \\
  \end{tabular}            
  \caption{\textit{Left panel:} CMD of the MSTO and red clump region of NGC 1783 together with overplotted isochrones at three different ages that cover the extent of the MSTO. \textit{Right panel:} Fitted SFH of NGC 1783 using only the MSTO region (blue solid line) and only the red clump region (red dashed line) along with 1 $\sigma$ error bars showing the statistical errors. The error bars result from bootstrapping the data (see text for details). }
   \label{fig:ngc1783_cmd}
\end{figure*}

\begin{figure*}
 \centering
 \begin{tabular}{cc}
  \includegraphics[width=9cm]{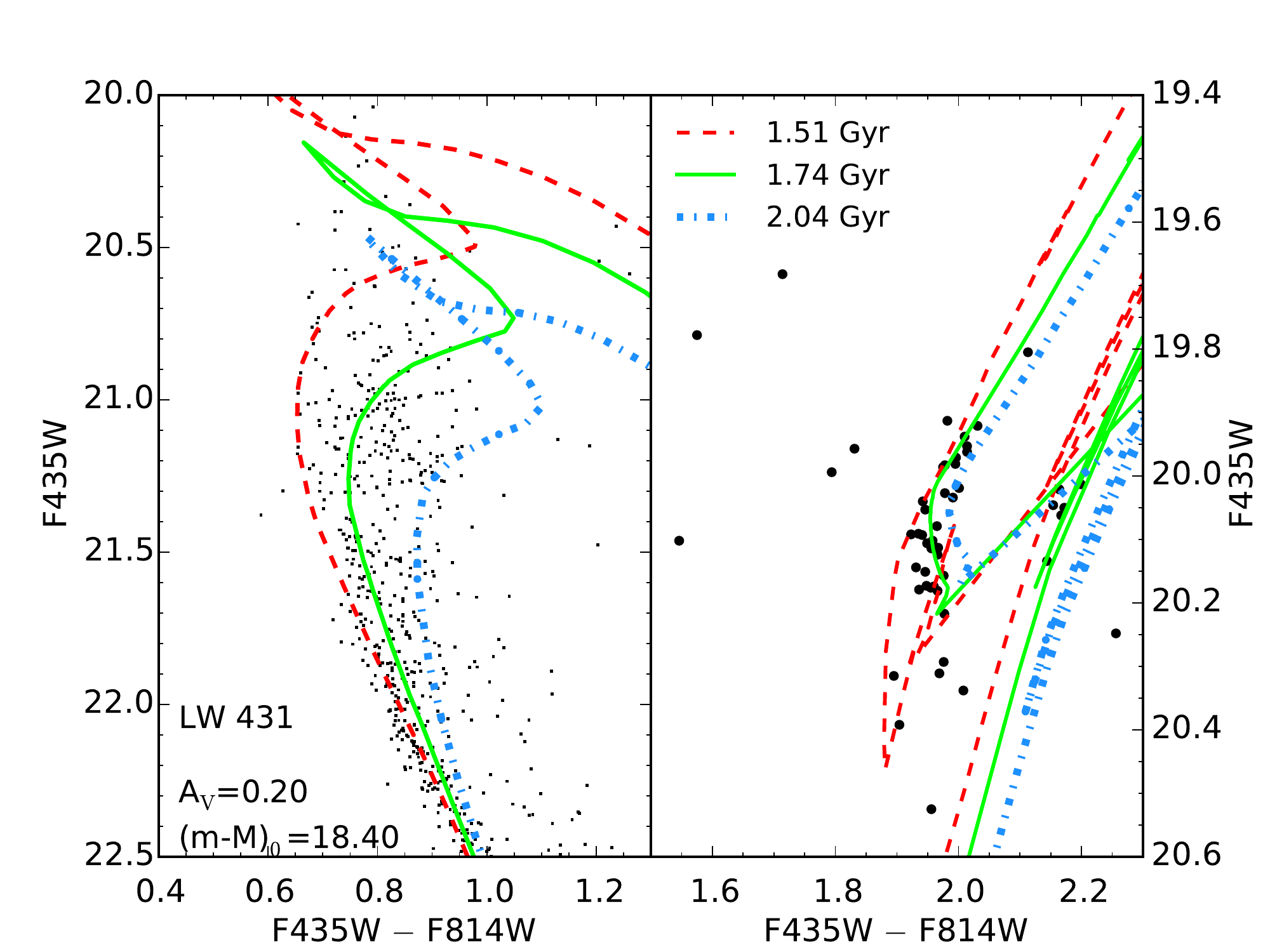} &
  \includegraphics[width=9cm]{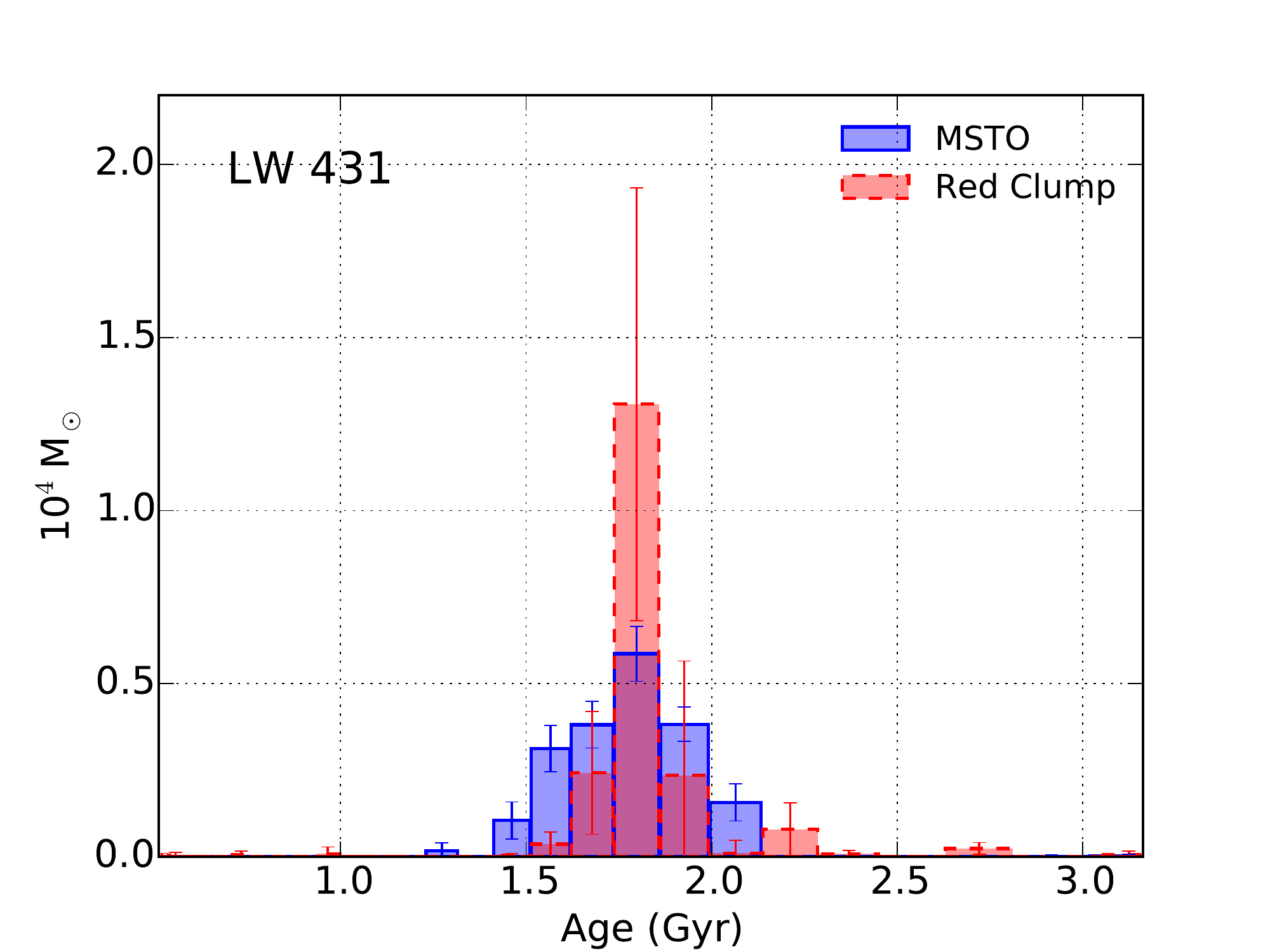}
   \\
  \end{tabular}            
  \caption{Same as Figure \ref{fig:ngc1783_cmd}, now for LW 431.}
   \label{fig:lw431_cmd}
\end{figure*}

\begin{figure*}
 \centering
 \begin{tabular}{cc}
  \includegraphics[width=9cm]{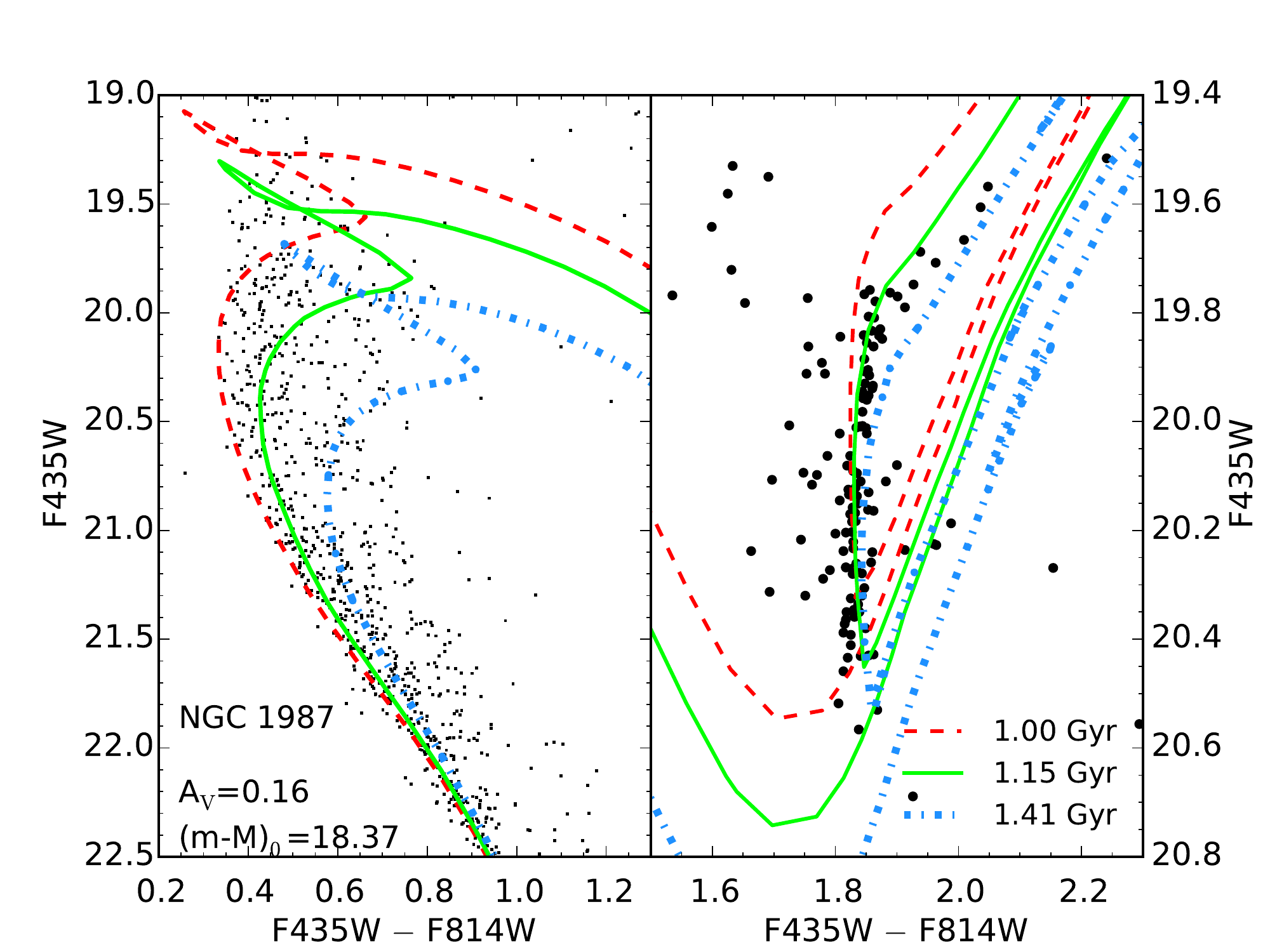} &
\includegraphics[width=9cm]{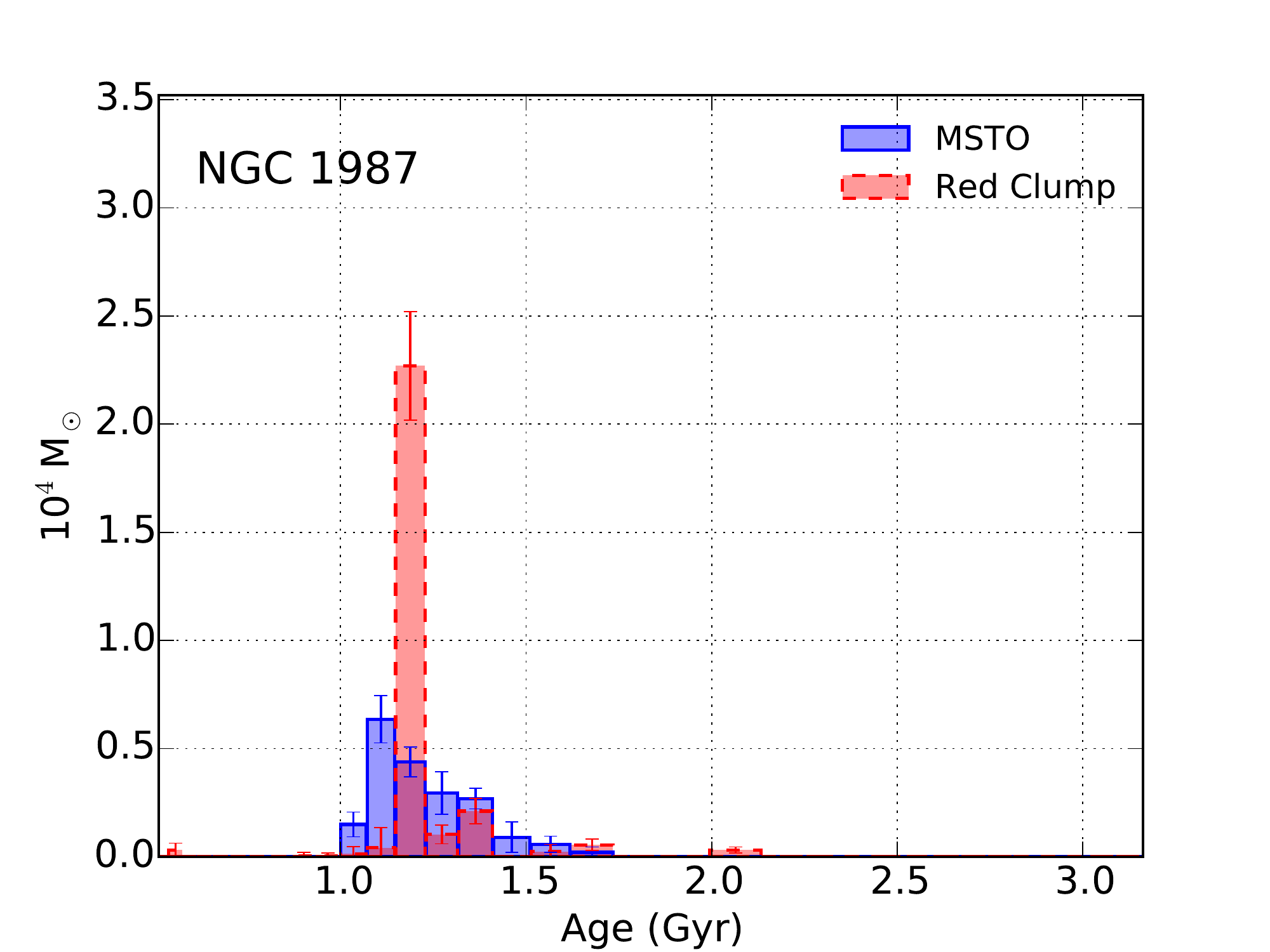}
   \\
  \end{tabular}            
\caption{Same as Figure \ref{fig:ngc1783_cmd}, now for NGC 1987.}  
   \label{fig:ngc1987_cmd}
\end{figure*}

\section{Results\label{sec:res}}

\subsection{The Star Formation History of the Main Sequence Turn-off and the Red Clump\label{sec:sfh}}

\begin{figure*}
 \centering
 \begin{tabular}{cc}
  \includegraphics[width=9cm]{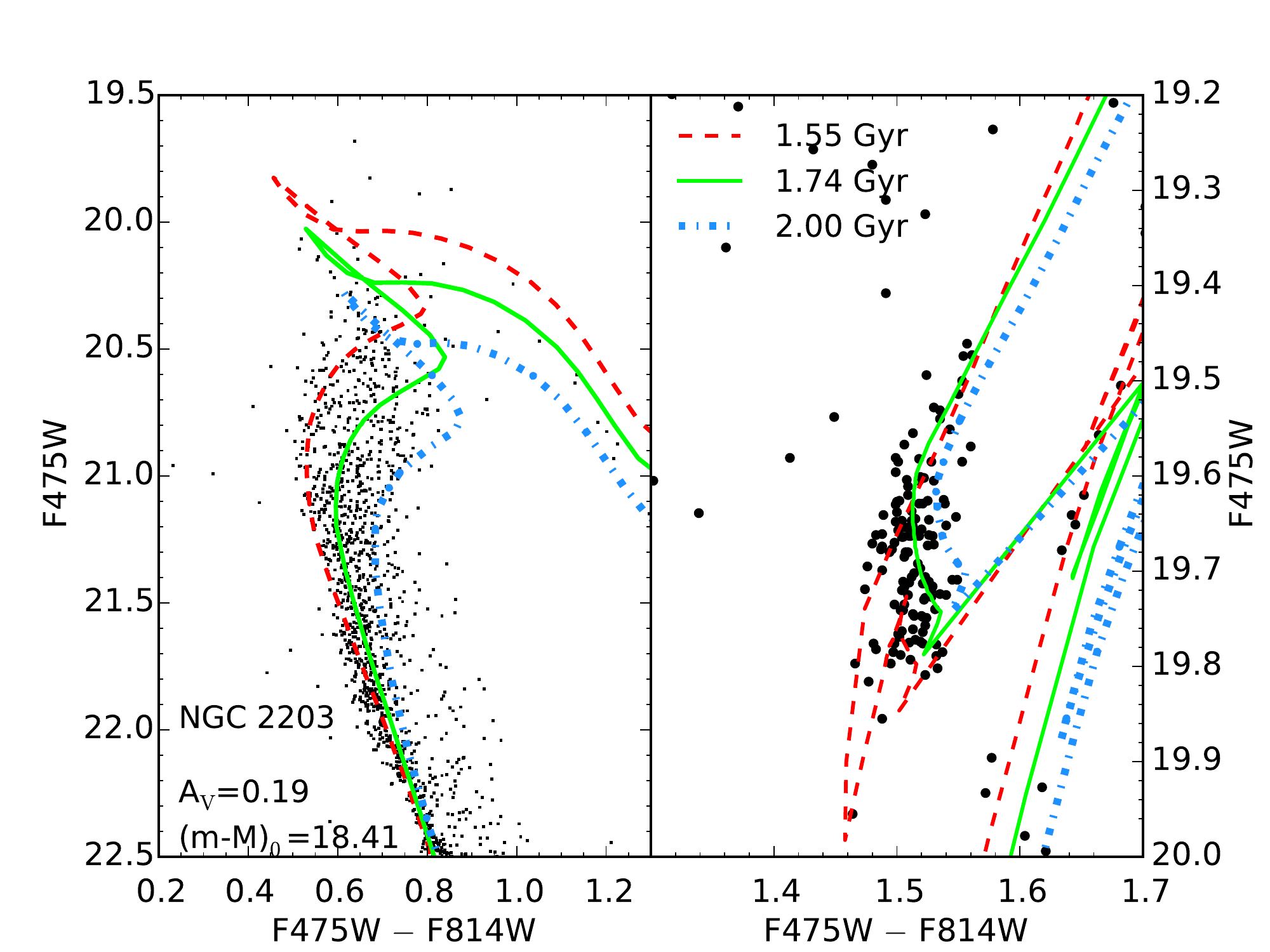} &
  \includegraphics[width=9cm]{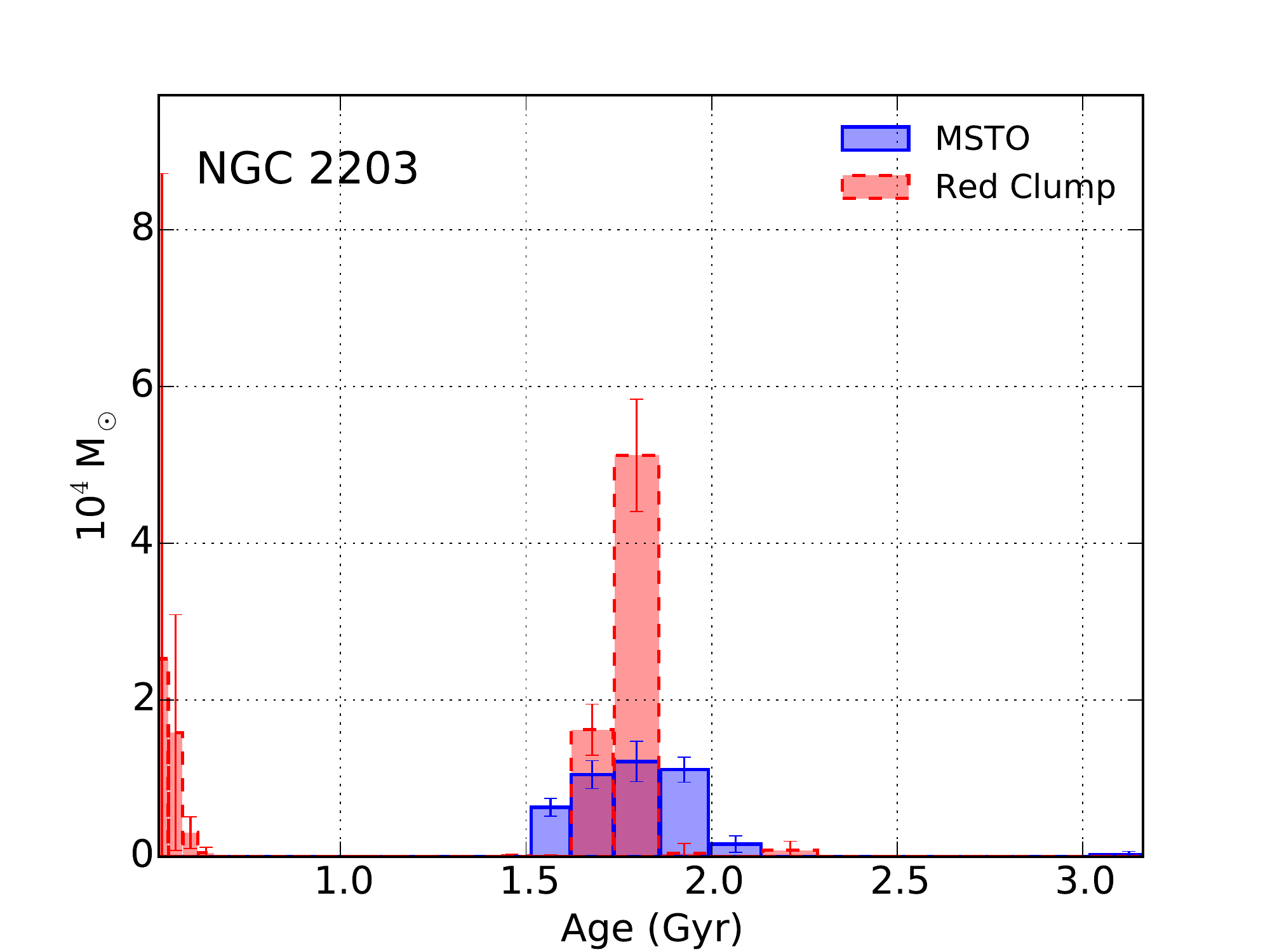}
   \\
  \end{tabular}            
  \caption{Same as Figure \ref{fig:ngc1783_cmd}, now for NGC~2203}
   \label{fig:n2203_cmd}
\end{figure*}

\begin{figure*}
 \centering
 \begin{tabular}{cc}
  \includegraphics[width=9cm]{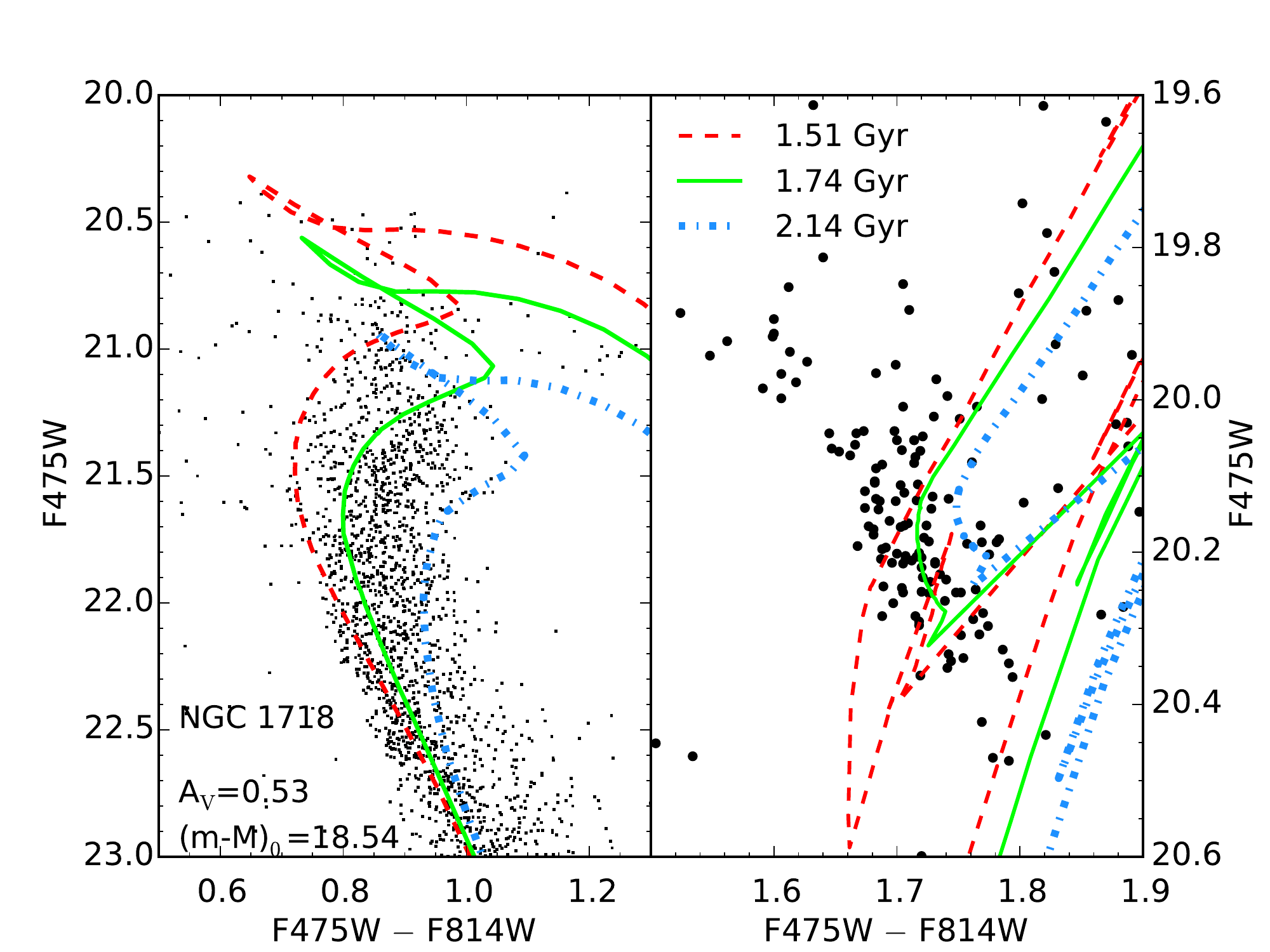} &
  \includegraphics[width=9cm]{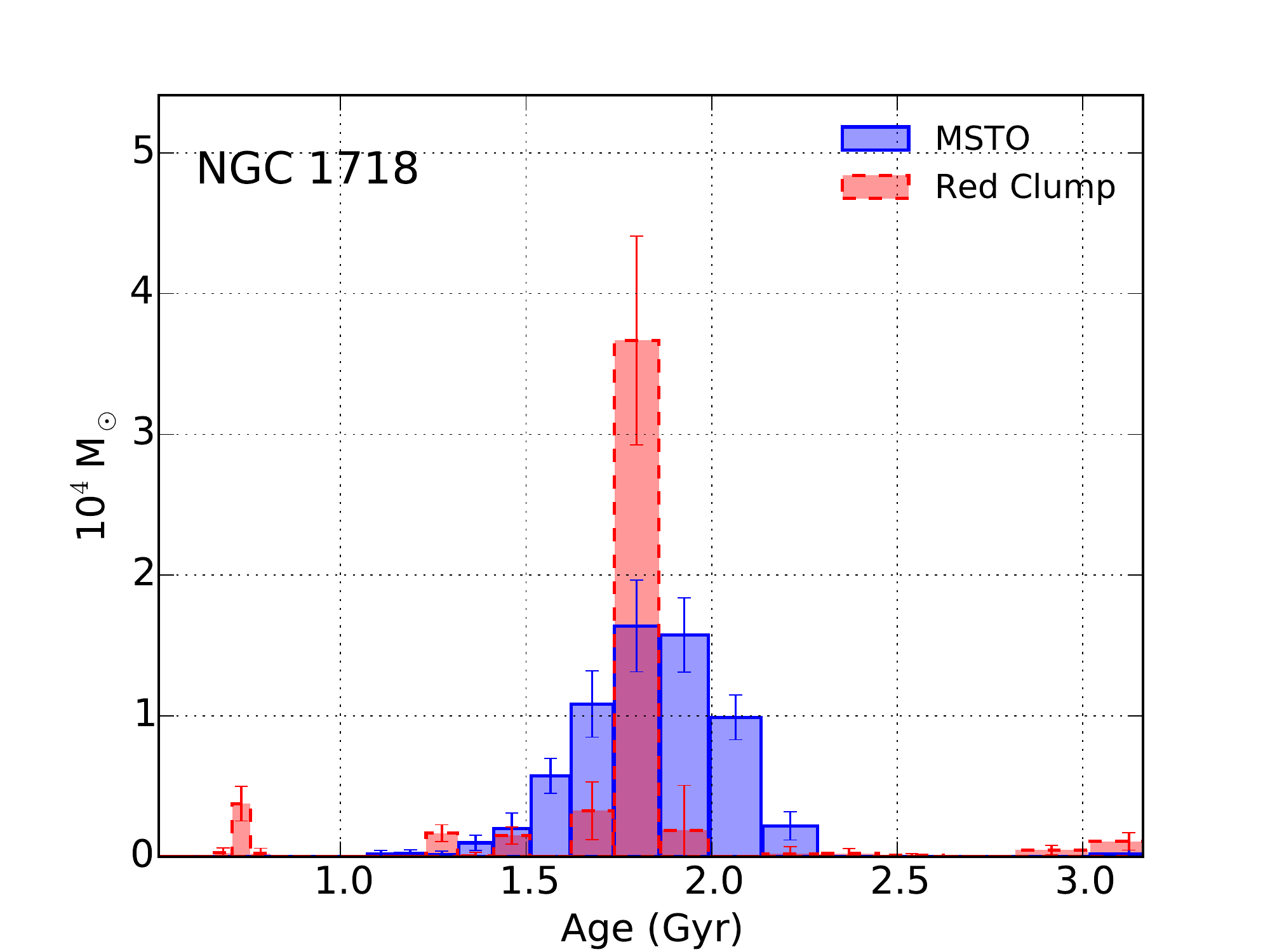}
   \\
  \end{tabular}            
  \caption{Same as Figure \ref{fig:ngc1783_cmd}, now for NGC 1718.}
   \label{fig:n1718_cmd}
\end{figure*}

\begin{figure*}
 \centering
 \begin{tabular}{cc}
  \includegraphics[width=9cm]{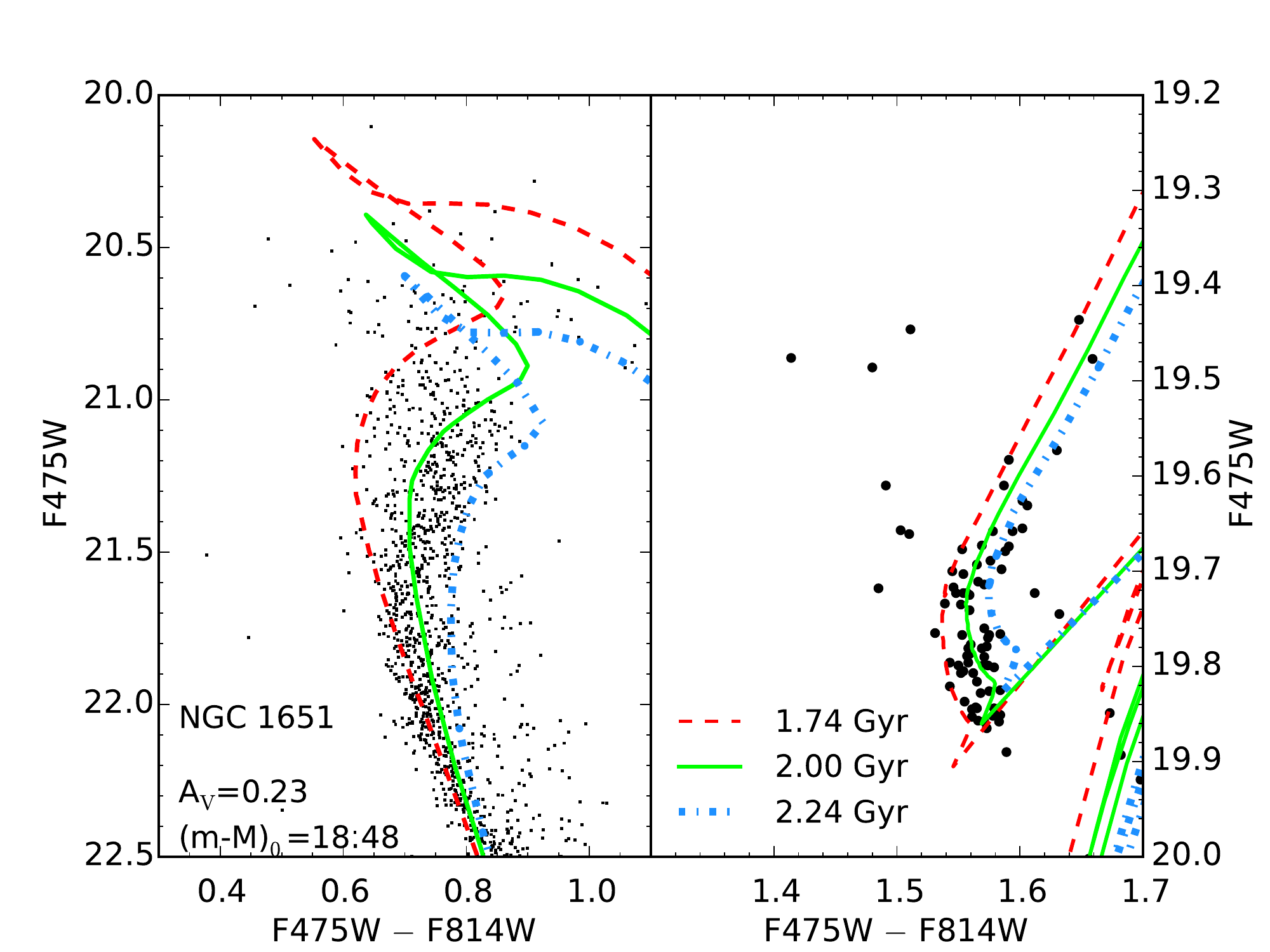} &
  \includegraphics[width=9cm]{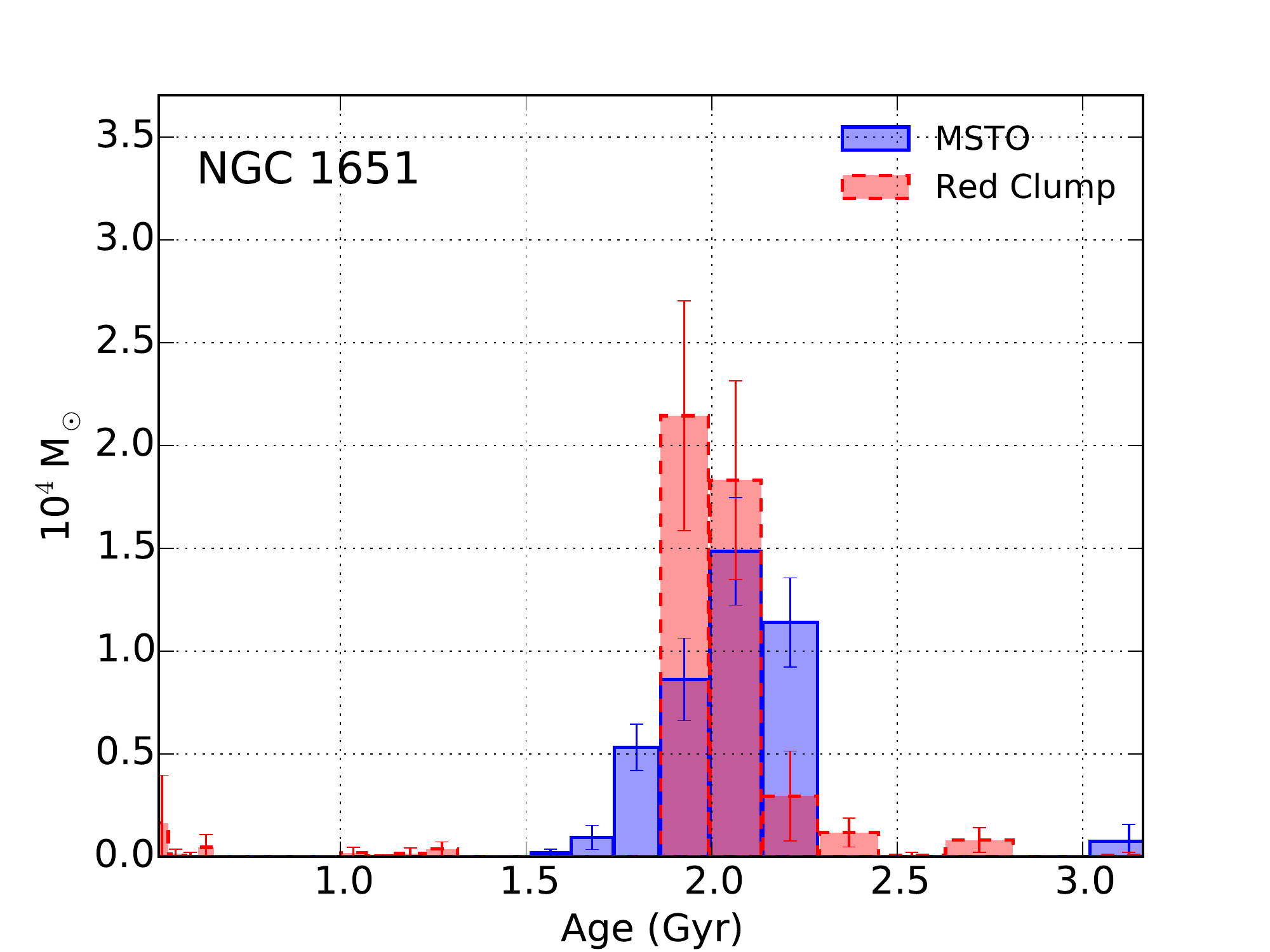}
   \\
  \end{tabular}            
  \caption{Same as Figure \ref{fig:ngc1783_cmd}, now for NGC 1651.}
   \label{fig:n1651_cmd}
\end{figure*}

\begin{figure*}
 \centering
 \begin{tabular}{cc}
  \includegraphics[width=9cm]{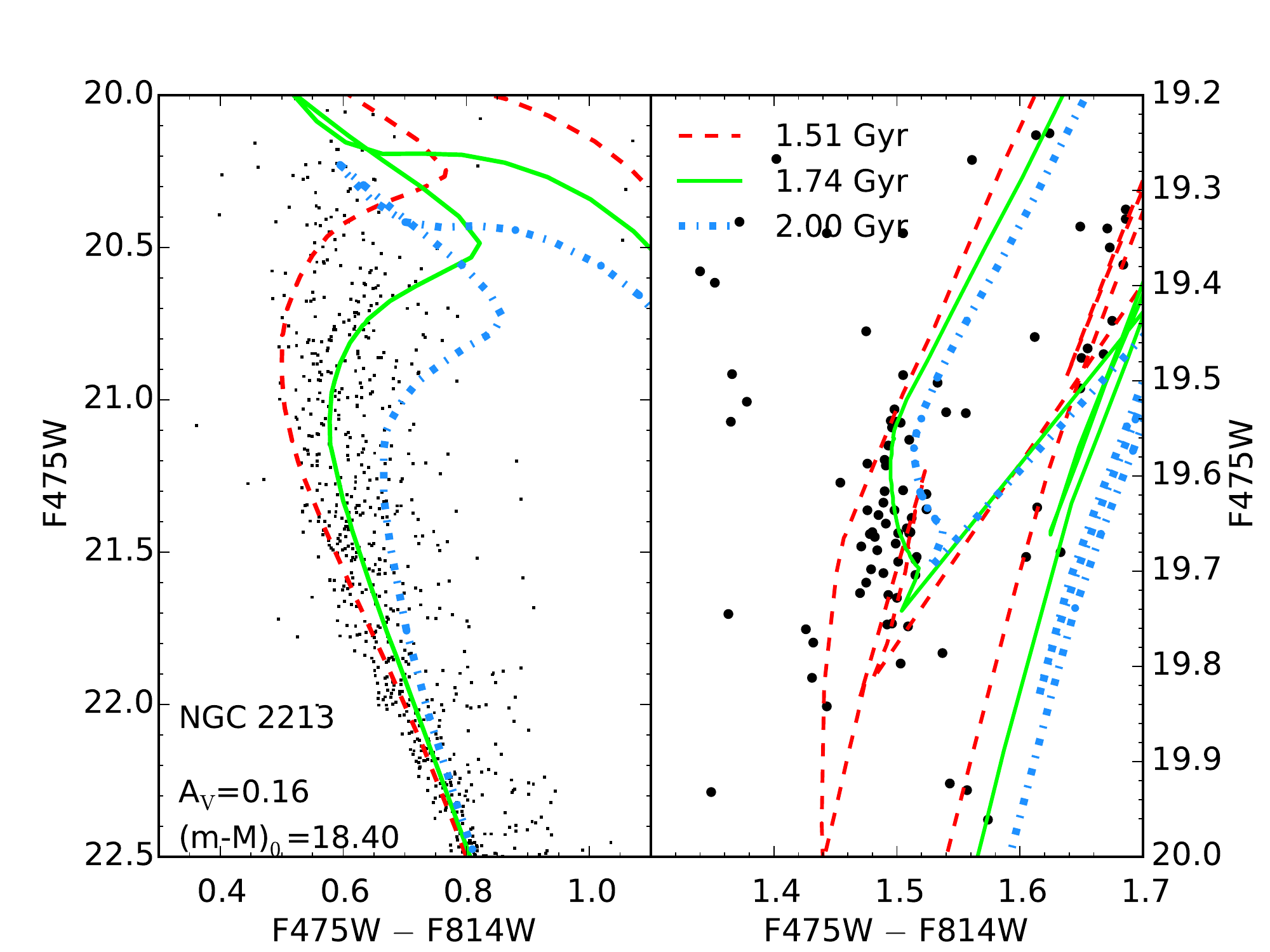} &
  \includegraphics[width=9cm]{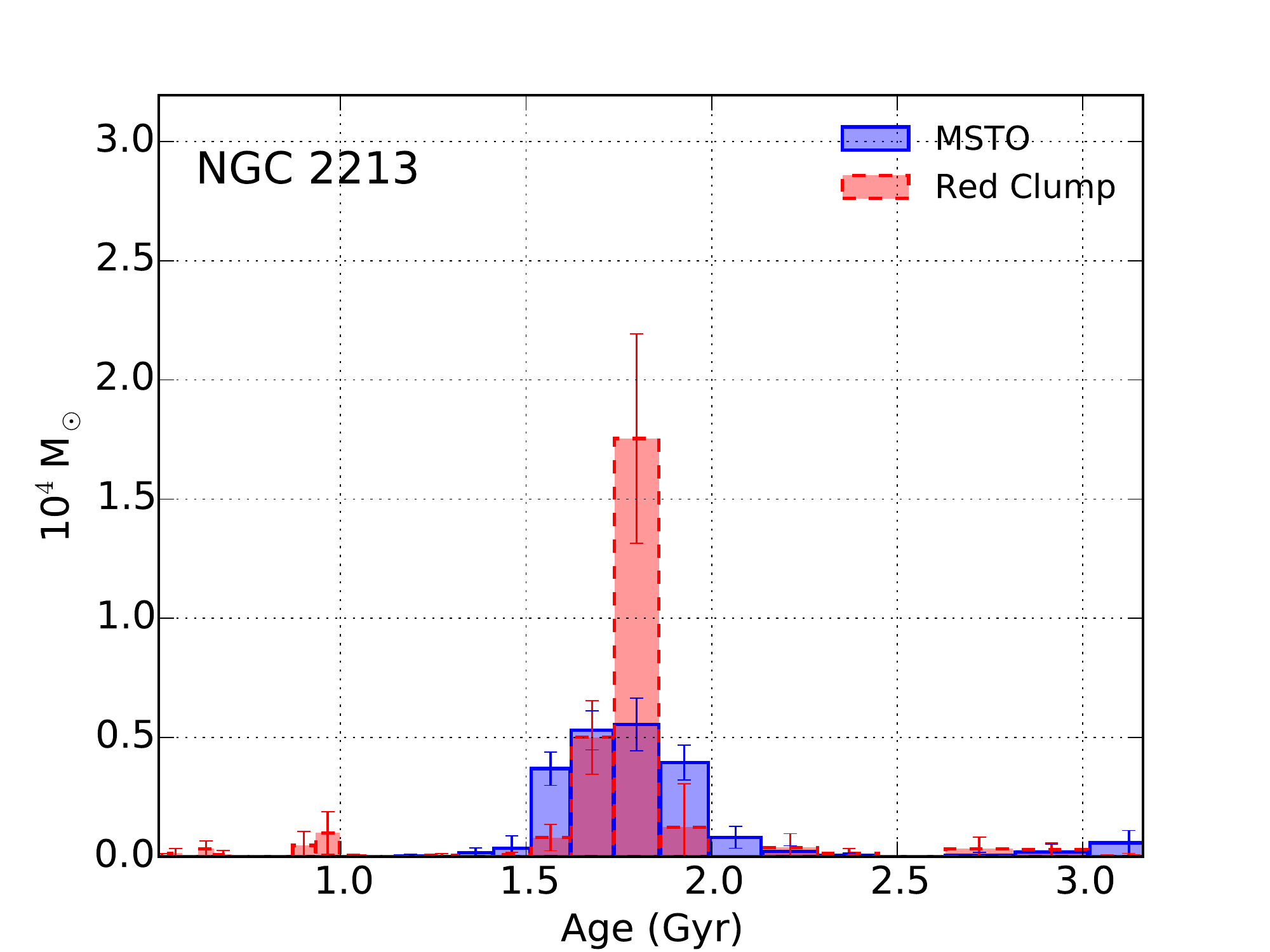}
   \\
  \end{tabular}            
  \caption{Same as Figure \ref{fig:ngc1783_cmd}, now for NGC 2213.}
   \label{fig:n2213_cmd}
\end{figure*}

\begin{figure*}
 \centering
 \begin{tabular}{cc}
  \includegraphics[width=9cm]{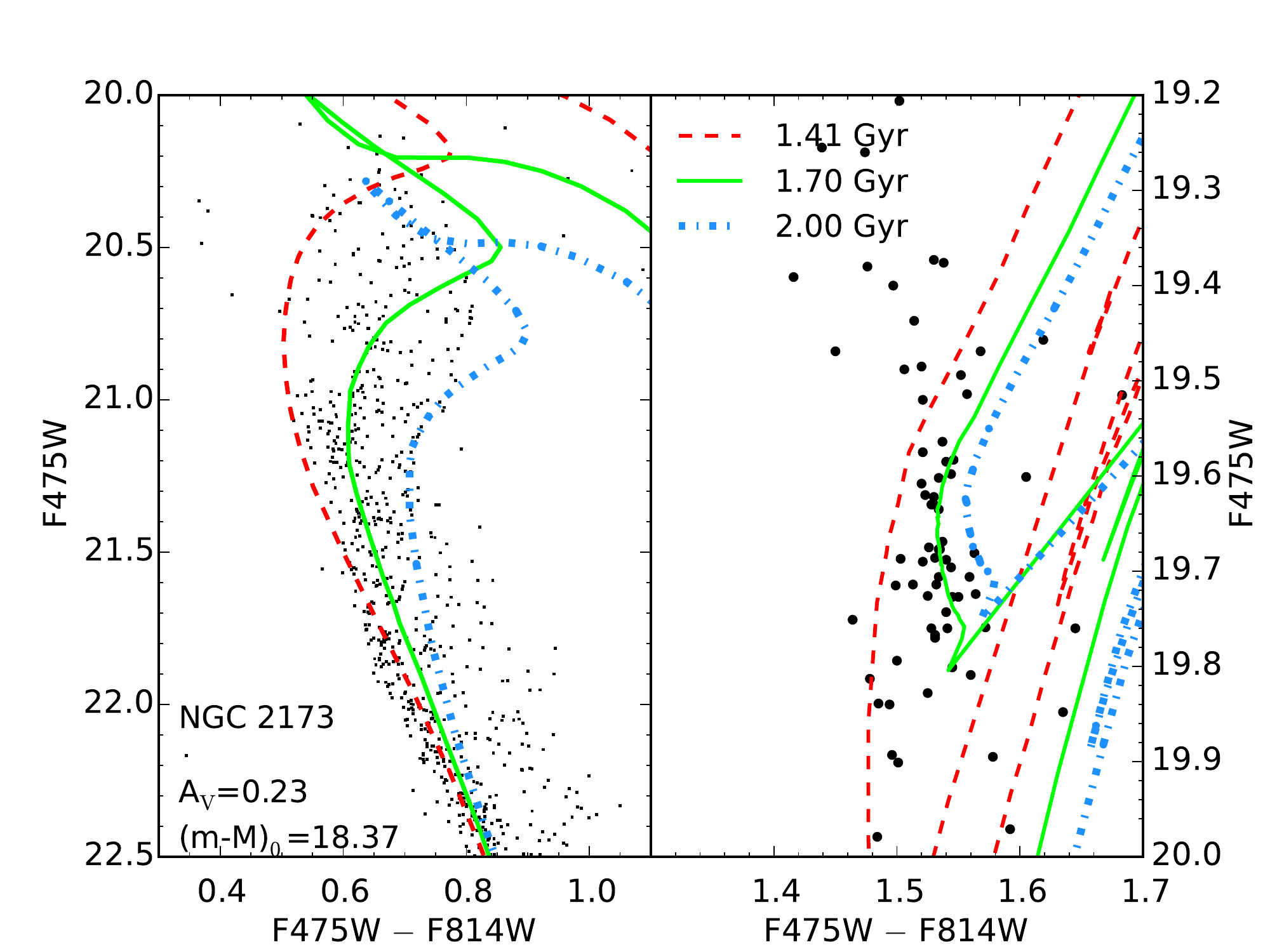} &
  \includegraphics[width=9cm]{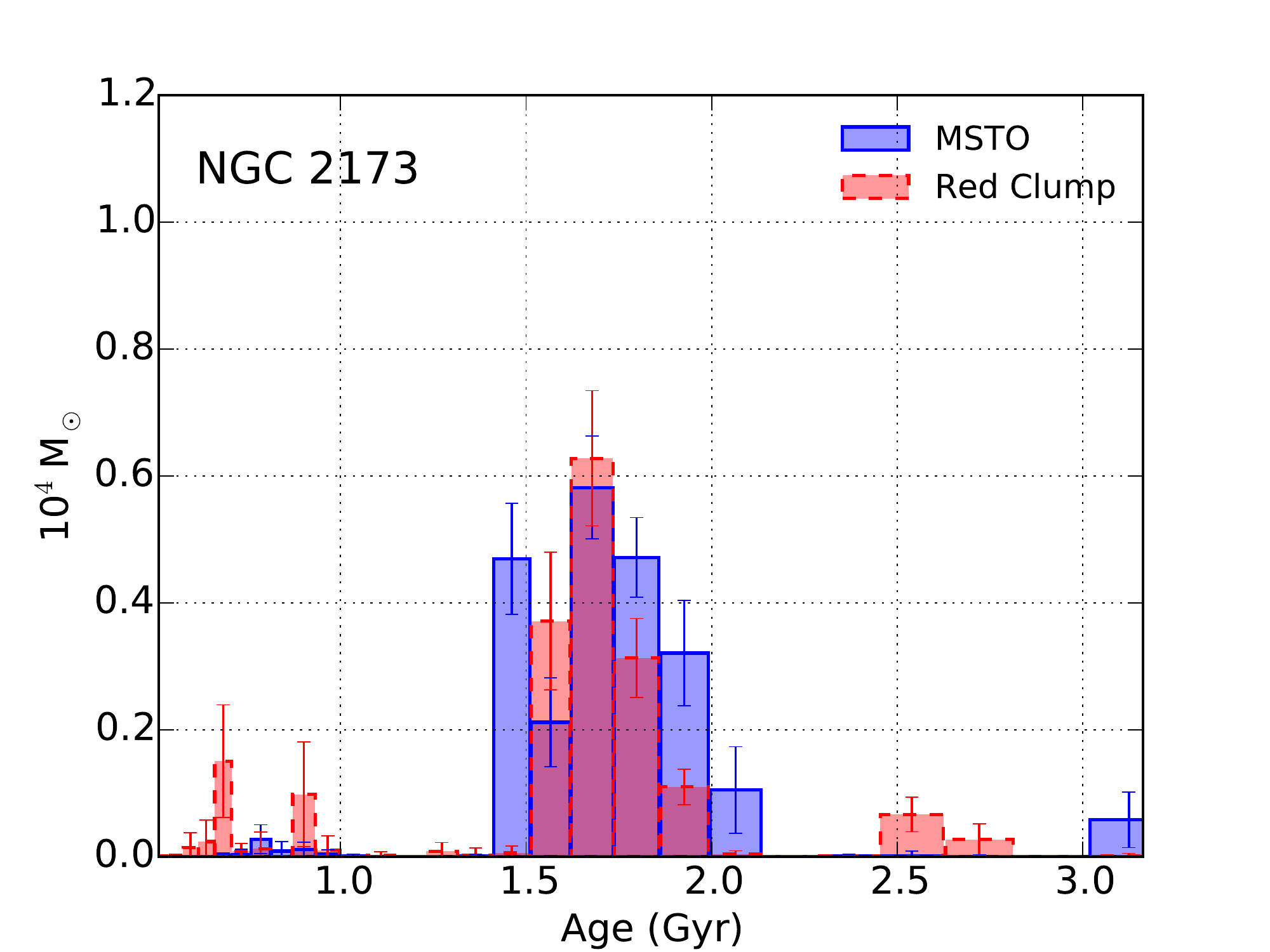}
   \\
  \end{tabular}            
 \caption{Same as Figure \ref{fig:ngc1783_cmd}, now for NGC 2173.}
   \label{fig:n2173_cmd}
\end{figure*}

\begin{figure*}
 \centering
 \begin{tabular}{cc}
  \includegraphics[width=9cm]{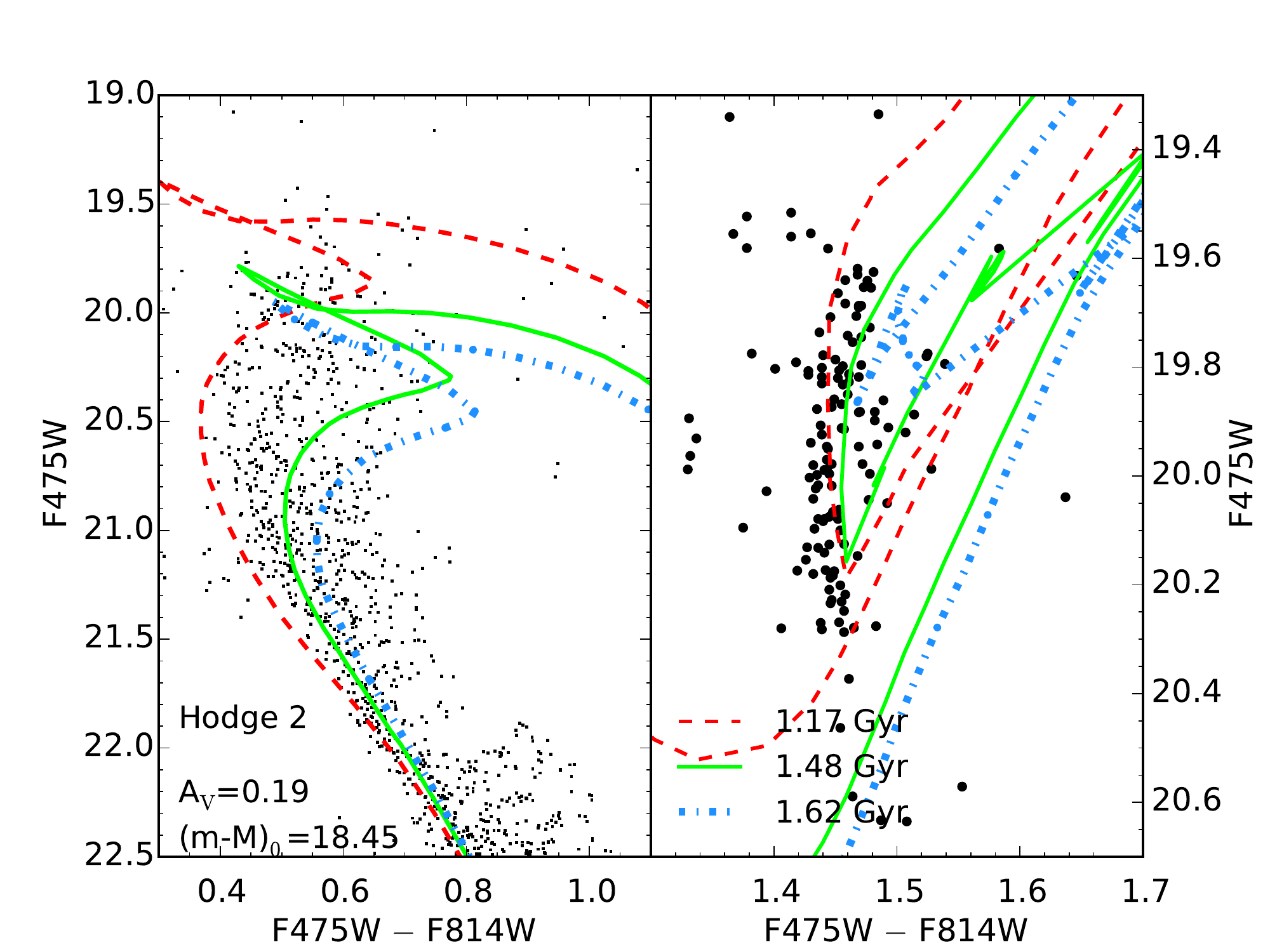} &
  \includegraphics[width=9cm]{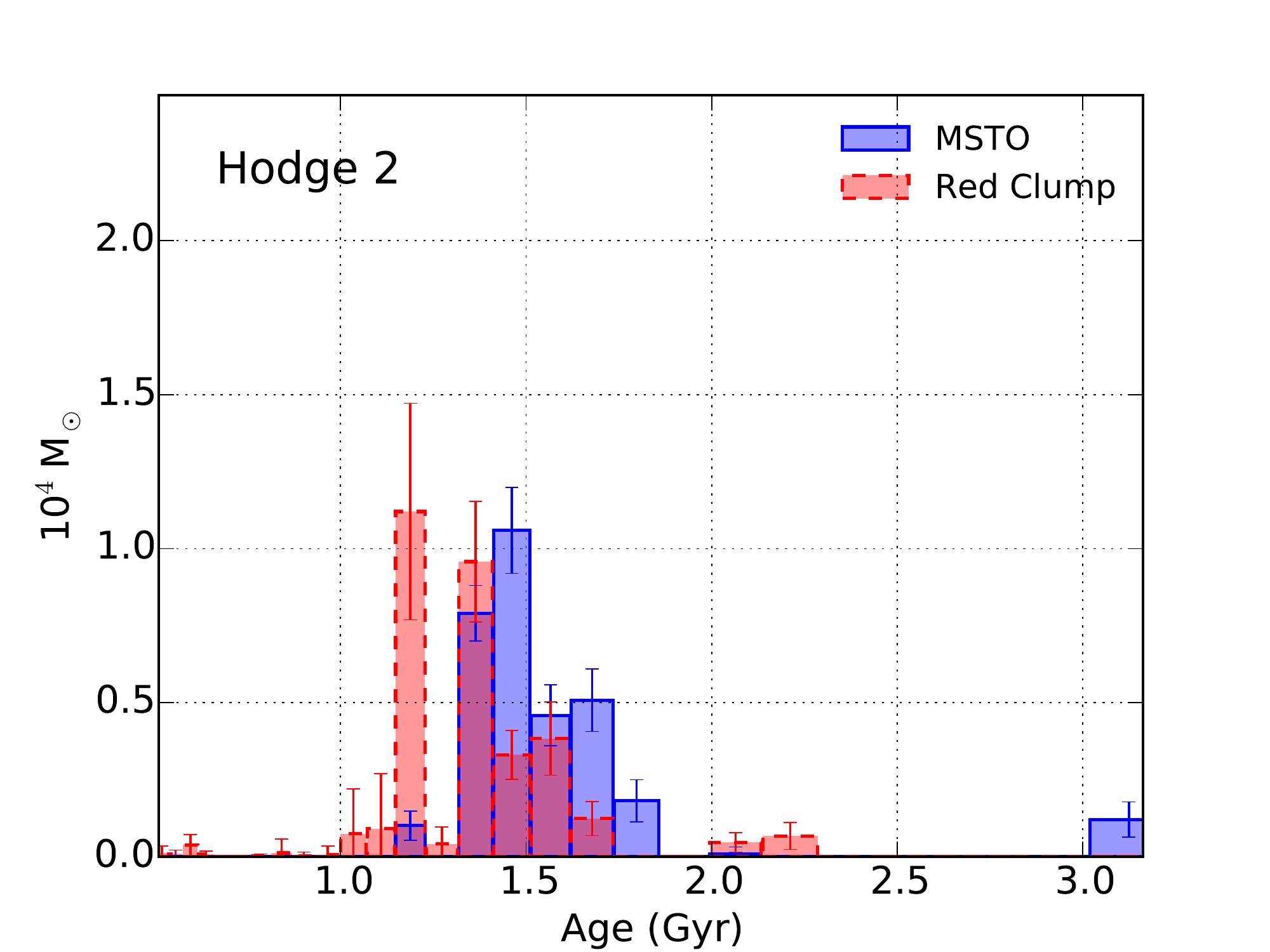}
   \\
  \end{tabular}            
\caption{Same as Figure \ref{fig:ngc1783_cmd}, now for Hodge 2.}   
   \label{fig:hodge2_cmd}
\end{figure*}

\begin{figure*}
 \centering
 \begin{tabular}{cc}
  \includegraphics[width=9cm]{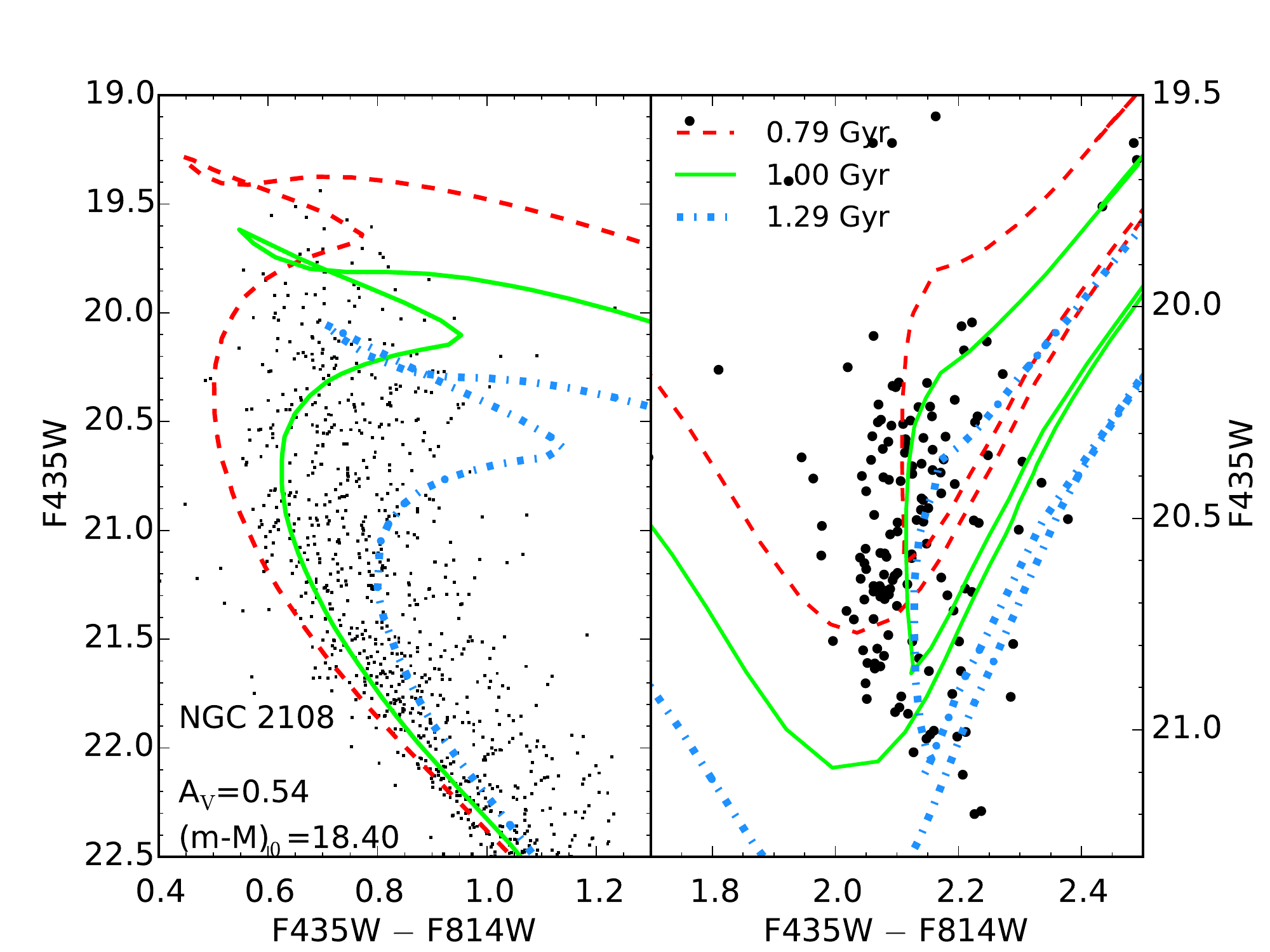} &
\includegraphics[width=9cm]{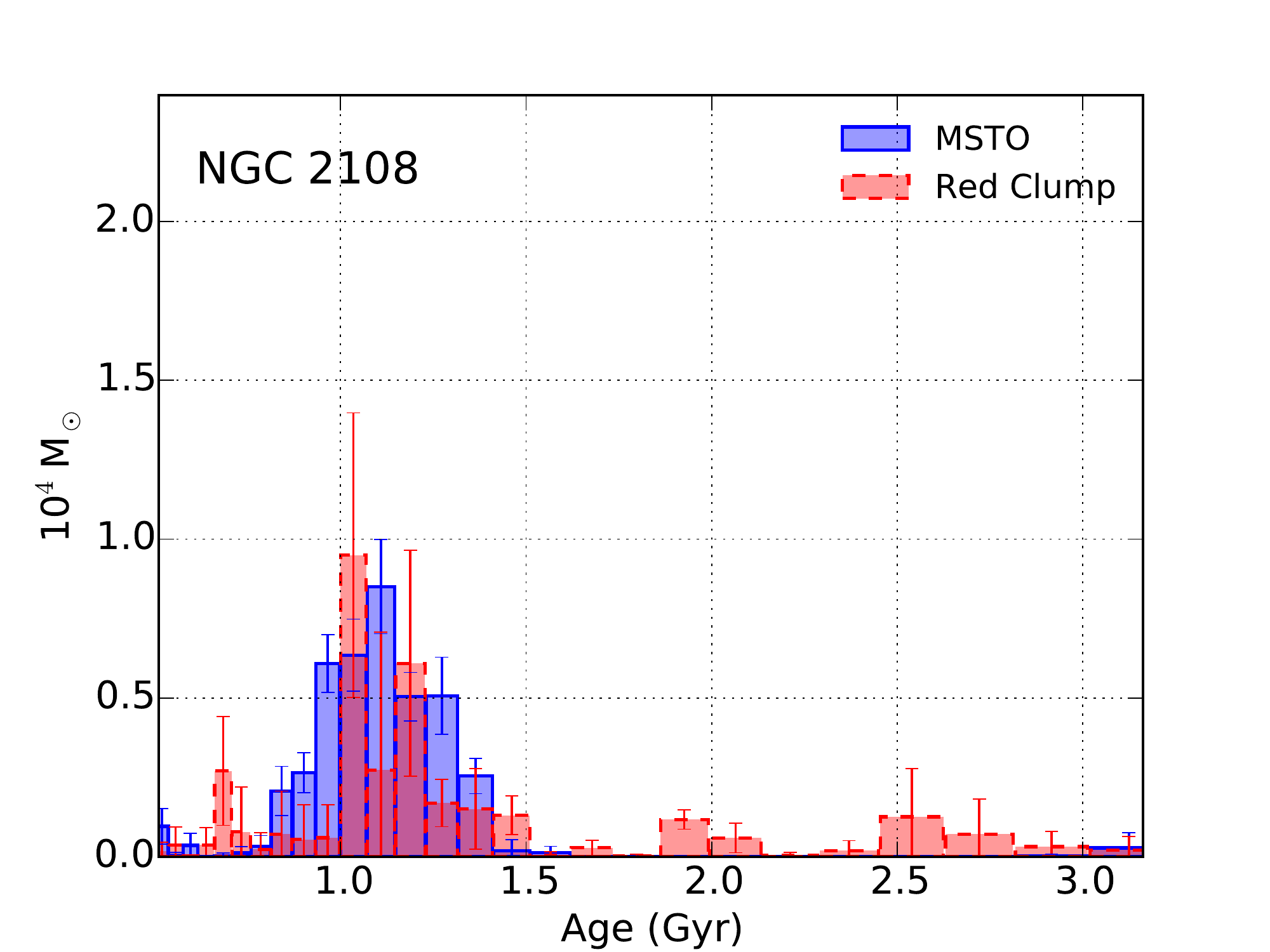}
   \\
  \end{tabular}            
  \caption{Same as Figure \ref{fig:ngc1783_cmd}, now for NGC 2108.}
   \label{fig:ngc2108_cmd}
\end{figure*}

\begin{figure*}
 \centering
 \begin{tabular}{cc}
  \includegraphics[width=9cm]{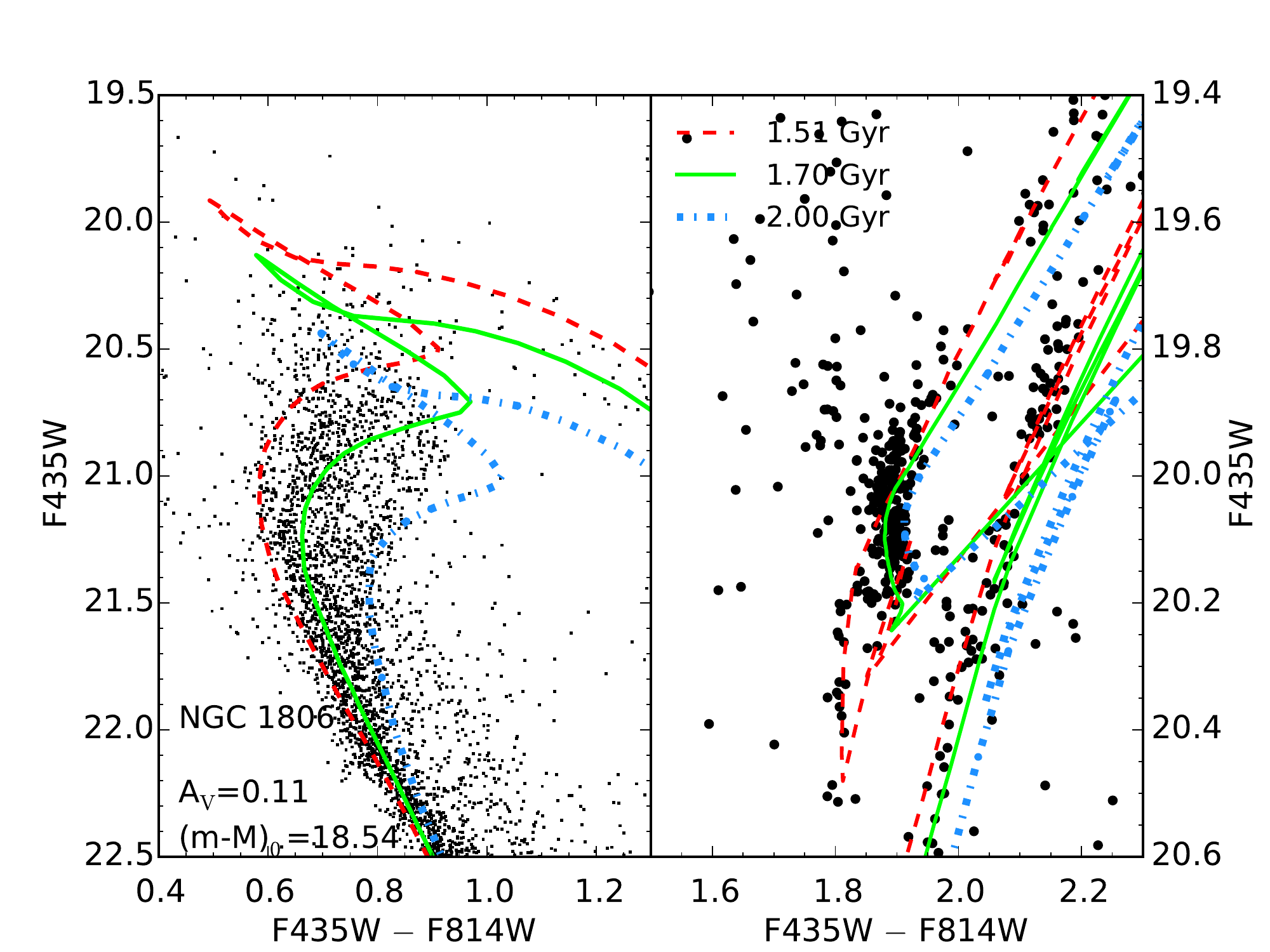} &
\includegraphics[width=9cm]{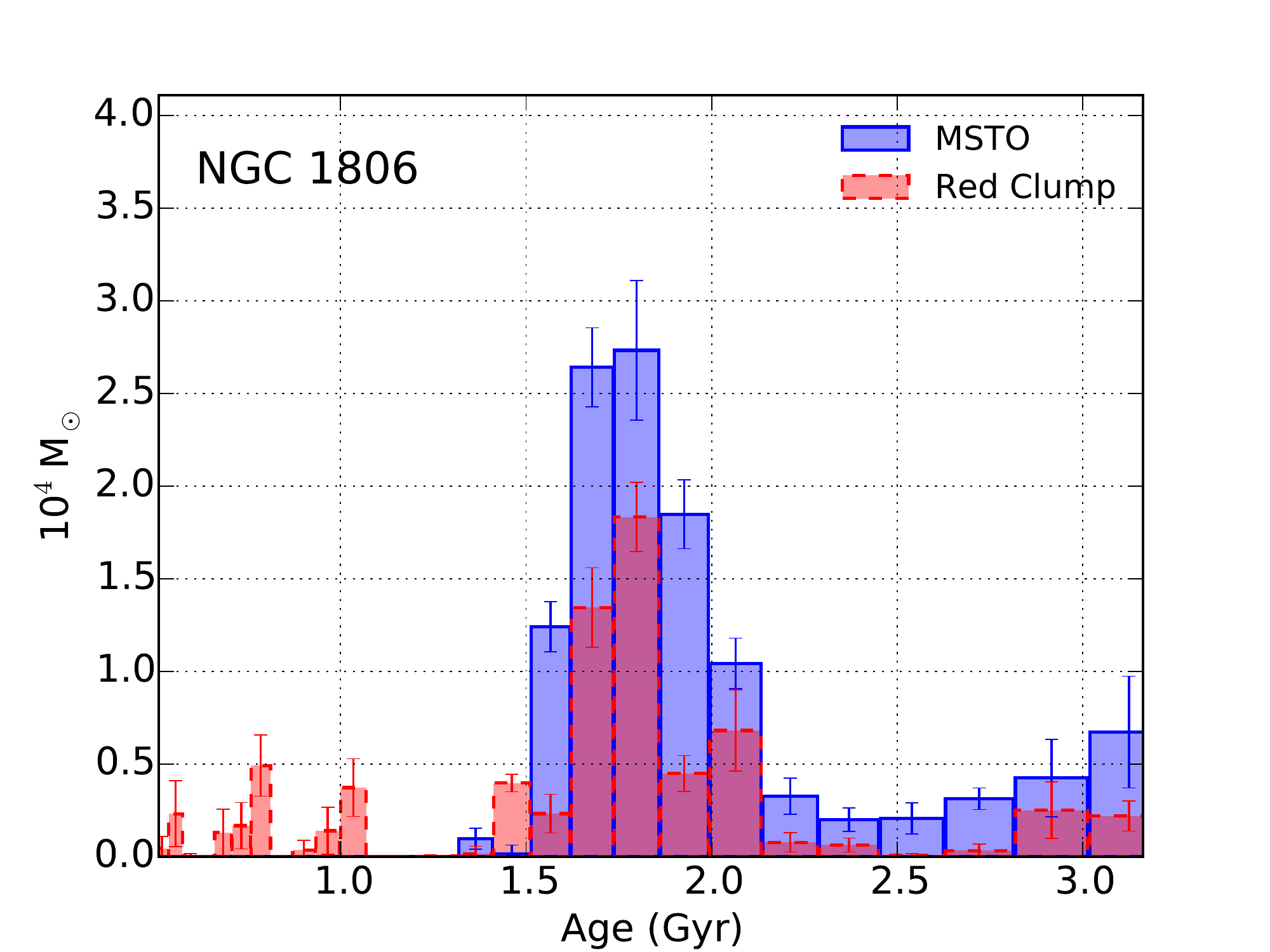}
   \\
  \end{tabular}            
  \caption{Same as Figure \ref{fig:ngc1783_cmd}, now for NGC 1806.}
   \label{fig:ngc1806_cmd}
\end{figure*}

\begin{figure*}
 \centering
 \begin{tabular}{cc}
  \includegraphics[width=9cm]{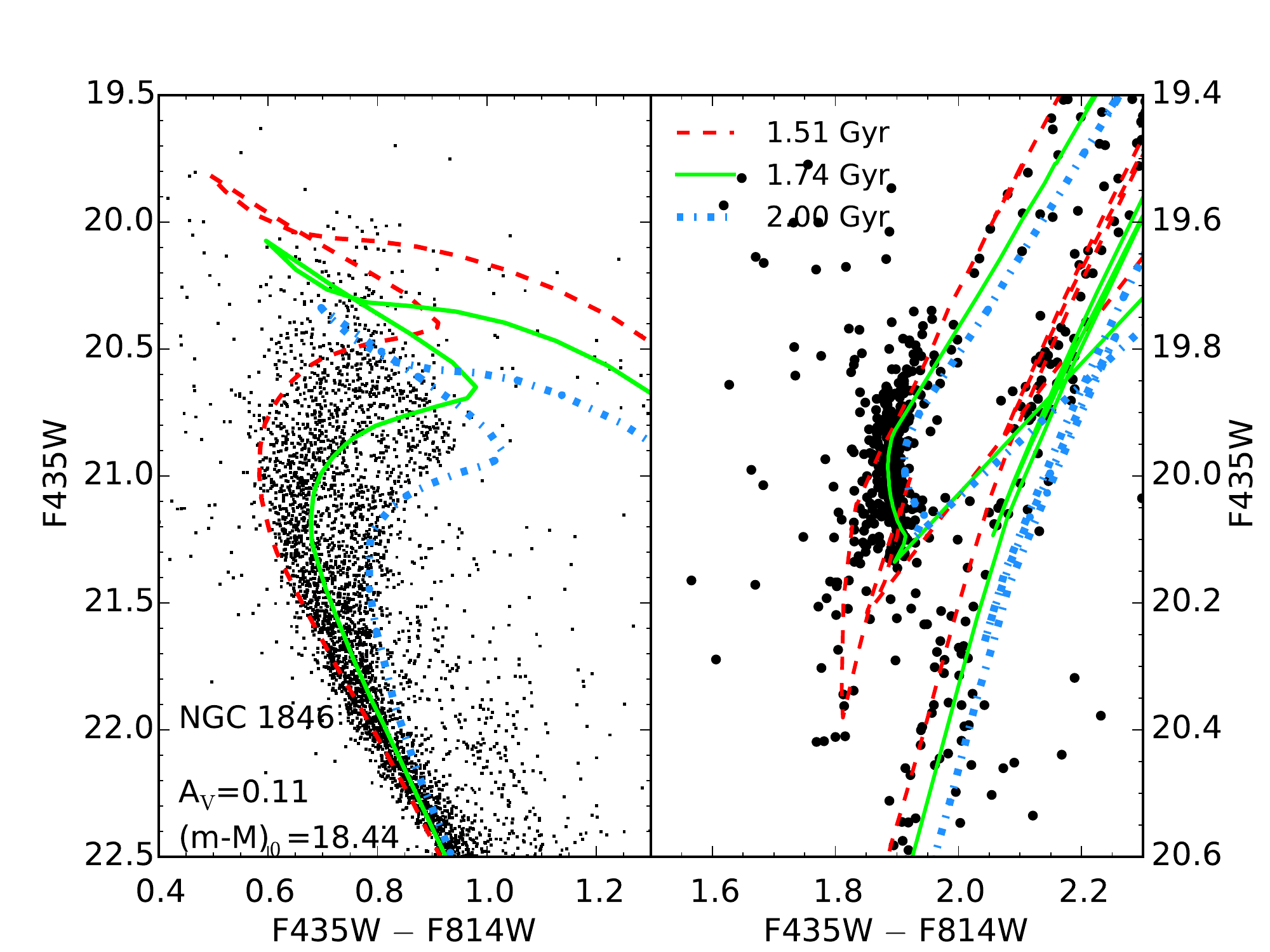} &
\includegraphics[width=9cm]{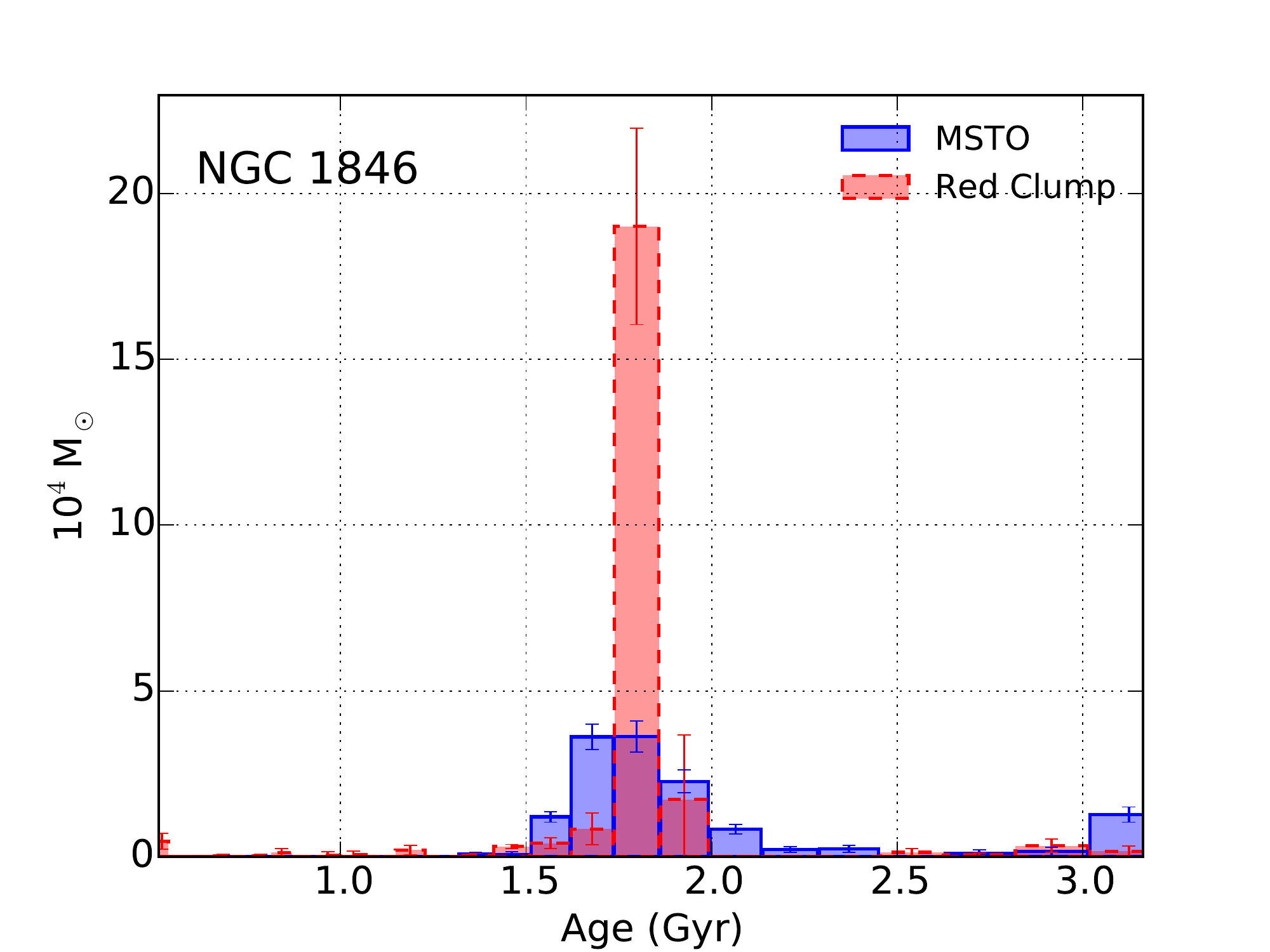}
   \\
  \end{tabular}            
  \caption{Same as Figure \ref{fig:ngc1783_cmd}, now for NGC 1846.}
   \label{fig:ngc1846_cmd}
\end{figure*}

The left panels of Figures \ref{fig:ngc1783_cmd} to \ref{fig:ngc1846_cmd} show the CMDs of each cluster in our sample, zoomed into two regions: The left of the two CMDs is centered at the turn-off region and the right one is centered around the red clump. Additionally, we show for every cluster the best fitting isochrones at three ages that span the extent of their turn-offs. The ages of the isochrones are given in the CMD showing the red clump. The parameters used to transform the isochrones from the absolute magnitude plane to the apparent magnitude plane (visual extinction, distance modulus) are stated in the lower left corner of the main sequence CMD. The typical age spreads of the isochrones describing the extended MSTO are of the order 400 to 500 Myr. The same isochrones are also shown in the plots showing the red clumps of the clusters. 
From a visual comparison we see that the red clumps of most of the clusters are compact in shape. However, it seems that some are not in agreement with a single age. Especially those clusters that have ages around 1.7 Gyr show an extension towards lower luminosities in their red clumps. These stars follow a younger isochrone. For the younger clusters (e.g. NGC~1987 and NGC~2108) the isochrones between $\sim$800 Myr and $\sim$1.2 Gyr coincide at the position of the red clump. Also NGC~2108 has a red clump which is quite extended in color. This might be a consequence of a remaining component of the differential reddening that was not removed completely.

The right panels of Figures \ref{fig:ngc1783_cmd} to \ref{fig:ngc1846_cmd} show the results of the SFH fitting. The blue histograms (blue solid line) correspond to the SFH of the MSTO region. The red histograms (red dashed line) show the age distribution inferred from the red clump region. 
All SFH plots show the expected 
statistical errors on the fitting, which we estimated from a series of bootstrapping. We created 20 bootstrap realizations for each cluster and fitted their SFH. The histograms show the mean of the results along with the 1 $\sigma$ error bars. 

We find that for most of the clusters in our sample the age spread resulting from the red clump region is smaller than the one from the MSTO. However, the red clumps of all but two clusters are not in agreement with a single age. The post main sequence age distributions of LW~431, NGC~2203,  NGC~1651, NGC~2213, NGC~2173 and NGC~1846 are narrower relative to the ones inferred from the width of the MSTO.  The red clumps of NGC~1783, Hodge~2, NGC~2108 and NGC~1806 show an extent which is comparable to the age spread of the MSTO. As we already pointed out, NGC~2108 is affected by differential extinction. In Section \ref{sec:diffext} we corrected the CMD for this effect, however, there is the possibility that it is not removed completely. Therefore some remaining variation in reddening across the cluster might introduce some scatter in the SFH of this cluster. This might also be the cause for the low level of fitted star formation at higher ages.
Only NGC~1718 and NGC~1987 show a single peak of star formation in the red clump, i.e. it seems to be consistent with a single isochrone. In Table \ref{tab:Cluster_Param} we summarize the ages and the maximum age spreads of the clusters inferred from the fitted SFH of the red clumps.

To assess the reliability of the fitted SFH we created for each cluster in our sample two artificial clusters, one with the fitted SFH from the MSTO and one with the age distribution inferred from the red clump region.  For this we made use of the function \texttt{repop} within StarFISH. The clusters were created with the same parameters that we used to fit the SFH. Also we normalized the number of stars to be the same in both the observed and the synthetic clusters. In Appendix A (Figure \ref{fig:repop}) we present the CMDs of all artificial clusters created. The two upper panels of each subplot show the clusters created from the fitted SFH of the red clump region, whereas the lower two panels display the artificial clusters using the recovered SFH from the MSTO. The CMDs have the same limits as the CMDs of their real counterparts shown in Figures \ref{fig:ngc1783_cmd} to \ref{fig:ngc1846_cmd} and the overlaid isochrones are also the same for a better comparison. The ages of the isochrones are given in the lower right panel of each cluster's subplot. 
We see that for the majority of clusters, the reconstructed MSTO and red clump from the fitted SFH of the respective CMD region are in good agreement with the observations (see upper right and lower left panel of each subplot in Figure \ref{fig:repop}). The fact that the extended MSTO in the simulations has no smooth structure but shows distinct peaks is due to our finite age resolution in the fitting (log($\Delta$age)=0.3) and the fact that StarFISH does not interpolate between isochrones. Each peak corresponds to an individual value in the histogram of the fitted SFH. 
We also notice that in clusters like NGC~1987, NGC~1718, NGC~2203 or NGC~1846 the MSTO resulting from the red clump fit is narrower than the observed ones and more concentrated.
The red clumps in clusters like NGC~1783 or NGC~1806 are more compact as would be expected from the fitted SFH. This suggests that the fit over-estimates the range of ages present in these clusters.
By comparing the CMDs of the two younger clusters Hodge~2 and NGC~2108 with the artificial clusters, it is noticeable that the reconstructed SFH of the red clump region is a poor fit to the observed data. Therefore, these last two clusters would not provide any meaningful contribution to our overall results.

\section{Correlations with Age\label{sec:goudfrooij15}}

\citet{Goudfrooij14} studied in detail the MSTO of a sample of 18 intermediate-age LMC and SMC clusters. Based on observed properties of the clusters, they found that there is a correlation between the mass of the clusters and the width of the MSTO, in terms of the full-width-half-maximum (FWHM). Moreover, the correlation becomes even stronger, when the early mass of the clusters is used. \citet{Goudfrooij14} assumed that the clusters underwent strong mass loss in the past and calculated the early masses of the clusters using correction factors that are a strong function of age. The existence of a correlation between the FWHM and 'corrected' initial mass of the clusters led \citet{Goudfrooij14} to conclude that a physical property of the cluster (i.e. mass or escape velocity) was the determinant whether clusters could host extended star-formation events.

\begin{figure}
 \begin{tabular}{c}
  \includegraphics[width=8cm]{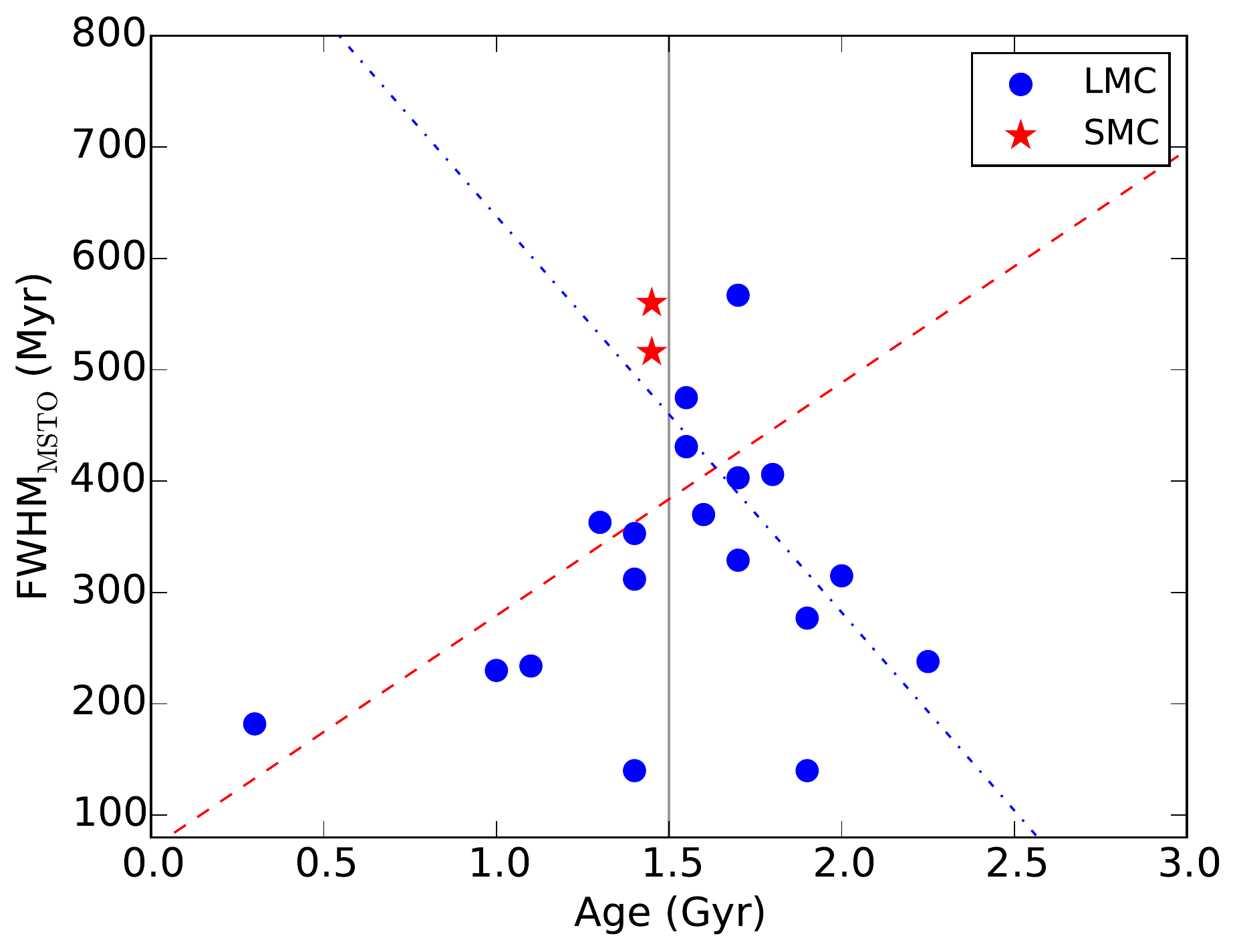} \\
  \includegraphics[width=8cm]{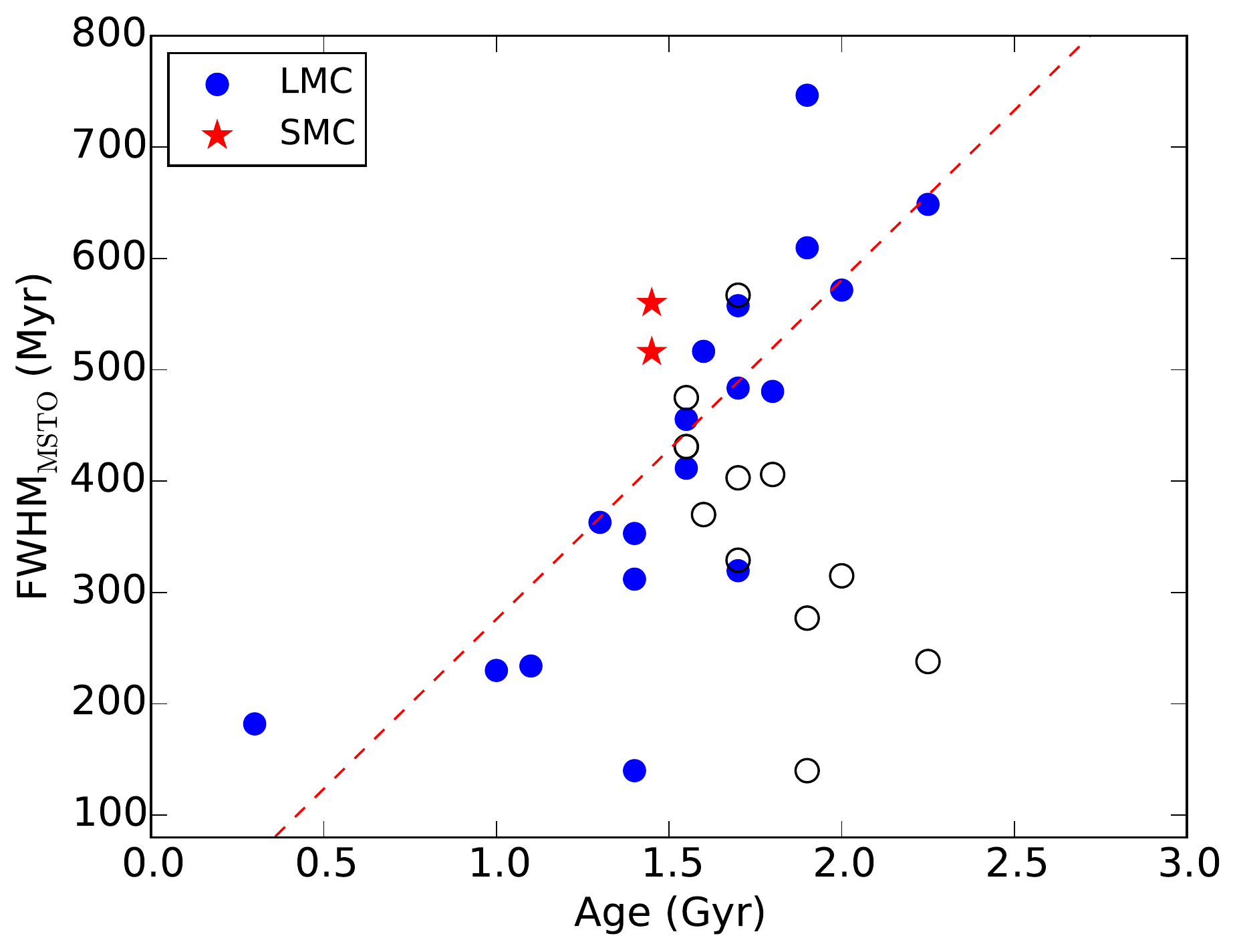} \\
 \end{tabular}
  \caption{\textit{Upper panel:} FWHM of the turn-off of clusters in the \citet{Goudfrooij14} sample as a function of their age. In this plot we also include the 300 Myr old cluster NGC~1856 which also shows an extended MSTO. Clusters in the LMC are marked with blue circles whereas the two SMC clusters are denoted with a red asterisk. The width of the MSTO seems to increase with the age of the cluster until an age of 1.5-1.7 Gyr and then decreases again. We divided the plot at 1.5 Gyr (vertical solid gray line) in two parts and performed a linear regression fit to the clusters in the two parts separately. The fits are indicated by the red dashed line (younger clusters) and the blue dash-dotted line (older clusters).\\
  \textit{Lower panel:} Same as the upper panel, but now we mirrored the older clusters at the horizontal plain that goes through the intersection point between the two liner regression fits in the upper panel. We did this to quantify the strength of the correlation of the FWHM with age. The original positions of the clusters are marked with black open circles. The red dashed line is a linear regression fit to the linearized cluster sample.}
   \label{fig:age_vs_fwhm}
\end{figure}

Here we look for additional correlations, 
such as the age of the clusters itself.
For this we took the values of the age and the FWHM of the clusters' turn-off as they were given in \citet{Goudfrooij14} and plotted the FWHM as a function of cluster age (see Figure \ref{fig:age_vs_fwhm} upper panel). In this plot we also included the 300 Myr old cluster NGC~1856, that also shows an extended turn-off (see Section \ref{sec:disc} for more detail). The blue filled circles mark the LMC clusters and the red asterisk symbols indicate the two SMC clusters. The turn-off width seems to be correlated with the cluster age in the sense that it first increases as the clusters get older, until an age of 1.5-1.7 Gyr. Then the extent of the MSTO decreases again. Such a behavior is expected if stellar rotation causes the spread in the MSTO \citep{BrandtHuang15a, Niederhofer15b}. However, there are two outliers to this relation, NGC~1795 at 1.4~Gyr and IC~2146 at 1.9~Gyr that have a compact MSTO structure. \citet{Goudfrooij14} additionally included these two low-mass clusters from the sample of \citet{Milone09} in their mass vs FWHM relation.

We divide now the diagram into two parts at 1.5 Gyr and have a closer look at the 'young' and 'old' branch separately (we note that the general result does not depend on the exact choice of the age at which we divide the data). To quantify the correlation we follow the procedure described in \citet{Goudfrooij14} and also performed a Kendall $\tau$ test for both of the branches. We performed this test two times, once excluding the outliers NGC~1795 and IC~2146 and once including them. The values in parenthesis give the results without the two clusters. We found that the two-sided $\tau$ value is 0.59 (0.82) for the younger clusters and -0.6 (-0.6) for the older clusters. The $p_{\tau}$ value of the test which gives the probability that the values are uncorrelated is 2.7 \% (0.5 \%) for the increasing branch and 0.7 \% (1.0 \%) for the decreasing one. For the FWHM vs early cluster mass relation, \citet{Goudfrooij14} found a $\tau$ of 0.82 and a $p_{\tau}$ of 1\%, and for the FWHM vs early escape velocity relation they obtained a $\tau$ value of 0.98 and a $p_{\tau}$ of 0.2\%. 

To quantify the strength of the overall correlation, we turned the '$\Lambda$' shaped function into a monotonically increasing function by mirroring the older clusters at the vertical line that goes through the intersection point of the two linear fits. This 'linearized' function is shown in the lower panel of Figure \ref{fig:age_vs_fwhm}. The original positions of the older clusters in the diagram are indicated by black open circles. The red dashed line is a linear regression fit to the 'linearized' data. We performed now again a Kendall $\tau$ test to the entire data. Also here, the values in parenthesis give the results excluding NGC~1795 and IC~2146. The Kendall $\tau$ value is now 0.65 (0.65) and the $p_{\tau}$ value gives a probability of 0.004 \% (0.009 \%) that a correlation is absent in the data. In terms of the scenario proposed by \citet{Goudfrooij14} where the width of the MSTO is caused by an age spread, such a correlation with age would not be expected. Although the Kendall $\tau$ test, which is not very sensitive to single outliers, suggests that there is a strong correlation between the FWHM and the age of the clusters, there are clusters that do not follow this relation. There is also the possibility that the found $\Lambda$ shaped curve is an envelope that gives an upper limit to the FWHM as a function of age. We note that in the stellar rotation scenario, the width of the MSTO depends on the rotation distribution within the cluster, which may vary between clusters.

\section{Discussion and Conclusions\label{sec:disc}}

In this study, we have analyzed the CMDs of a sample of 12 intermediate-age LMC clusters which all show an extended MSTO. We have fitted the SFH in two portions of the CMD, a) centered on the extended MSTO and b) focused on the red clump, SGB and lower red giant branch. 

To reconstruct the SFH of the 12 clusters in our sample we made use of the PARSEC 1.2S isochrone set. Our main result is that for ten out of 12 clusters the fitted SFH inferred from the red clump region is not compatible with the range of stellar ages resulting from the MSTO but it is narrower. 
However, the red clumps also show an extended SFH and are not in agreement with a coeval stellar population. Only in NGC~1806 (and to a lesser extent NGC~1783) are the inferred SFHs derived from the red clump and MSTO region comparable. However, as can be seen in Figure \ref{fig:repop} the observed red clump has a more compact structure than would be expected from the fitted SFH, suggesting that we over-estimate its age spread. We also found that for most of the clusters, the peak of the reconstructed red clump and MSTO ages coincide. There are two exceptions to this, Hodge~2 and NGC~2108. In the first cluster, we find an additional peak in the red clump SFH that has no counterpart in the MSTO. In the second one, NGC~2108, the age distribution of the red clump shows two peaks, one on each side of the peak of the MSTO SFH.

The results that we obtained in this work differ from the findings by \citet{Li14} and \citet{BastianNiederhofer15}. These studies analyzed the morphology of the SGB and the red clump of NGC~1651, NGC~1806 and NGC~1846 and compared it to what would be expected if the MSTO would be due to an age spread. Both studies adopted the \citet{Marigo08} isochrones. They concluded that a) SGB and red clump position and morphology were not consistent with the inferred age spreads from the MSTO and b) the SGB and the red clump follow the youngest isochrone going through the turn-off. However, our findings based on the PARSEC isochrone set suggest on the one hand that the red clump in these cluster is more extended than would be expected from a single age and on the other hand that the peak age of the red clump and MSTO agree with each other.

The results are in line with the recent study by \citet{Goudfrooij15}, in which the analysis of the MSTO is coupled with the study of the SGB morphology in the clusters NGC~1651, NGC~1783, NGC~1806 and NGC~1846. In contrast to the studies by \citet{Li14} and \citet{BastianNiederhofer15}, they argue that the position of the SGB should not be used (due to uncertainties in the stellar models) but rather that the width is a better indicator of age spreads. The authors find that the width of the SGB of these clusters does agree with an extended SFH using the \citet{Marigo08} set of isochrones. However, Li et al. (in preparation) find that the width of the SGB of NGC~411 is significantly lower than expected from the MSTO width, if an age spread is present.

Additionally, \citet{Goudfrooij15} studied the red clump of NGC~1806 using the new, finer mass resolution, PARSEC isochrones. The authors find that the red clump is more extended than expected for a single age population, with a width similar to that found from the MSTO. This is in good agreement with the results presented here. \citet{Goudfrooij15} also highlighted
an elongated extension that lies below the otherwise compact red clump. This additional 'tail' was interpreted as a 'double red clump' similar to what was found in the SMC clusters NGC~411 and NGC~419 (\citealt{Girardi09} and \citealt{Girardi13}). These stars are believed to be slightly more massive, and therefore younger stars that are just massive enough to avoid the state of degenerate core He-burning.
Overall, our results agree with the findings by \citet{Goudfrooij15} in terms that the range of ages in the red clump is extended and not compatible with a coeval population. However, our fitting of the SFH suggests that the age spread in the red clump in most clusters in our sample is smaller than what would be expected from the MSTO. We also note that stellar rotation is expected to broaden the red clump to some degree.

In conclusion we can state that the results we found in this work are ambiguous and do neither confirm nor exclude large age spreads in intermediate-age LMC clusters. Due to the uncertainties in the stellar models, especially at post-MS phases, an analysis of the CMDs of extended MSTO clusters alone seems insufficient to test the nature of the extended MSTO. For this reason, independent and complementary observations should be used to test whether age spreads are present within massive clusters. There exists already a variety of observations that are at odds with the predictions of an age spread scenario. 
In the following we list the arguments that disfavour an age spread as cause of the extended MSTO regions.

\subsection{Other Considerations on Age Spreads within Clusters}

If age spreads are present in these intermediate-age LMC clusters, as a consequence young clusters with similar properties should have age spreads or signs of star formation, as well. However, no evidence of ongoing star formation has been found in young massive clusters in the LMC and other galaxies (e.g. \citealt{Bastian13b, Cabrera-Ziri14}). \citet{BastianSilva13} and \citet{Niederhofer15a} found no evidence for age spreads comparable in duration to that suggested for the intermediate-age clusters in a sample of 14 young massive LMC clusters. There could be some spreads present in the young clusters (the case of NGC~1856 will be discussed in more detail below), however, these spreads would be much smaller that those inferred for the intermediate-age clusters, and the spread is a function of cluster age \citep{Niederhofer15b}.

Also, the formation of a second generation of stars requires a gas reservoir inside the clusters out of which the new stars can be formed. Several studies have searched for gas inside young and massive clusters and up to now have not found any (e.g. \citealt{BastianStrader14, Cabrera-Ziri15}; see also \citealt{Longmore15}). \citet{Bastian14} and \citet{Hollyhead15} showed that young massive clusters expel their gas to over 200 pc in just a few Myr. It is not clear how this gas should be re-accreted to the cluster again. 

\citet{Goudfrooij14} found a correlation between the width of the MSTO and the escape velocity of the cluster at young ages after applying large correction factors to the current values.
To calculate these early escape velocities, \citet{Goudfrooij14} adopted the models by \citet{D'Ercole08} which were designed for highly extended and tidally limited clusters within a strong tidal field. However, it is not clear if such models would be expected to apply in the LMC and SMC, as the the tidal field (especially for the SMC clusters far out in the galaxy) in these galaxies is much weaker than in the inner regions of the Milky Way, where the original models were applied. \citet{Goudfrooij14} concluded that there is a velocity threshold beyond which the cluster is able to retain the slow wind ejecta of evolved stars to form a second generation of stars inside the cluster. They determined this threshold to be at 12-15~km~s$^{-1}$.  In the model of \citet{Goudfrooij14} the clusters are expected to be 3-4 times more massive in the past. But the clusters appear to be well contained within their tidal radii
\citep{Glatt11}, so their mass loss is expected to be very low, unless their current positions have only recently been acquired. 

Finally, if age spreads were present in the clusters and the second generation of stars would have been formed (at least partially) out of the ejecta of AGB stars, then chemical abundance spread would be expected inside the cluster. However, \citet{Mucciarelli08, Mucciarelli11, Mucciarelli14} analyzed the extended MSTO clusters NGC~1651, NGC~1783, NGC~1806, NGC~1846 and NGC ~2173 and did not find evidence for significant abundance spreads. 

\subsection{An Alternative Explanation}

As an alternative scenario to age spreads, \citet{BastianDeMink09} proposed the idea that the broadening of the MSTO could be caused by rotating stars. However, their results were disputed by \citet{Girardi11}. Later, \citet{Yang13} showed in their study that a MSTO broadening indeed would be expected for stellar rotation at the ages of the extended MSTO clusters. The models of \citet{Yang13} predict that the turn-off of rotating stars is redder and fainter for ages between 0.8 and 2.2 Gyr and therefore appear older than non-rotating stars. This effect, however, reverses at ages older than about 2.4 Gyr, where rotating MSTO stars are bluer and brighter than their non-rotating counterparts. Whereas stellar rotation seems to be able to broaden the turn-off region in the way it is observed, it is not clear how it affects the SGB and the red clump. 
 
\citet{BrandtHuang15a} used the \textsc{syclist} rotating stellar models \citep{Georgy13} to compute synthetic photometry of stars of varying rotation rates. They were able to reproduce the extended MSTO features in clusters between 1 and 2 Gyr with stars at different rotation velocities and viewing angles. Also,
\citet{BrandtHuang15b} analyzed the two younger (800 Myr) nearby Hyades and Praesepe open clusters. Both clusters show an extended MSTO feature that seems not to be in agreement with a coeval population (e.g. \citealt{Eggen98}). Because the clusters have only stellar masses of $\sim$400 $\mathrm{M_{\odot}}$ (Hyades, \citealt{Perryman98}) and $\sim$500 $\mathrm{M_{\odot}}$ (Praesepe, \citealt{KrausHillenbrand07}), they would not fit the \citet{Goudfrooij14} scenario. \citet{Piatti15} have found extended main sequences in a sample of four low mass intermediate-age LMC clusters ($<$5000 M$_{\odot}$), suggesting that cluster mass does not determine whether a cluster can host an extended MSTO.

\citet{BrandtHuang15b} showed that the spread at the turn-off can be explained with stellar rotation without the need of a prolonged SFH. Similarly, \citet{Niederhofer15b} compared rotating and non-rotating isochrones from the \textsc{syclist} models. They found that stars which rotate at different velocities in a coeval population can produce an extended MSTO feature. The predicted spread in the turn-off is proportional to the age of the cluster and seems to follow closely the observations.

\subsection{Young Clusters}

There might also be other effects present in the CMDs of clusters that are not in agreement with an SSP and which we do not completely understand.
\citet{Milone13b} studied the young ($\sim$150 Myr) moderately massive (5.0~$\times~10^3\mathrm{M_{\odot}}$, \citealt{Baumgardt13}) cluster NGC~1844 in the LMC using HST photometry. The main sequence of this cluster shows broadened features which are shown to be an intrinsic property of the cluster.  The authors tried to explain the broadening using various approaches, including stellar rotation and multiple populations, both including different ages or spreads in metallicity, He content and C-N-O abundance. While \citet{Milone13b} excluded multiple stellar populations to cause the broadened main sequence, other explanations, like stellar rotation, also yield unsatisfying results in their simple approach.  
This work shows that there are still observations of CMDs that currently are not well reproduced by stellar models.

\citet{Milone15} and \citet{Correnti15} found evidence that NGC~1856 which is only 300 Myr old has an extended MSTO, as well. This is the first detection of a broadened turn-off in a cluster younger than 1 Gyr. The studies found that the width of the turn-off would correspond to an age spread of 180 Myr \citep{Correnti15} or 150 Myr \citep{Milone15}. 
 Additionally, \citet{Milone15} discovered that the main sequence of NGC~1856 splits in two parts beneath the turn-off which could be due to a spread in age or due to chemical abundance variations. 
\citet{D'Antona15} analyzed the CMD of NGC~1856 concentrating on the split of the main-sequence and the extended turn-off region. They found that these features can also be explained with a rapidly rotating and a non-rotating population both having the same age.
These detections provide evidence that also young clusters are not SSPs, however their deviations are smaller than the ones from intermediate-age clusters. Therefore it is tempting to claim that what ever causes this anomaly, changes with the age of the cluster.

\subsection{Summary}

Taking all facts together, it is still unclear what causes the extended MSTO. With 8-10m class ground-based facilities it will be possible to test the rotating stars scenario and measure directly the projected rotation velocity $vsin(i)$ of stars along the spread of the MSTO. Such measurements will give a definitive answer how much rotating stars contribute to the spread in color and magnitude at the end of their main sequence lifetime.

\begin{acknowledgements} 

We are grateful to Paul Goudfrooij and Antonino Milone for providing their photometry and catalogs.
We thank the anonymous referee for useful comments and suggestions that helped to improve the manuscript.
We thank Aaron Dotter for his useful comments.
This research was supported by the DFG cluster of excellence "Origin and Structure of the Universe".
NB is partially funded by a Royal Society University Research Fellowship.
V. K-P. is greatly appreciated to Jay Anderson  for sharing with us his 
ePSF library PSF fitting software. 	
Support for this work was provided by NASA through grant number AR-12642 from the Space Telescope Science Institute, which is operated by AURA, Inc., under NASA contract NAS 5-26555.	

\end{acknowledgements}

\begin{appendix}
\section{Artificial Clusters\label{sec:artificial clusters}}

\begin{figure*}
 \centering
 \begin{tabular}{lll}
  \includegraphics[width=5.75cm]{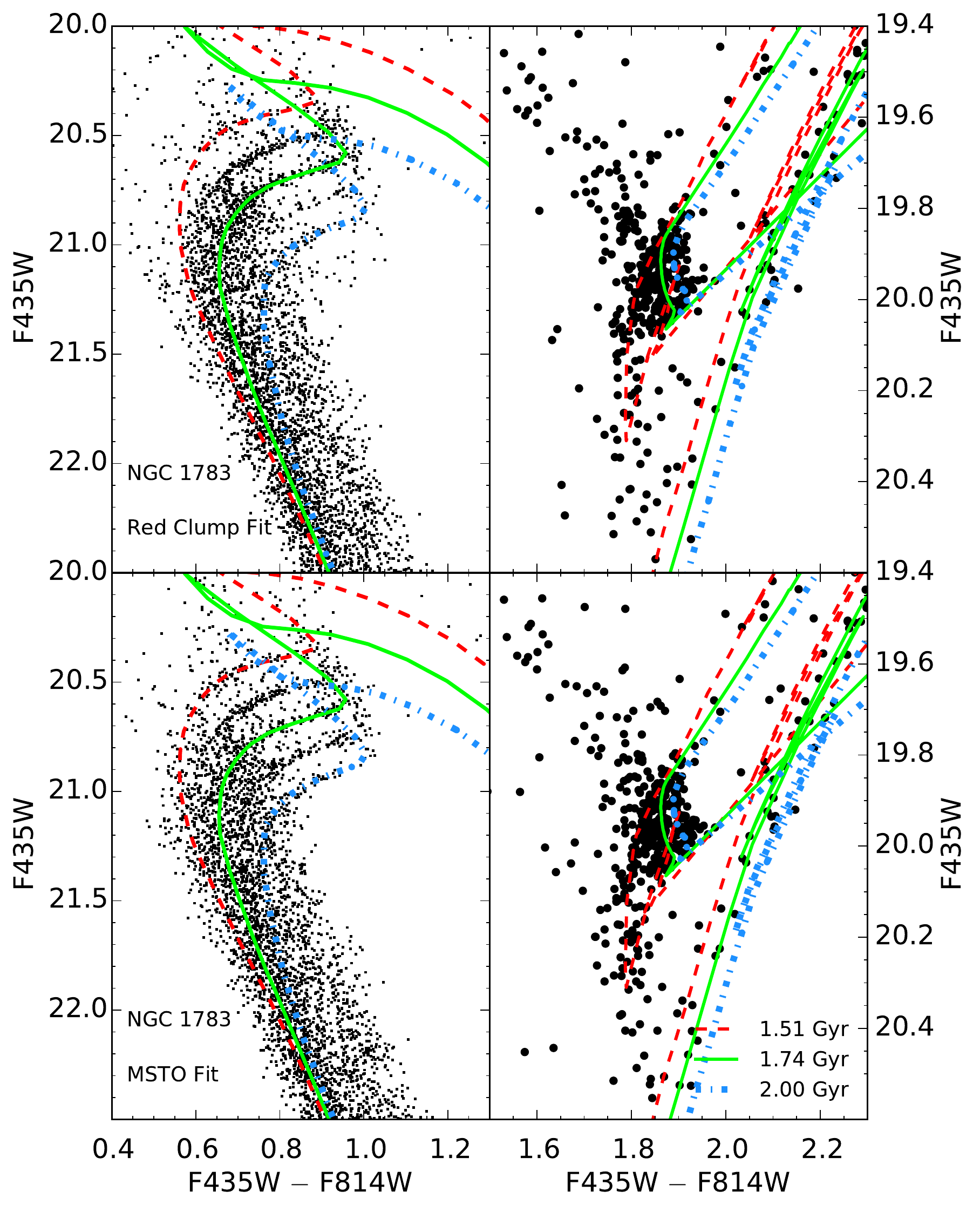} &
\includegraphics[width=5.75cm]{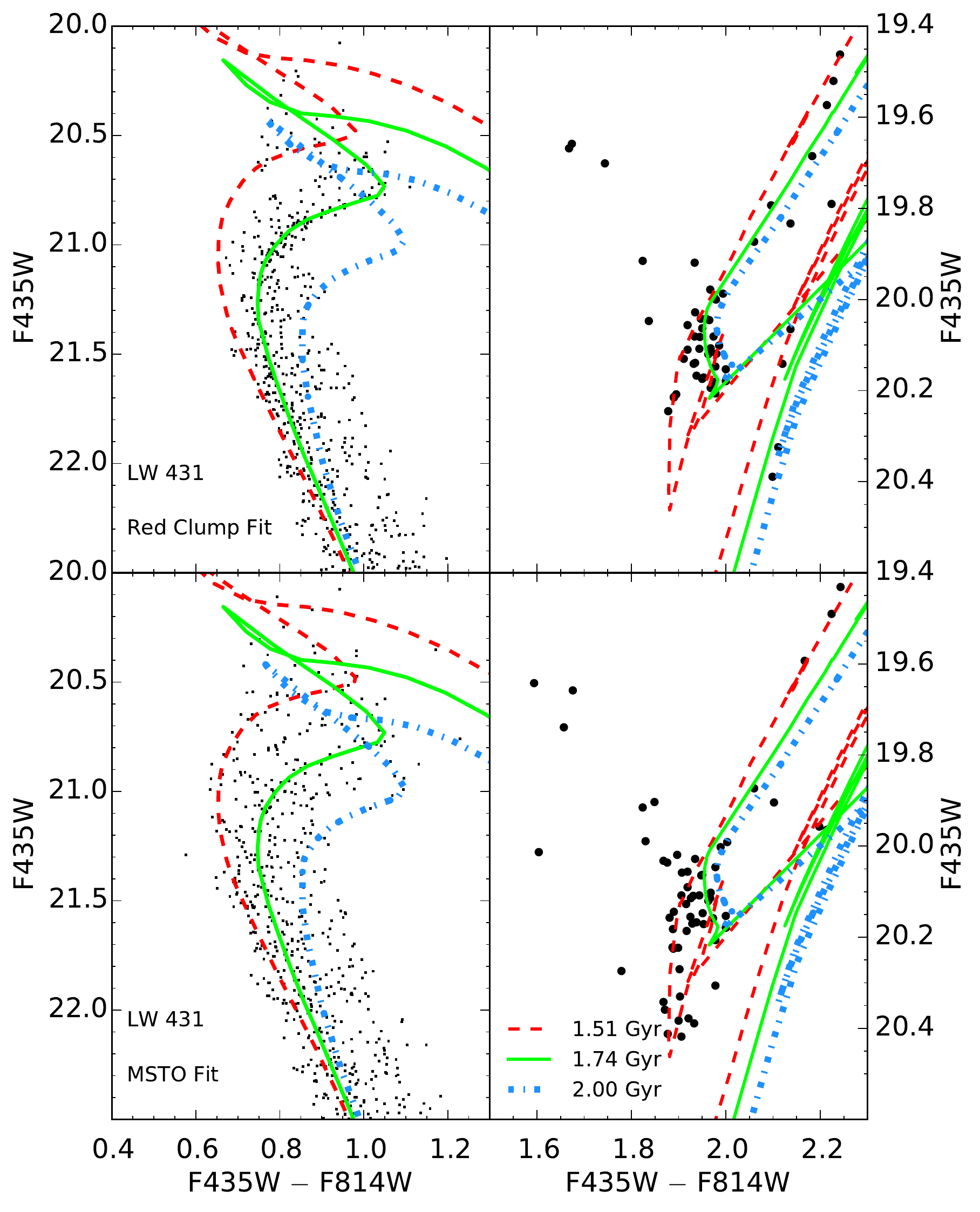} &
\includegraphics[width=5.75cm]{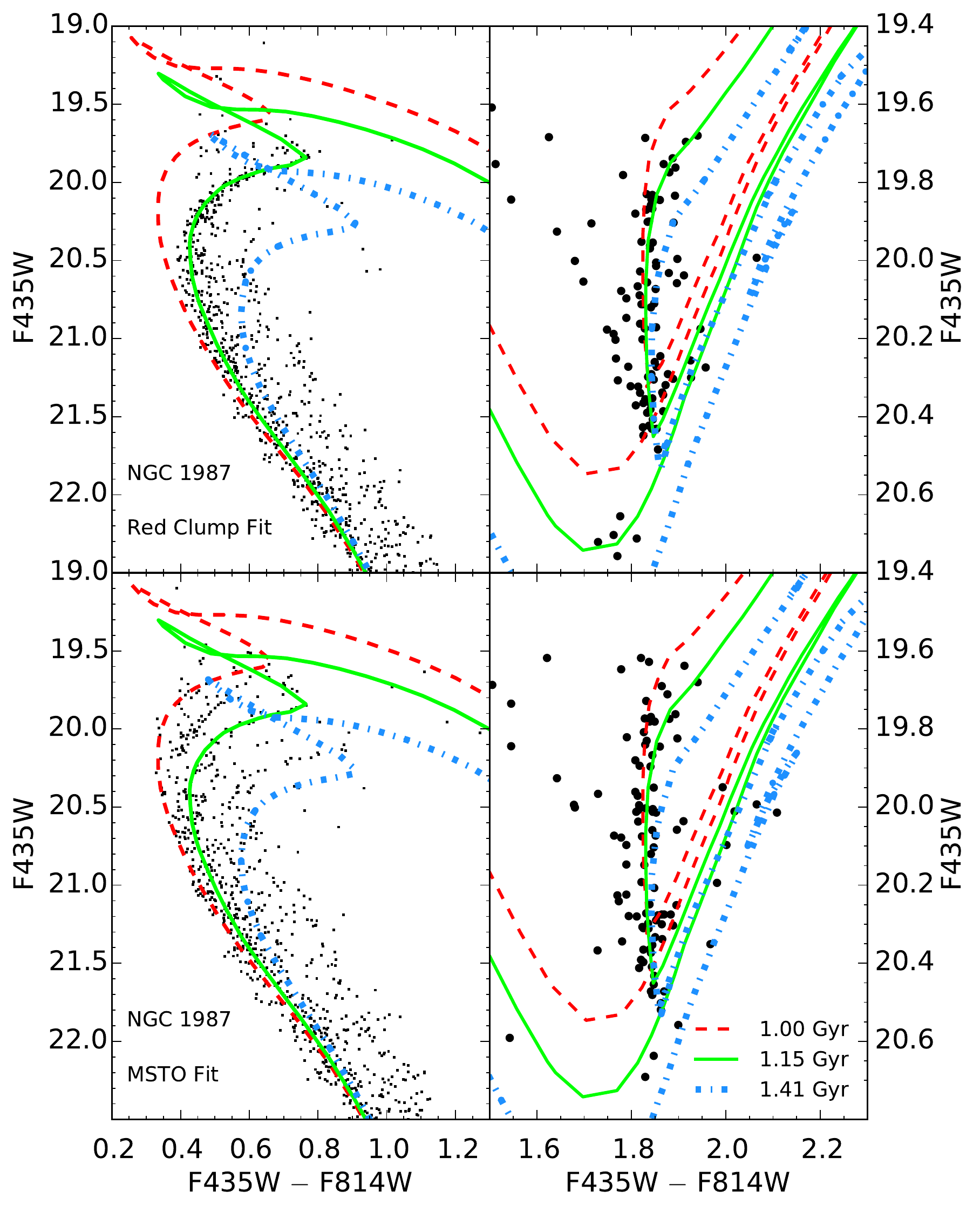} \\
\includegraphics[width=5.75cm]{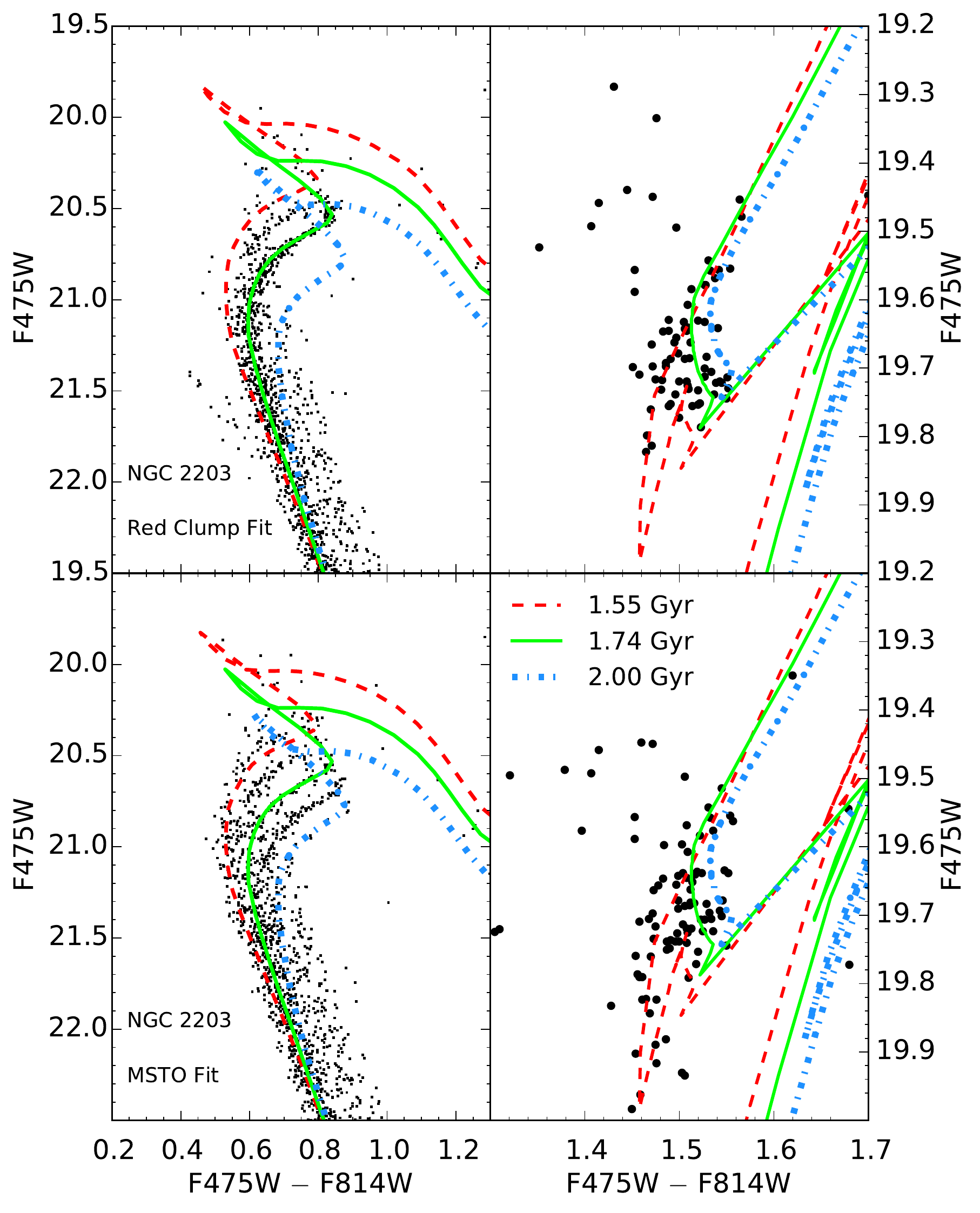} &
\includegraphics[width=5.75cm]{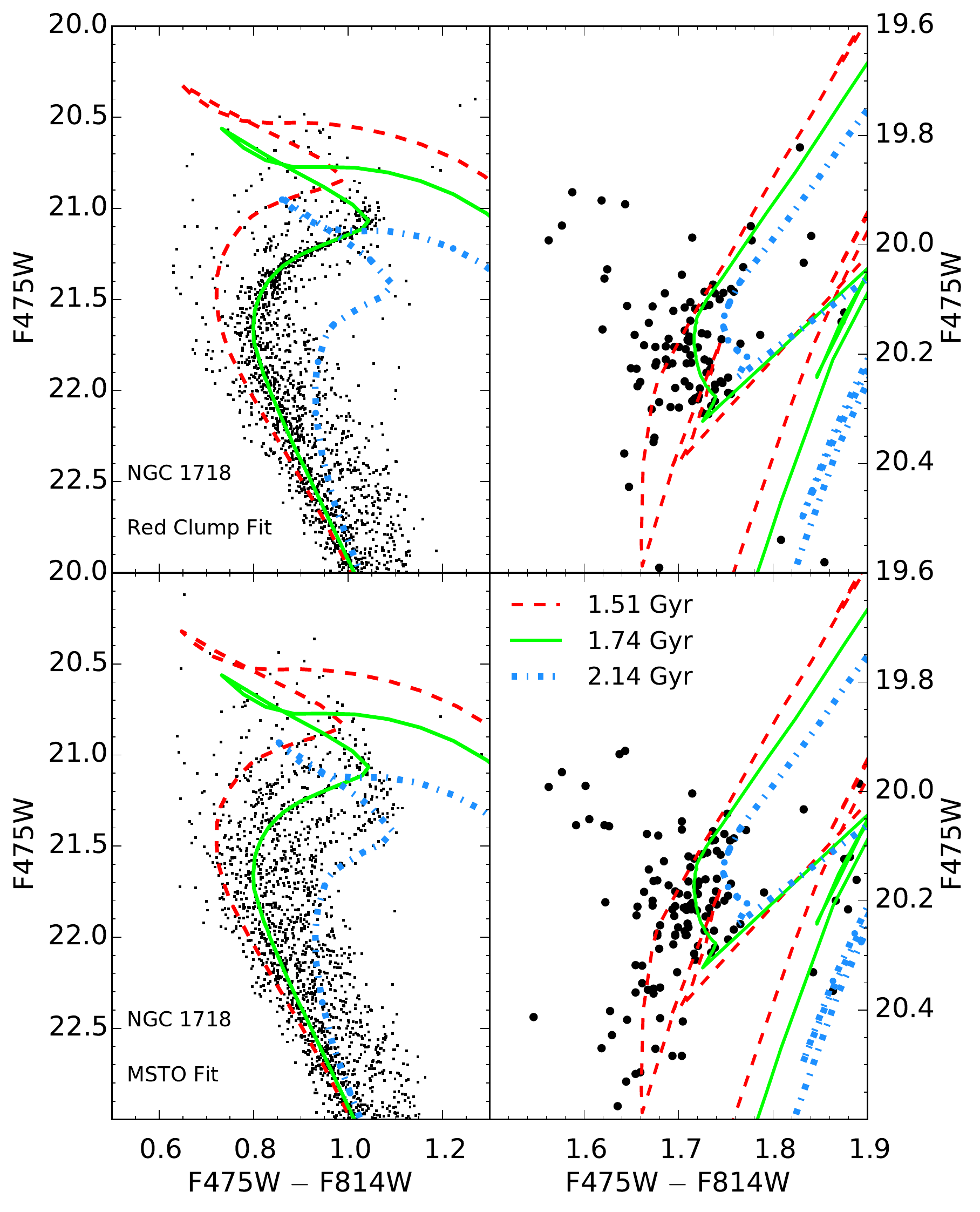} &
\includegraphics[width=5.75cm]{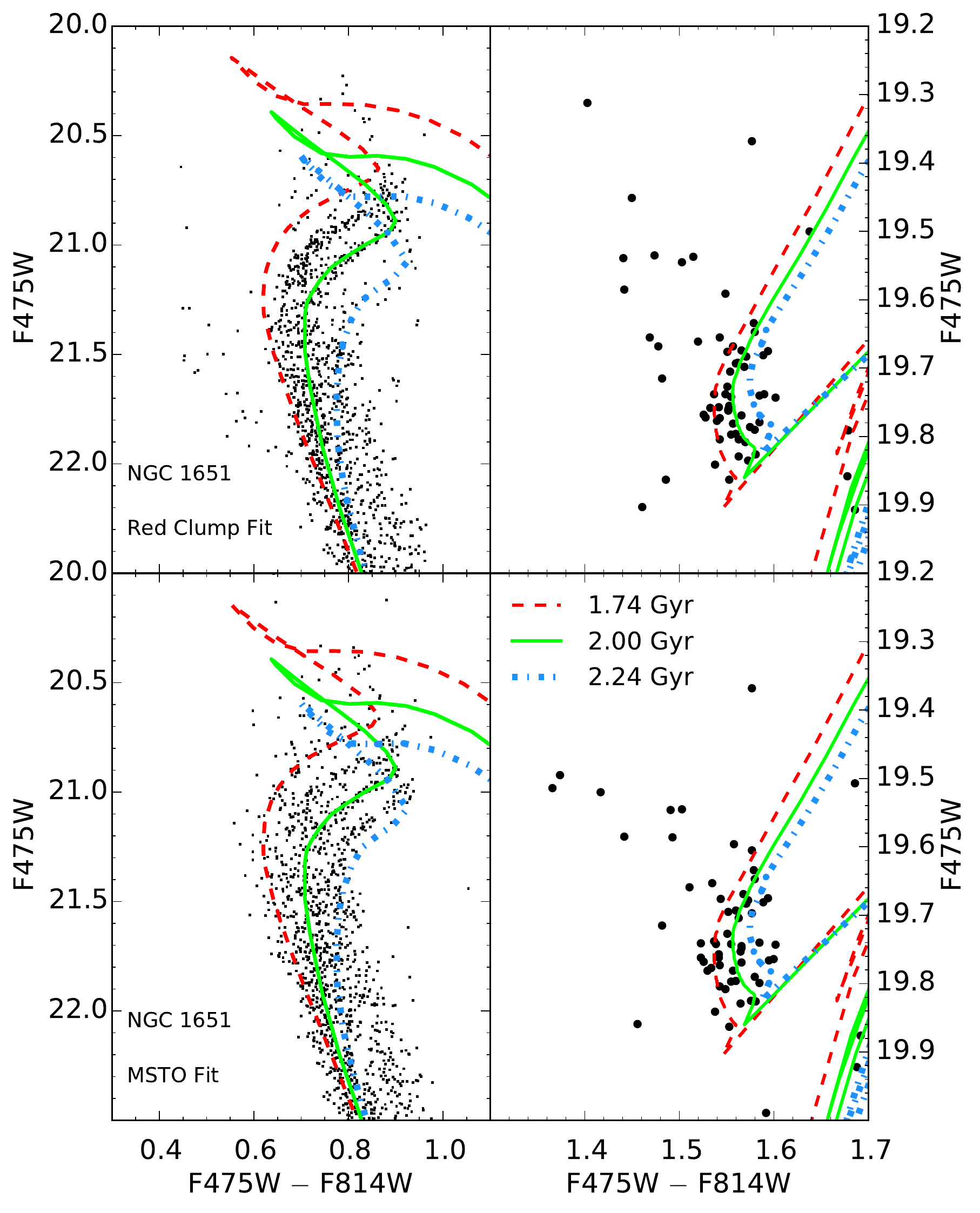} \\
   \\
  \end{tabular}            
  \caption{CMDs of artificial clusters constructed from the fitted SFHs of our sample of LMC clusters. The upper two panels of each subplot show clusters created out of the fitted SFH of the red clump region, whereas the lower panels are synthetic clusters with the recovered SFH from the MSTO region. All CMDs have the same limits as the CMDs of the real counterparts shown in Figures \ref{fig:ngc1783_cmd} to \ref{fig:ngc1846_cmd}. The superimposed isochrones are also the same.}
   \label{fig:repop}
\end{figure*}

\begin{figure*}
 \centering
 \begin{tabular}{lll}
\includegraphics[width=5.75cm]{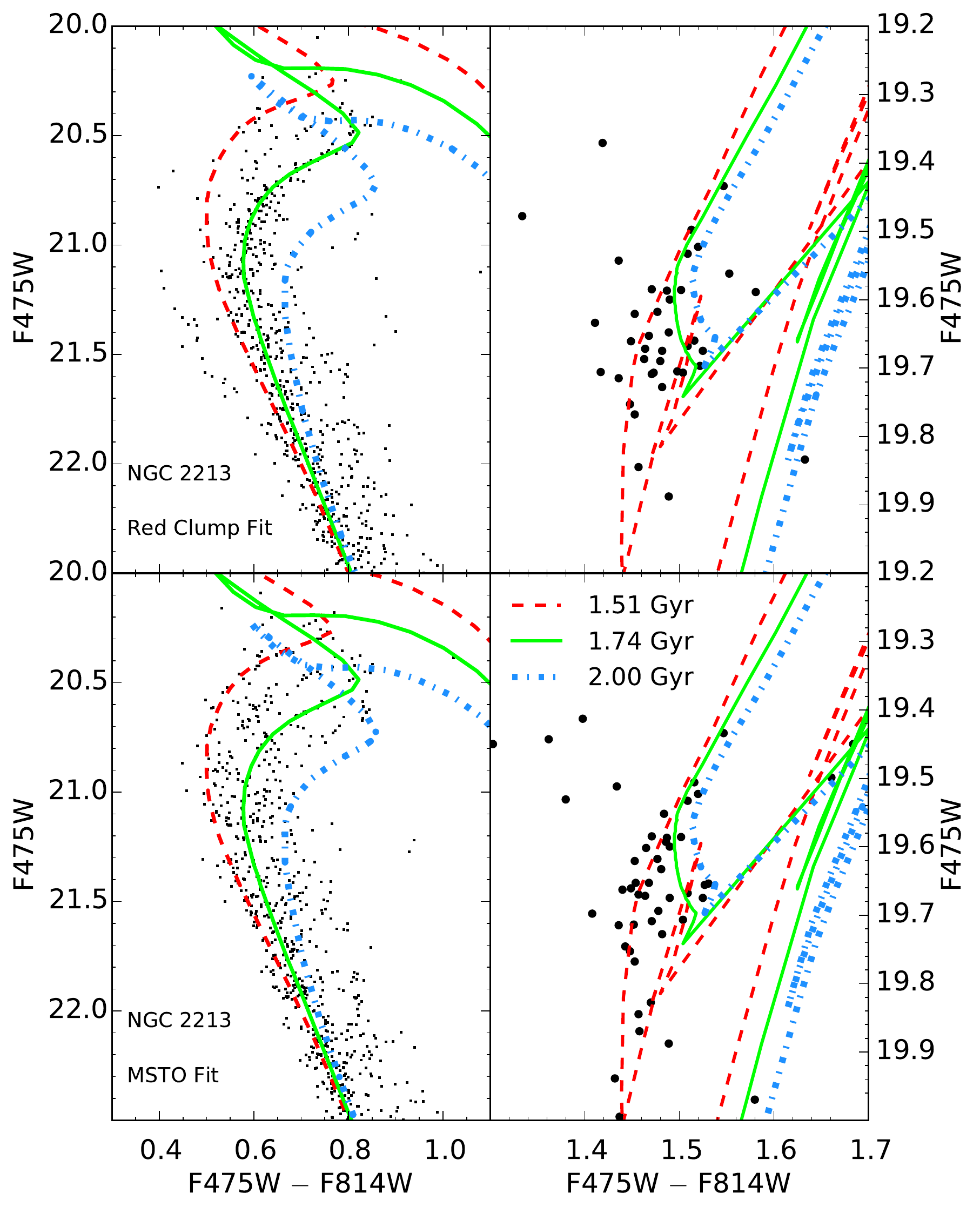} &
\includegraphics[width=5.75cm]{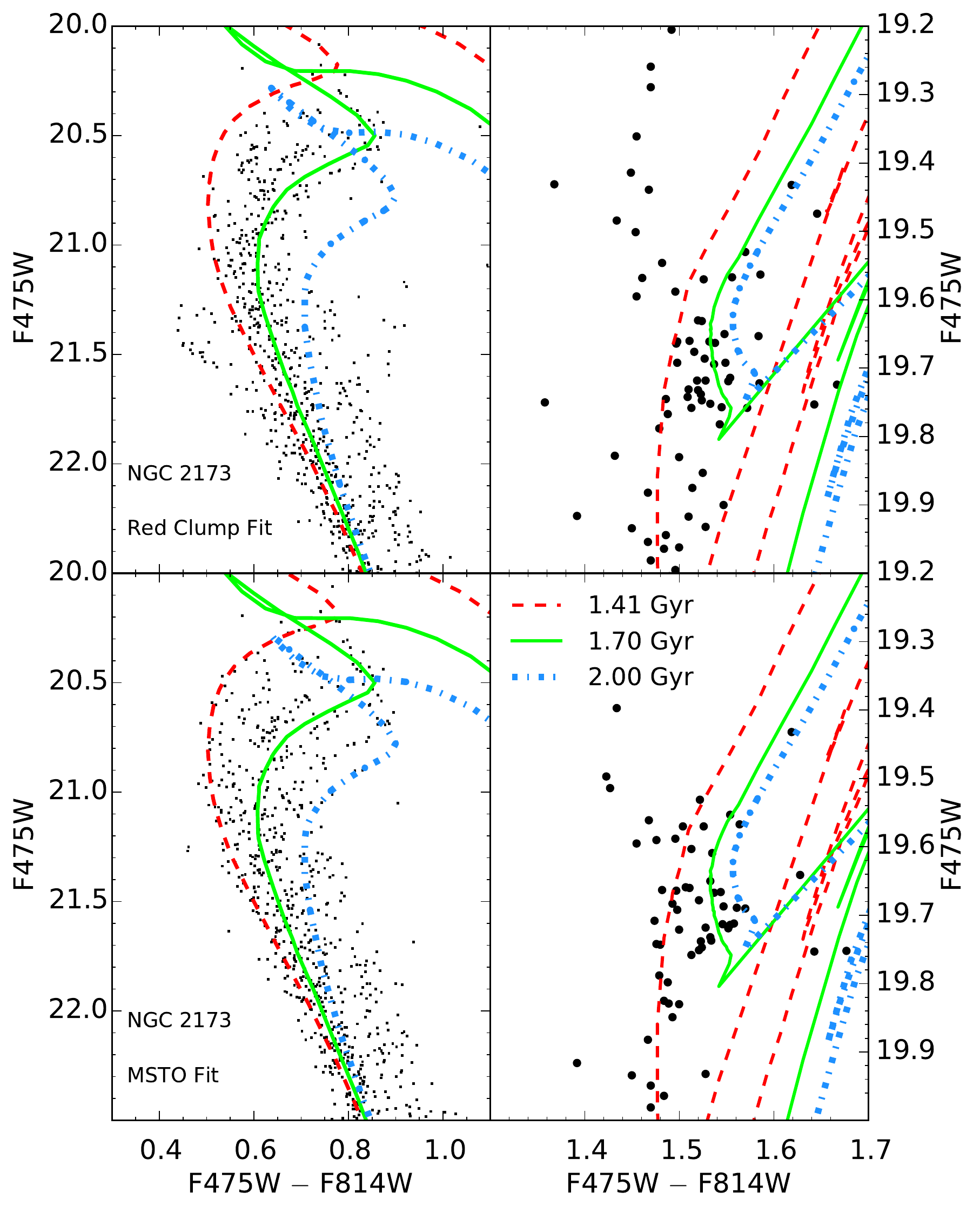} &
\includegraphics[width=5.75cm]{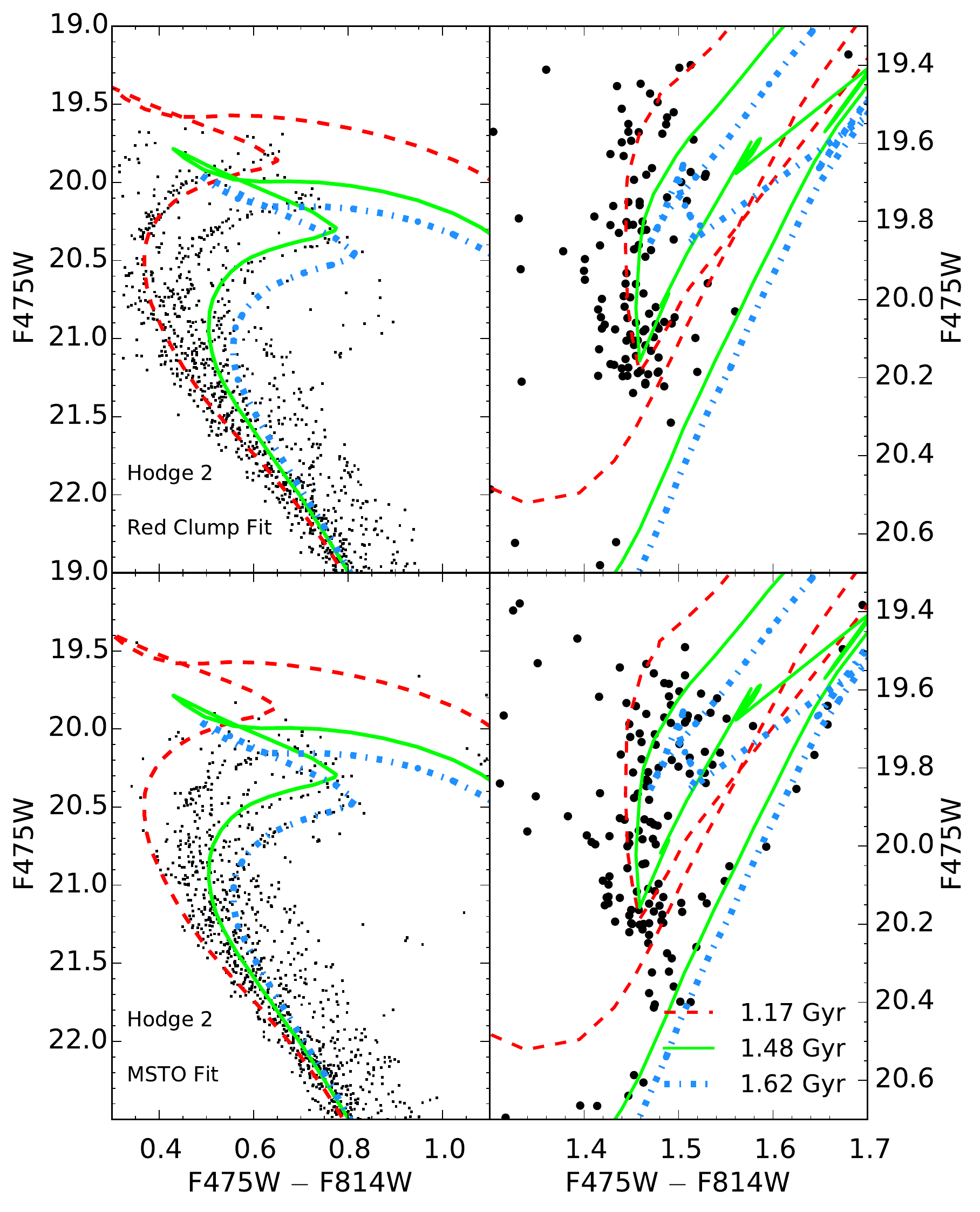} \\
\includegraphics[width=5.75cm]{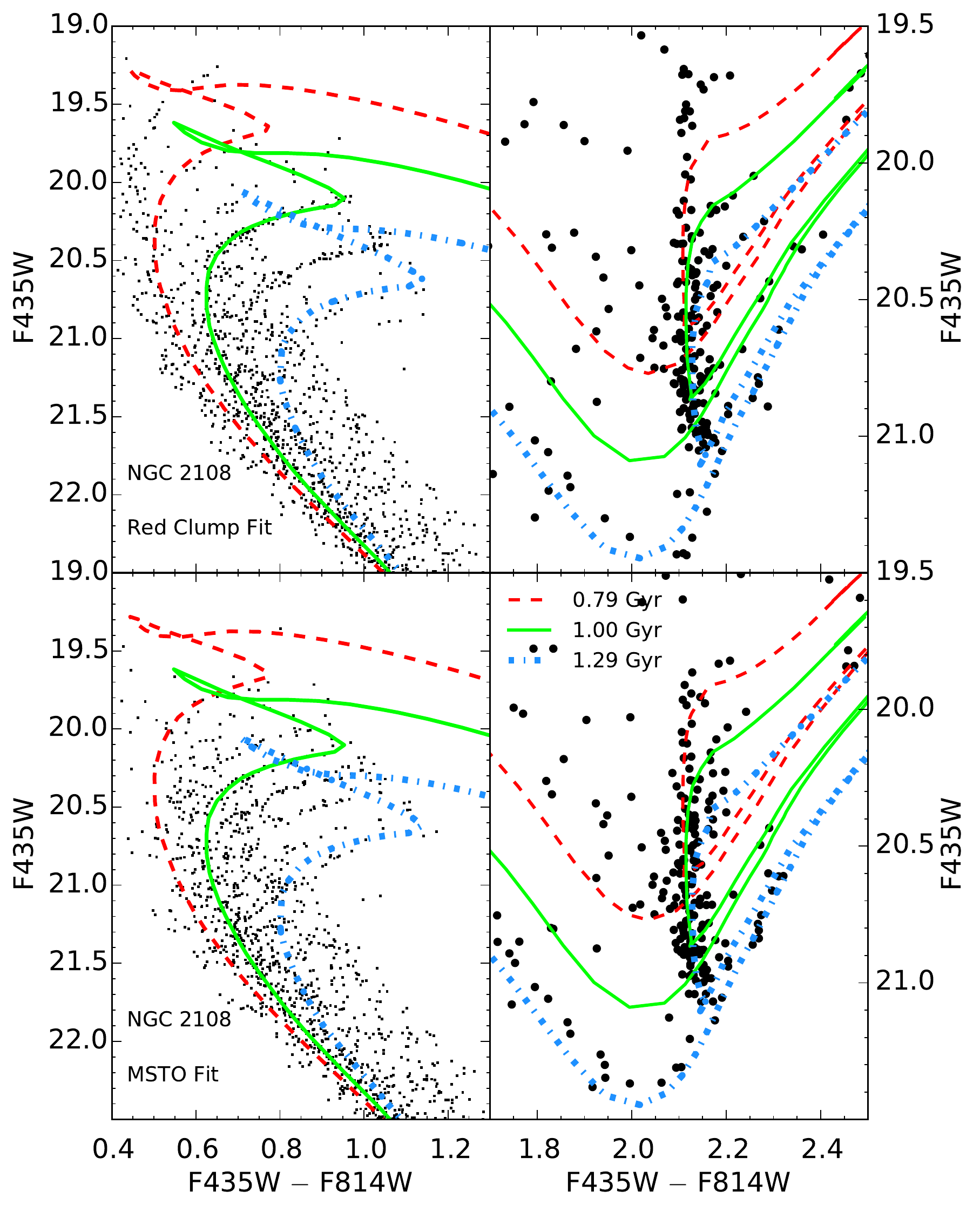} &
\includegraphics[width=5.75cm]{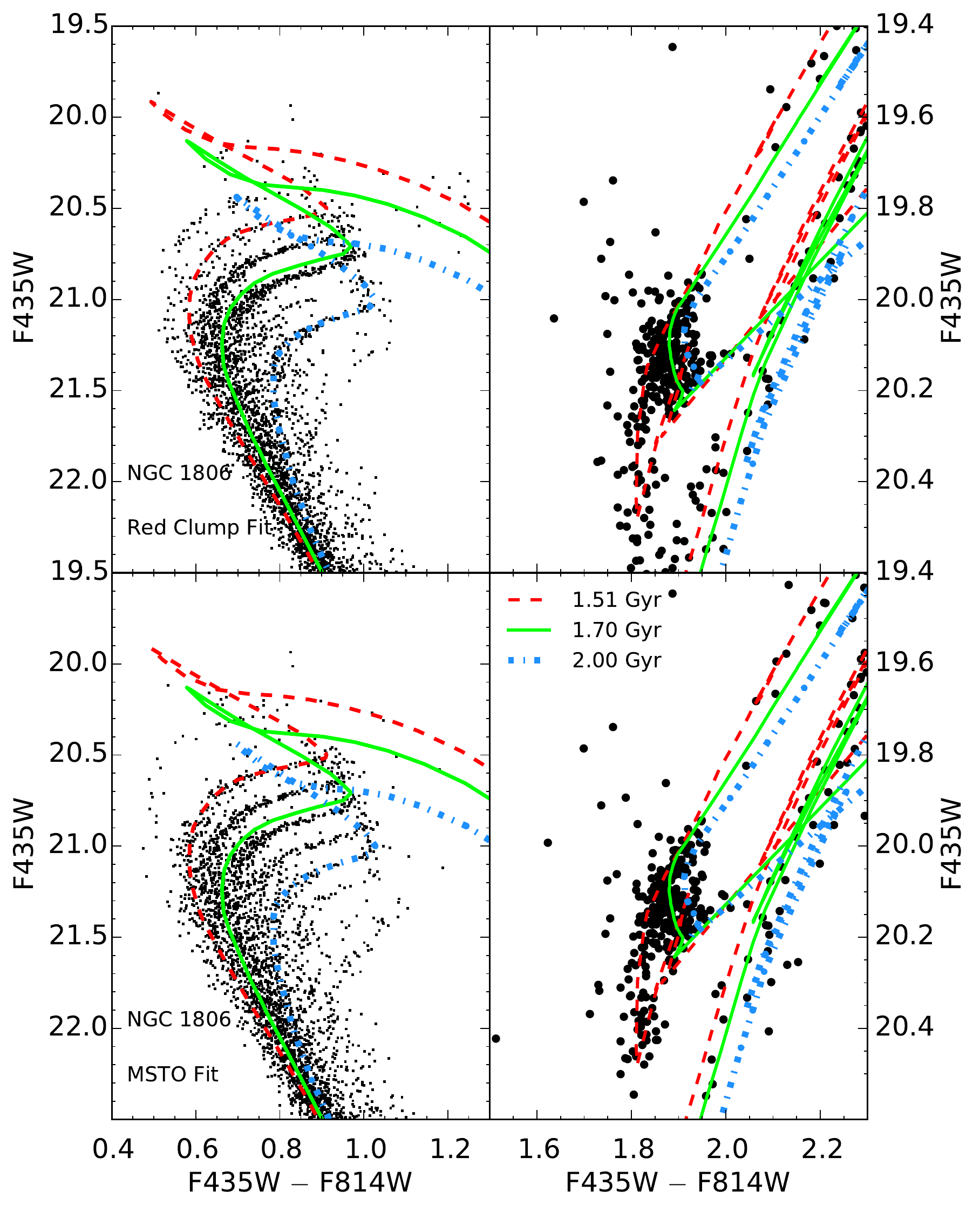} &
\includegraphics[width=5.75cm]{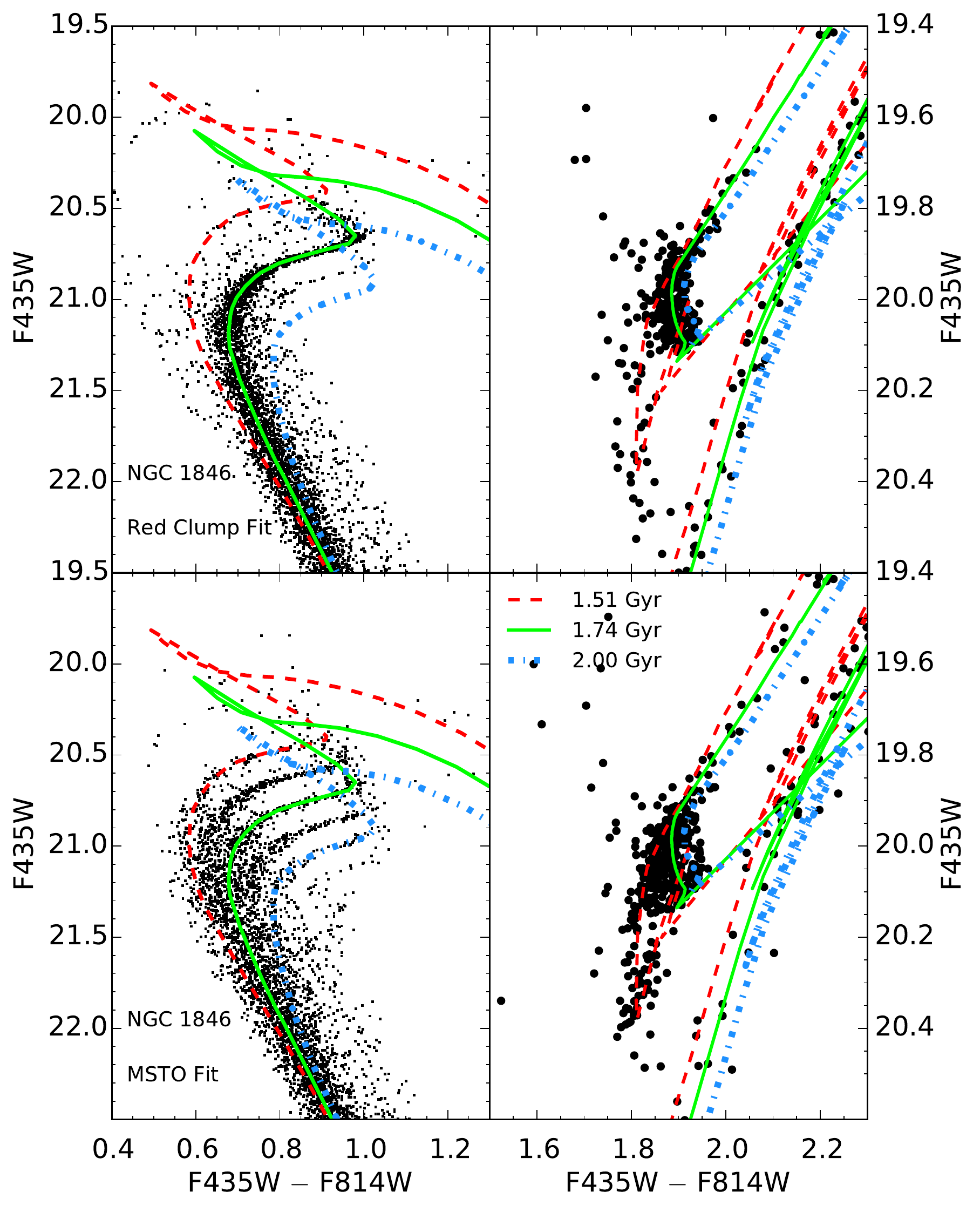} 
   \\
  \end{tabular}            
  \caption*{Figure \ref{fig:repop} continued}
\end{figure*}

\end{appendix}

\end{document}